\documentclass[twocolumn,showpacs,preprintnumbers,amsmath,amssymb]{revtex4-1}
\usepackage{graphicx}% Include figure files
\usepackage{dcolumn}% Align table columns on decimal point
\usepackage{bm}% bold math
\usepackage{braket}
\usepackage[usenames,dvipsnames]{color}% use as \textcolor{red}{******} 

\begin{document}

%%%%%%%%%%%%%%%%%%%%
%%%%%%%%%%%%%%%%%%%%
\title{Exciton-phonon cooperative mechanism of the triple-$\bm{q}$ charge-density-wave \\ and antiferroelectric electron polarization in TiSe$_2$}

\author{Tatsuya Kaneko$^1$, Yukinori Ohta$^2$,  and Seiji Yunoki$^{1,3,4}$}
\affiliation{
$^1$Computational Condensed Matter Physics Laboratory, RIKEN, Wako, Saitama 351-0198, Japan\\
$^2$Department of Physics, Chiba University, Chiba 263-8522, Japan\\
$^3$Computational Quantum Matter Research Team, RIKEN Center for Emergent Matter Science (CEMS), Wako, Saitama 351-0198, Japan\\
$^4$Computational Materials Science Research Team, RIKEN Center for Computational Science (R-CCS), Kobe, Hyogo 650-0047, Japan}
%$^4$Computational Materials Science Research Team, RIKEN Advanced Institute for Computational Science (AICS), Kobe, Hyogo 650-0047, Japan}

\date{\today}

\begin{abstract}  
We investigate the microscopic mechanisms of the charge-density-wave (CDW) formation in a monolayer TiSe$_2$ 
using a realistic multiorbital $d$-$p$ model with electron-phonon coupling and intersite Coulomb (excitonic) 
interactions. 
First, we estimate the tight-binding bands of Ti $3d$ and Se $4p$ orbitals in the monolayer TiSe$_2$ on the basis 
of the first-principles band-structure calculations.  We thereby show orbital textures of the undistorted band structure 
near the Fermi level.  
Next, we derive the electron-phonon coupling using the tight-binding approximation and show that the softening 
occurs in the transverse phonon mode at the M point of the Brillouin zone.  The stability of the triple-$\bm{q}$ 
CDW state is thus examined to show that the transverse phonon modes at the M$_1$, M$_2$, and M$_3$ 
points are frozen simultaneously.  
Then, we introduce the intersite Coulomb interactions between the nearest-neighbor Ti and Se atoms that lead 
to the excitonic instability between the valence Se $4p$ and conduction Ti $3d$ bands.  Treating the 
intersite Coulomb interactions in the mean-field approximation, we show that the electron-phonon and excitonic 
interactions cooperatively stabilize the triple-$\bm{q}$ CDW state in TiSe$_2$.   
We also calculate a single-particle spectrum in the CDW state and reproduce the band folding spectra observed in 
photoemission spectroscopies.  
Finally, to clarify the nature of the CDW state, we examine the electronic charge density distribution 
and show that the CDW state in TiSe$_2$ is of a bond-type and induces a vortex-like antiferroelectric 
polarization in the kagom\'e network of Ti atoms.  
\end{abstract}

\maketitle

%%%%%%%%%%%%%%%%%%%%
%%%%%%%%%%%%%%%%%%%%
\section{Introduction}

Transition-metal dichalcogenides (TMDs)~\cite{WY69,CSEetal13} are representative materials that show the 
charge-density-wave (CDW) states~\cite{Mo86,Ro11}.  
The majority of the group-IV (Ti, Zr, Hf) TMDs are simple $d^{0}$ semiconductors, in which the Fermi level is located 
between the valence chalcogen $p$ and conduction transition-metal $d$ bands~\cite{CSEetal13}.  
However, 1$T$-TiSe$_2$, one of the group IV TMDs, is either a slightly band-overlap semimetal or a small band-gap 
semiconductor~\cite{ZF78,FGH97}, which is the only material that shows the CDW transition among this 
group \cite{WBMetal76,DMW76,HZHetal01}.  
Thus, in contrast to the conventional (nesting induced) CDWs in low-dimensional solids~\cite{Gr88,Gr00} or 
to the CDWs in the $d^1$ TMDs~\cite{WDM74,WDM75}, a peculiar mechanism of the CDW formation should be 
expected in the $d^0$ TMD, 1$T$-TiSe$_2$.  
Furthermore, in 1$T$-TiSe$_2$, it is known that the emergence of superconductivity (SC) with melting of 
the CDW is caused by 
intercalation~\cite{MZDetal06,MLOetal07,LWCetal07,MWZetal10,GXHetal10,LYPetal16,SKJetal16,SNHetal17}, 
applying pressures~\cite{KSBetal09,JCGetal14}, or 
carrier doping~\cite{LOLetal16,LXTetal16}.  
Therefore, clarifying the origin of the CDW is significant also for the elucidation of the mechanism of its SC.  

Because the electronic band structure of TiSe$_2$ is located near the semimetal-semiconductor phase 
boundary, its CDW phase has been investigated as a candidate for the excitonic phase~\cite{Wi77,TMSetal78}.  
This phase is also referred to as an excitonic insulator state, where a spontaneous hybridization between 
the orthogonal valence and conduction bands occurs by the interband Coulomb 
interaction to open the band gap~\cite{KK65,De65,JRK67,Ko67,HR68,HR68-2}.  
Studies of the excitonic phases have recently been developed in terms of localized orbital models appropriate 
for strongly correlated electron systems~\cite{SEO11,KSO12,EKOetal14,KA14-1,KA14-2,Ku15,NWNetal16} and adaptation of the excitonic theory for real materials is desired.
Because TiSe$_2$ as well as another candidate 
material~Ta$_2$NiSe$_5$~\cite{DCFetal86,WSTetal09,KTKetal13,SWKetal14,YDO16,LKLetal17} 
are among transition-metal compounds, 
the orbital textures and Coulomb interactions between the local orbitals may be essential factors in considering 
their electronic properties.  In fact, photoemission spectroscopies and related theoretical analyses have 
suggested that the excitonic mechanism can be applied for the CDW formation in TiSe$_2$ 
\cite{PHBetal00,KMCetal02,CMCetal07,LHQetal07,QHWetal07,ZOWetal07,MCCetal09,MSGetal10,
MSGetal10-2,MBJetal11,CCHetal12,MMAetal12,MMAetal12-2,MMHetal15}. 

The phononic mechanism (or the band-type Jahn-Teller mechanism) of the CDW formation
has also been suggested~\cite{Hu77,WL77}, where the CDW transition around $T=200$~K associated with 
the 2$\times$2$\times$2 periodic lattice displacement (PLD)~\cite{WBMetal76,DMW76,HZHetal01} is essentially 
explained by the electron-phonon coupling.  
Microscopic theory of the phononic mechanism was developed by Motizuki and 
coworkers~\cite{YM80,TM80,MSYetal81,MYT81,SYM84,SYM85,Mo86} using the realistic crystal and electronic structures 
of 1$T$-TiSe$_2$.  The realization of the 2$\times$2$\times$2 PLD was thereby explained quantitatively.  
Recent first-principles phonon calculations~\cite{CM11,BCM15,DBS15,FHYetal16,SHTetal17,HBBetal17} 
have also predicted consistent results with those of Motizuki {\it et al.}.  
Experimentally, the lattice dynamics and phonon softening corresponding to the superlattice formation 
have been studied by the Raman and infrared spectroscopy~\cite{HWKetal77,SMUetal80,SKCetal03,BKKetal08,GKWetal12,DRHetal17}, 
as well as by the inelastic neutron and  x-ray scattering experiments~\cite{SDCetal76,WSWetal78,Ja79,WRCetal11,MRHetal16}.  

Thus the two different driving forces for the CDW formation, i.e., excitonic and phononic forces, have been 
suggested in TiSe$_2$, of which the determination is still controversial.  Recent theoretical studies have 
also suggested that the electron-phonon coupling and excitonic interactions cooperatively stabilize 
the CDW state~\cite{vWNS10-2,ZFBetal13,WSY15,KZFetal15}.  
These studies, however, do not assume the electron-phonon couplings with realistic phonon modes 
corresponding to the experimentally observed PLD.  The studies by Motizuki {\it et al.} and first-principles 
phonon calculations, on the other hand, do not assume the excitonic ordering induced by the interband 
Coulomb interaction.  
In addition, local orbital textures of the CDW in TiSe$_2$ have not been investigated in detail.
Therefore, to elucidate the origin and local structure of the CDW and PLD in TiSe$_2$, it is highly desired to 
develop a quantitative microscopic theory based on a realistic model that reflects the actual crystal and 
electronic orbital structures in TiSe$_2$, taking into account both the phononic and excitonic interactions.  

Motivated by these developments in the field, here we investigate the microscopic 
mechanisms and electronic structures of the CDW phase in a monolayer TiSe$_2$ on the basis of the 
realistic multi-orbital $d$-$p$ model, where both the electron-phonon coupling and intersite Coulomb 
interactions are taken into account.  We thereby clarify both the phononic and excitonic 
mechanisms of the CDW transition. Although we assume the monolayer TiSe$_2$ for simplicity, 
our theoretical study will provide helpful interpretations of recent experiments on monolayer as well as  
few-layer TiSe$_2$~\cite{PGZetal15,CCFetal15,SNSetal16,CCWetal16,FHCetal17}. 

First, we construct the tight-binding bands of the Ti 3$d$ and Se 4$p$ orbitals in the monolayer 
TiSe$_2$ using the first-principles band-structure calculations.  
From the obtained energy bands in the undistorted crystal structure, we show orbital components 
of the bands and deduce the effective electronic structure near the Fermi level.  
Next, we derive the electron-phonon coupling in the tight-binding approximation for the transverse 
phonon modes, of which the softening has been observed experimentally~\cite{HZHetal01}.
Then, taking into account the electron-phonon coupling only, we show the softening of the transverse 
phonon mode at the M point of the Brillouin zone (BZ).  We thus discuss the instability toward the triple-$\bm{q}$ 
CDW state, where the transverse phonon modes at the M$_1$, M$_2$, and M$_3$ points are frozen 
simultaneously.  
Furthermore, we introduce the intersite Coulomb interaction between the nearest-neighbor Ti and 
Se atoms that induces the excitonic instability between the valence Se 4$p$ and conduction Ti 
3$d$ bands. We investigate the roles of the excitonic interaction in the triple-$\bm{q}$ CDW state 
using the mean-field approximation for the intersite Coulomb interactions.  
We thus show that the electron-phonon and excitonic interactions cooperatively stabilize 
the triple-$\bm{q}$ CDW state in TiSe$_2$.  
We can also show that the calculated single-particle spectrum in the 
CDW state can reproduce the band folding spectrum observed in photoemission spectroscopies.  
Finally, we examine the nature of the CDW state by calculating the change in the electron density 
distribution and predict that the CDW state in TiSe$_2$ is of a bond-centered-type, rather than a 
site-centered-type, and induces a vortex-like antiferroelectric polarization in the kagom\'e network 
of Ti atoms.  

The rest of this paper is organized as follows.  
In Sec.~II, we derive the effective eleven-orbital $d$-$p$ model for the monolayer TiSe$_2$ taking 
into account both the electron-phonon coupling and intersite Coulomb interactions.  
In Sec.~III, we show the effective electronic structure near the Fermi level in the undistorted crystal structure.  
In Sec.~IV, we present the phonon softening and instability toward the triple-$\bm{q}$ CDW state without 
taking into account the intersite Coulomb interactions.  
In Sec.~V, we briefly review the mean-field approximation for the excitonic ordering and discuss 
the roles of the Coulomb interaction for the triple-$\bm{q}$ CDW in TiSe$_2$.  
In Sec.~VI, we show the single-particle spectrum and charge density distribution in the CDW state.  
Discussions and summary are given in Sec.~VII.
Details of the calculations 
are provided in Appendices A--E.  

%%  Fig. 1  %%
\begin{figure}[t]
\begin{center}
\includegraphics[width=\columnwidth]{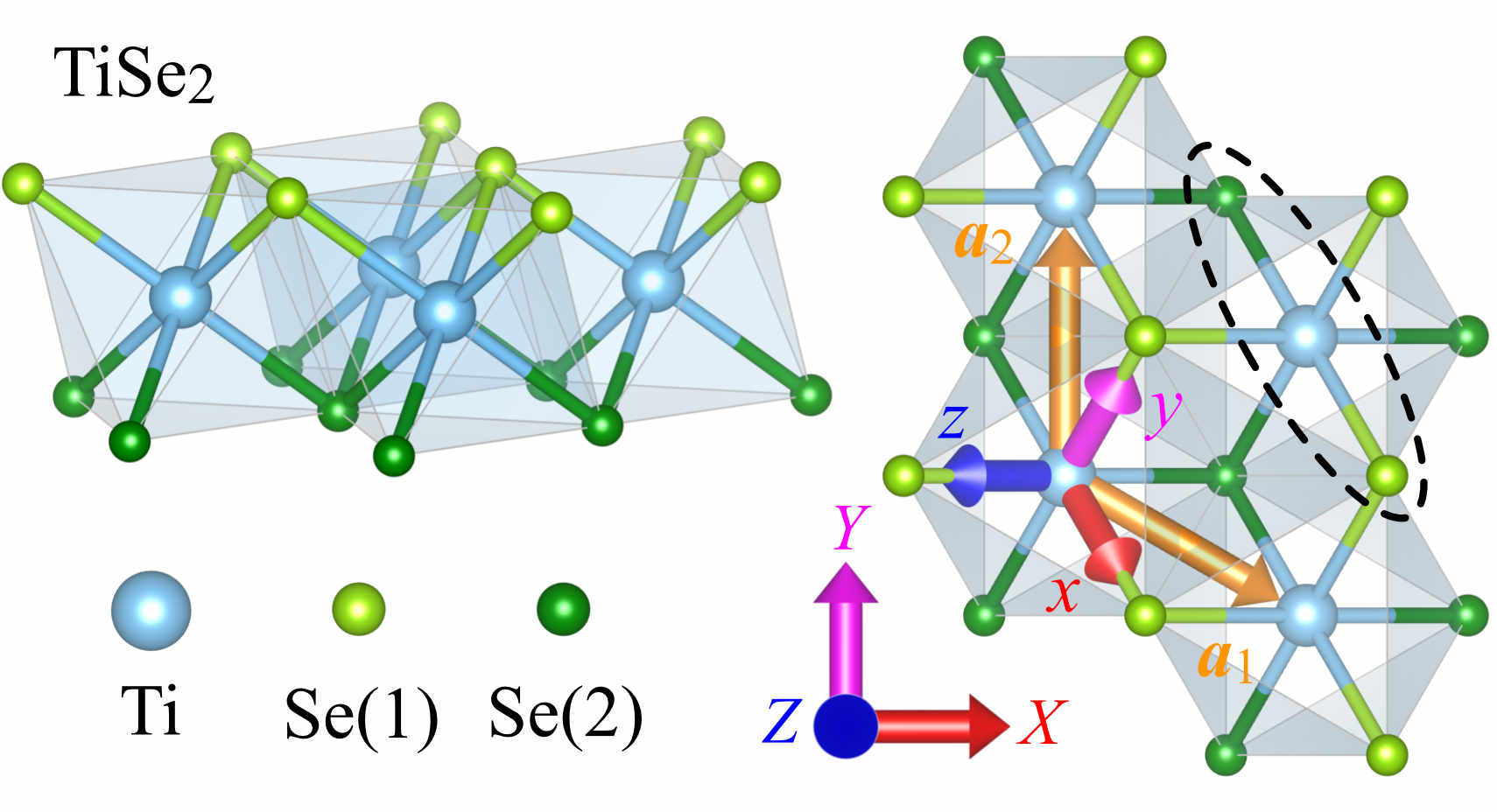}
\caption{
Schematic representations of the crystal structure of the monolayer TiSe$_2$.  
$X$-$Y$-$Z$ are the global coordinate axes and $x$-$y$-$z$ are the local coordinate axes in the 
TiSe$_6$ octahedron.  Dashed ellipse is the unit cell taken in this paper.  
$\bm{a}_1$ and $\bm{a}_2$ are the primitive translation vectors.}
\label{TiSe2_cs_fig1}
\end{center}
\end{figure}
%%%%%%%%

%%%%%%%%%%%%%%%%%%%%
%%%%%%%%%%%%%%%%%%%%
\section{Model}

First, let us construct the effective eleven-orbital $d$-$p$ model for the monolayer TiSe$_2$ 
taking into account the electron-phonon coupling and interband Coulomb interactions.  
The model enables us to consider both the phononic and excitonic mechanisms of the CDW 
transition.  The crystal structure, tight-binding bands, electron-phonon coupling, and Coulomb 
interactions in TiSe$_2$ are discussed in the following subsections.

%%%%%%%%%%%%%%%%%%%%
\subsection{Crystal structure} \label{TiSe2_effm_cs}
\begin{figure}[tb]
\begin{center}
\includegraphics[width=\columnwidth]{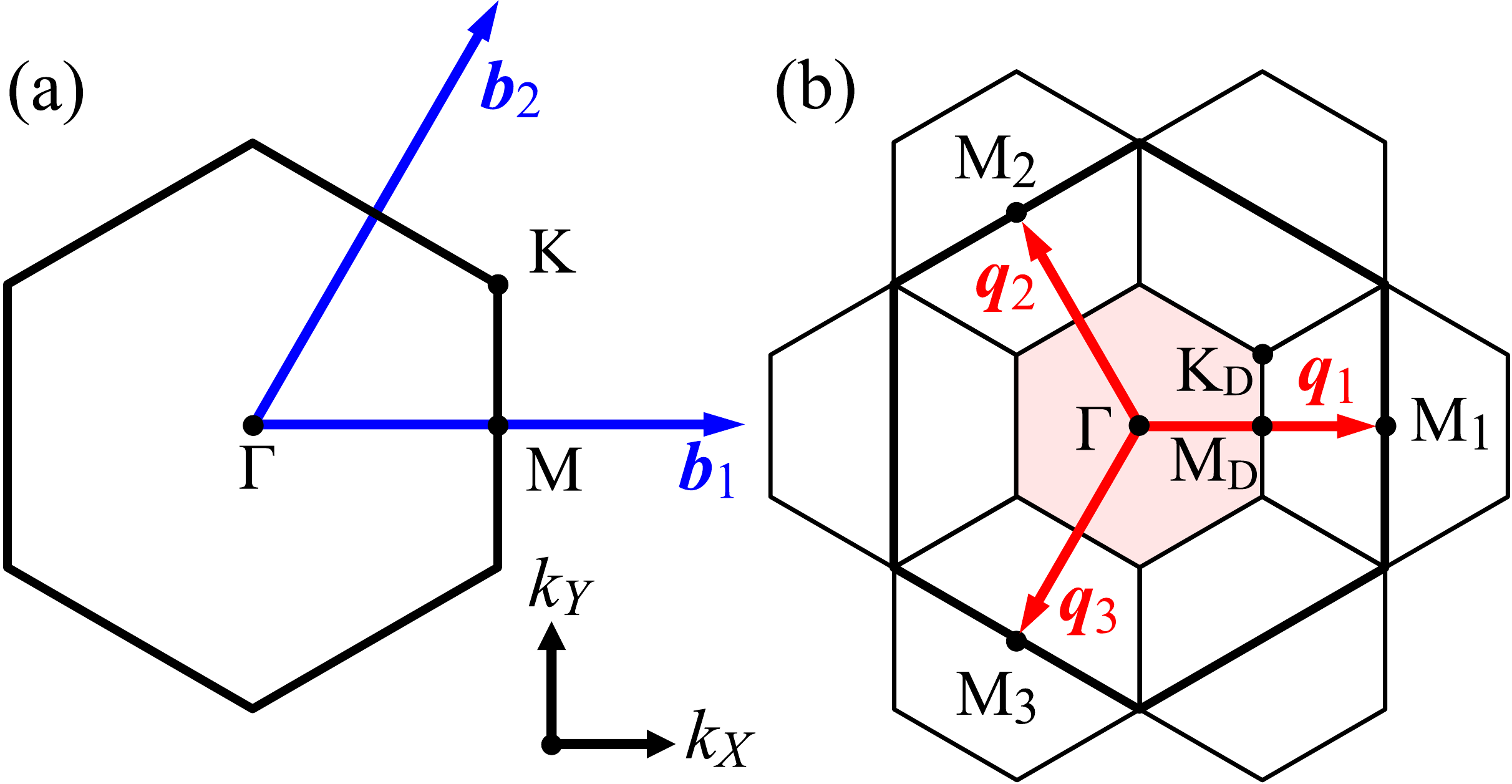}
\caption{
(a) The BZ of the monolayer TiSe$_2$, where $\bm{b}_1$ and $\bm{b}_2$ are the reciprocal primitive vectors.  
(b) Three M-points and reduced BZ (RBZ). $\bm{q}_1$, $\bm{q}_2$, and $\bm{q}_3$ corresponding to the modulation wave vectors of the CDW in TiSe$_2$.   
Red shaded area indicates the RBZ in the 2$\times$2 superlattice structure. }
\label{TiSe2_cs_fig2}
\end{center}
\end{figure}

The crystal structure of the monolayer 1$T$-TiSe$_2$ is illustrated in Fig.~\ref{TiSe2_cs_fig1}~\footnote{The crystal structures and the isosurfaces are  visualized using VESTA, K. Momma and F. Izumi, J. Appl. Crystallogr. {\bf 44} 1272 (2011).}.  
We assume the lattice constant $a = 3.54$~\AA~\cite{Ri76} and use the primitive translation 
vectors $\bm{a}_1=(\sqrt{3}a/2,-a/2)$ and $\bm{a}_2=(0,a)$ shown in Fig.~\ref{TiSe2_cs_fig1}.  
The unit cell contains one Ti ion and two Se ions, Se(1) and Se(2).  
The position of the Ti and Se ions in the unit cell are $\bm{\tau}_{\rm Ti} = (0, 0, 0)$ and 
$\bm{\tau}_{\rm Se1} =-\bm{\tau}_{\rm Se2} = (a/2\sqrt{3}, -a/2, z_{\rm Se})$ with 
$z_{\rm Se} = 1.552$~\AA, where we apply the atomic position optimization in the WIEN2k code~\cite{WIEN2k} 
to determine $z_{\rm Se}$ 
\footnote{
We optimize the internal coordinates of Se ions in the trigonal structure (space group $P\bar{3}m1$) 
with the lattice constants $a=3.54$ and $c=10$~\AA, where $c=10$~\AA~is large enough to get rid of 
the interlayer connections.  In the self-consistent calculations, we adopt the GGA~\cite{PBE96} 
and use 88 $k$-points in the irreducible part of the BZ, assuming the muffin-tin radii 
($R_{\rm MT}$) of 2.28 (Ti) and 2.28 (Se) bohr and the plane-wave cutoff of $K_{\rm max}=7/R_{\rm MT}$.  
}.  
We also illustrate the BZ of the monolayer TiSe$_2$ in Fig.~\ref{TiSe2_cs_fig2}, 
where the reciprocal primitive vectors are given by $\bm{b}_1=(4\pi / \sqrt{3}a,0)$ and 
$\bm{b}_2=(2\pi / \sqrt{3}a,2\pi / a)$.  

%%  Fig. 2  %%
%%%%%%%%

%%%%%%%%%%%%%%%%%%%%
\subsection{Tight-binding bands}
We use the energy bands in the tight-binding (TB) approximation as a noninteracting band structure.  
The Hamiltonian of the TB bands is given by 
\begin{align}
\mathcal{H}_{e} = 
 \sum_{\bm{k}} \sum_{\mu \ell, \nu m} t_{\mu \ell, \nu m} (\bm{k}) c^{\dag}_{\bm{k},\mu \ell} c_{\bm{k},\nu m}, 
\label{TiSe2_effm_eq1}
\end{align}
where $c^{(\dag)}_{\bm{k},\mu \ell}$ is the annihilation (creation) operator of an electron in orbital $\ell$ 
of atom $\mu$ at momentum $\bm{k}$.  We do not write the spin index explicitly in this paper.  
$t_{\mu \ell, \nu m} (\bm{k})$ is the Fourier transform of the transfer integral 
\begin{align}
 t_{\mu \ell, \nu m} (\bm{k}) = \sum_{\bm{R}_n} t_{\mu \ell, \nu m} (\bm{R}_n)e^{-i \bm{k}\cdot \bm{R}_n}. 
 \label{TiSe2_effm_eq2}
\end{align}
$t_{\mu \ell, \nu m} (\bm{R}_n)$ is the transfer integral between the atomic orbitals $\mu \ell$ 
and $\nu m$ at $\bm{R}_n = n_1\bm{a}_1 + n_2\bm{a}_2$, where $n_1$ and $n_2$ are integers.  
The energy levels of the atomic orbitals are given by $t_{\mu \ell, \mu \ell} (\bm{R}_n = \bm{0})=\varepsilon_{\mu\ell}$.  

From the first-principles band calculations~\cite{ZF78,FGH97}, it is known that the band structure of TiSe$_2$ 
is given by six bands based on the Se 4$p$ orbitals below the Fermi level and five bands based on the Ti 3$d$ 
orbitals above the Fermi level.  Therefore we consider the total eleven orbitals from the five 3$d$ orbitals 
in the Ti atom and three 4$p$ orbitals in the Se(1) and Se(2) atoms. 
A TiSe$_6$ octahedron has the $D_{3d}$ point-group symmetry and therefore we define the local coordinate 
axes $x$-$y$-$z$ from the global coordinate axes $X$-$Y$-$Z$ using the rotational transformation~\cite{KST69}
\begin{align}
\left(
\begin{array}{c}
x  \\  y  \\  z
\end{array}
\right)
=
\left(
\begin{array}{r r r}
 {1}/{\sqrt{6}} &-{1}/{\sqrt{2}} & {1}/{\sqrt{3}}  \\
 {1}/{\sqrt{6}} & {1}/{\sqrt{2}} & {1}/{\sqrt{3}}  \\
-{2}/{\sqrt{6}} &  0                        & {1}/{\sqrt{3}} 
\end{array}
\right)
\left(
\begin{array}{c}
X \\ Y \\ Z
\end{array}
\right)
\label{TiSe2_effm_eq3}
\end{align}
as shown in Fig.~\ref{TiSe2_cs_fig1}.  
In the local coordinate axes $x$-$y$-$z$, we define the $d_{xy}$, $d_{yz}$, $d_{zx}$, $d_{x^2-y^2}$, 
and $d_{3z^2-r^2}$ orbitals in the Ti atom and $p_x$, $p_y$, and $p_z$ orbitals in the Se(1) and Se(2) atoms. 

%%  Fig. 3  %%
\begin{figure}[tb]
\begin{center}
\includegraphics[width=0.8\columnwidth]{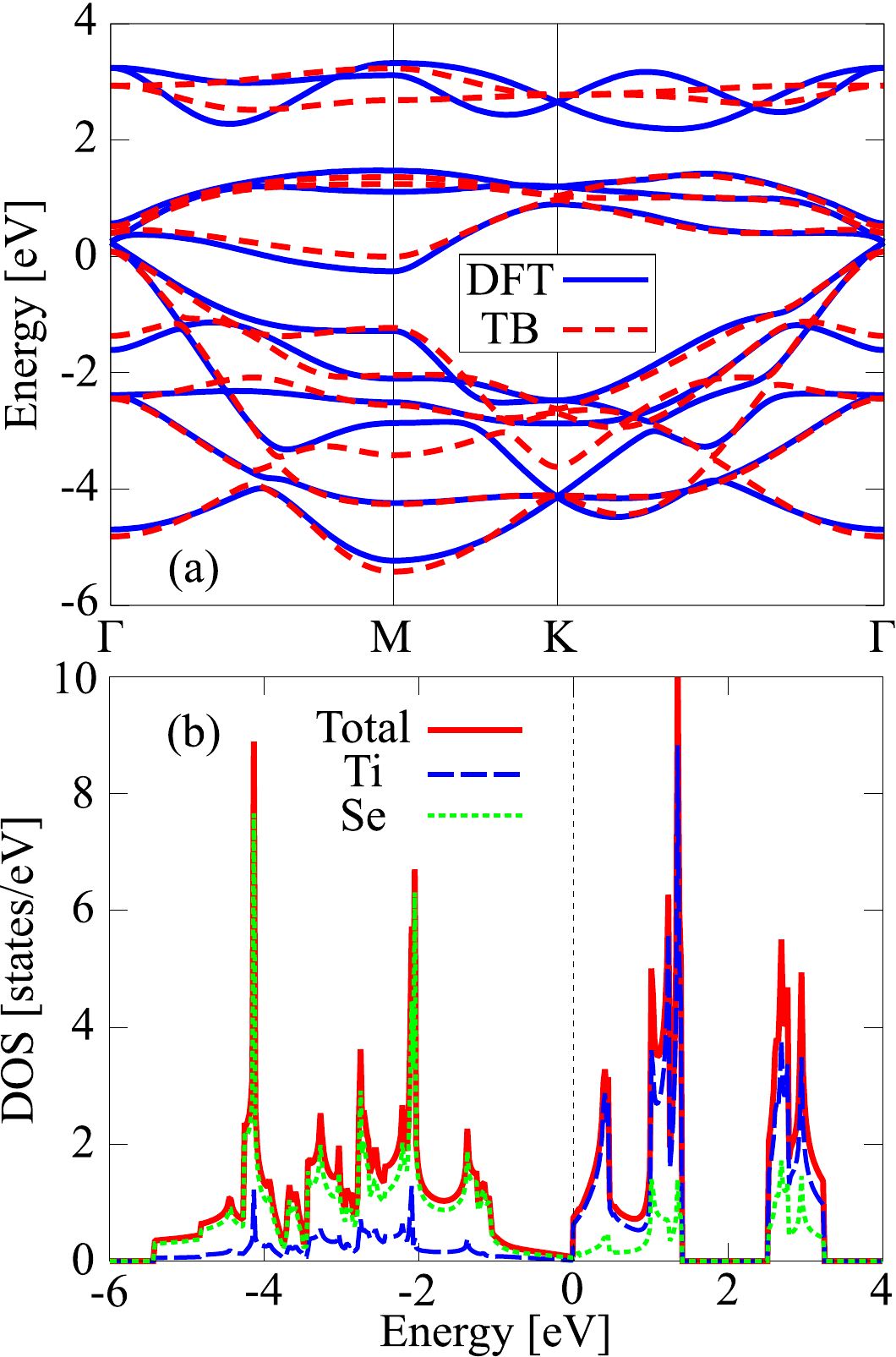}
\caption{
(a) Tight-binding (TB) bands of the monolayer TiSe$_2$ compared with the density-functional-theory (DFT) 
based bands used as the reference bands in the fitting of the TB parameters.  The Fermi energy of the TB 
bands is set to zero, and thus the Fermi energy of the DFT bands is located at $-0.19$ eV.  
(b) Density of states (DOS) of the TB bands of the monolayer TiSe$_2$.  The partial densities 
of states of the Ti (dashed line) and Se (dotted line) orbitals are also shown.}
\label{TiSe2_tb_fig1}
\end{center}
\end{figure}
%%%%%%%%
 
 %% Table 1 %%
\begin{table}[t]
\caption{
Slater-Koster transfer integrals determined by fitting to the first-principles DFT bands shown in 
Fig.~\ref{TiSe2_tb_fig1}(a).  In this fitting, the energy levels of the Ti $d\gamma$ ($d_{x^2-y^2}$, $d_{3z^2-r^2}$), 
Ti $d\varepsilon$ ($d_{xy}$, $d_{yz}$, $d_{zx}$), and Se $p$ ($p_x$, $p_y$, $p_z$) orbitals 
satisfy $\varepsilon_{d\gamma} - \varepsilon_{d\varepsilon} = 1.112$ eV and 
$\varepsilon_{d\varepsilon} - \varepsilon_{p} = 2.171$ eV.  Note that 
$t(pd\sigma)=-t(dp\sigma)$ and $t(pd\pi)=-t(dp\pi)$.}
\begin{center}
{\renewcommand\arraystretch{1.4}
\begin{tabular}{l c r c l c r }
\hline \hline 
\multicolumn{7}{c}{Transfer Integral  [eV]} \\
\hline 
\; $t(pd\sigma)$ &\;=\;&$-$1.422       &&\quad $t(pp\sigma)_1$ &\;=\;& 0.709 \;\\
\: $t(pd\pi)       $ &\;=\;&  0.797          &&\quad $t(pp\pi)_1$ &\;=\;&$-$0.103 \; \\
\; $t(dd\sigma)$ &\;=\;&$-$0.347       &&\quad $t(pp\sigma)_2$ &\;=\;& 0.592 \; \\
\; $t(dd\pi)       $ &\;=\;&   0.119         &&\quad $t(pp\pi)_2$ &\;=\;&$-$0.009 \; \\
\; $t(dd\delta)  $ &\;=\;&$-$0.030       &&\quad  & & \\
\hline\hline 
\end{tabular}
}
\end{center}
\label{TiSe2_table1}
\end{table}
%%%%%%%%

Since a TiSe$_6$ has octahedral structure, we consider the energy levels 
$\varepsilon_{d\gamma}$, $\varepsilon_{d\varepsilon}$, and $\varepsilon_{p}$ of the Ti $d\gamma$ ($d_{x^2-y^2}$, $d_{3z^2-r^2}$), 
Ti $d\varepsilon$ ($d_{xy}$, $d_{yz}$, $d_{zx}$), and 
Se $p$ ($p_x$, $p_y$, $p_z$) orbitals, respectively~\cite{YM80,SYM85,Mo86,MBCetal11}.  
The transfer integrals $t_{\mu \ell, \nu m} (\bm{R}_n)$ are obtained by the Slater-Koster scheme~\cite{SK54} 
as the nine transfer integrals, $t(pd\sigma)$, $t(pd\pi)$, $t(dd\sigma)$, $t(dd\pi)$, $t(dd\delta)$, 
$t(pp\sigma)_1$, $t(pp\pi)_1$, $t(pp\sigma)_2$, and $t(pp\pi)_2$, where 
$t(pd\sigma)$ and $t(pd\pi)$ are the transfer integrals between the nearest-neighbor (NN) Ti 3$d$ and Se 4$p$ orbitals, 
and $t(dd\sigma)$, $t(dd\pi)$, and $t(dd\delta)$ are the transfer integrals between the NN Ti-Ti 3$d$ orbitals.  
The subscripts 1 and 2 in $t(pp\sigma)$ and $t(pp\pi)$ indicate the transfer integrals between the NN Se(1)-Se(1) [Se(2)-Se(2)] 4$p$ orbitals 
and between the NN Se(1)-Se(2) 4$p$ orbitals, respectively.  
The Slater-Koster transfer integrals are evaluated by a least-square fitting of the TB bands to 
the first-principles DFT bands along the high symmetry lines $\Gamma$-M-K-$\Gamma$.  
In the DFT band calculation 
\footnote{
In the DFT band calculation, we use the optimized crystal structure obtained in Sec.~\ref{TiSe2_effm_cs}.  
In the self-consistent calculations, we use 416 $k$-points in the irreducible part of the BZ, 
assuming the muffin-tin radii ($R_{\rm MT}$) of 2.41 (Ti) and 2.41 (Se) bohr and the plane-wave cutoff 
of $K_{\rm max}=7/R_{\rm MT}$.  
}, 
we use the full-potential linearized augmented-plane-wave method with the generalized gradient 
approximation (GGA)~\cite{PBE96} for electron correlations implemented in the WIEN2k code~\cite{WIEN2k}.  
As the initial values of the parameters in the least-square fitting procedure, we use the Slater-Koster 
transfer integrals [$t(pd\sigma)$ etc.]~roughly estimated from the TB bands obtained via the maximally 
localized Wannier functions~\cite{wien2wannier,wannier90}.  
We use 252 $\bm{k}$-points along the $\Gamma$-M-K-$\Gamma$ lines in our least-square fitting.  

The optimized values of the transfer integrals are summarized in Table~\ref{TiSe2_table1}.  The obtained TB bands 
are compared with the original DFT bands in Fig~\ref{TiSe2_tb_fig1}(a) to find a good agreement, indicating 
that our TB band structure can capture the overall character of the first-principles DFT band structure.  
We find that the valence bands are composed mainly of the Se $p$ ($p_x$, $p_y$, $p_z$) orbitals 
and the conduction bands near the Fermi level are composed mainly of the Ti $d\varepsilon$ ($d_{xy}$, $d_{yz}$, $d_{zx}$) orbitals.  
The Ti $d\gamma$ bands are located well above the $d\varepsilon$ bands due to the 
crystal field splitting $\varepsilon_{d\gamma} > \varepsilon_{d\varepsilon}$.  
The valence-band top is located at the $\Gamma$ point of the BZ and the conduction-band 
bottom is located at the M points of the BZ.  
The valence-band maximum and conduction-band minimum are +0.081~eV and -0.007~eV, respectively, from the Fermi level, resulting in the %slightly overlapped 
semimetallic band structure located in the vicinity of a zero-gap semiconducting state.

%%%%%%%%%%%%%%%%%%%%
\subsection{Electron-Phonon Coupling} \label{TiSe2_effm_ep}

To discuss the lattice displacements in TiSe$_2$, we introduce the electron-phonon coupling, following 
the method of Motizuki $\textit{et al.}$.~\cite{YM80,YM80,TM80,MSYetal81,MYT81,SYM84,SYM85,Mo86}. 
The electron-phonon coupling strengths are given by the changes in the transfer integrals with respect 
to the lattice displacements $\delta \bm{R}_{i\mu}$ from their equilibrium positions $\bm{R}_{i\mu}$.  
In the reciprocal lattice space, $\delta \bm{R}_{i\mu}$ is given by 
\begin{align}
\delta \bm{R}_{i\mu} &= \frac{1}{\sqrt{N}} \sum_{\bm{q}}e^{i\bm{q}\cdot\bm{R}_i}  \bm{u}_{\bm{q},\mu} 
\notag \\
&= \frac{1}{\sqrt{NM_{\mu}}} \sum_{\bm{q}}e^{i\bm{q}\cdot\bm{R}_i}  \bm{\varepsilon}(\bm{q},\mu) Q_{\bm{q}},
\label{TiSe2_effm_eq4}
\end{align}
where $\bm{u}_{\bm{q},{\mu}}$ is the lattice displacement in $\bm{q}$-space given by 
$\bm{u}_{\bm{q},\mu} =  \left(\bm{\varepsilon}(\bm{q},\mu)/\sqrt{M_{\mu}} \right)Q_{\bm{q}}$.  
The displacement $\bm{u}_{\bm{q},\mu}$ is characterized by the normal coordinate $Q_{\bm{q}}$ 
and polarization vector $\bm{\varepsilon}(\bm{q},\mu)$ of a particular phonon mode at $\bm{q}$, 
where $M_{\mu}$ is the mass of atom $\mu$.  
Details of the derivation and general form of the electron-phonon coupling are summarized 
in Appendix~\ref{app_ep}.  
In this approach, the Hamiltonian of the electron-phonon coupling is given by 
\begin{align}
\mathcal{H}_{ep} = \frac{1}{\sqrt{N}}\sum_{\bm{k},\bm{q}} \sum_{\mu \ell, \nu m}  g_{\mu \ell, \nu m} (\bm{k},\bm{q}) 
Q_{\bm{q}}c^{\dag}_{\bm{k},\mu \ell} c_{\bm{k}-\bm{q},\nu m}
\label{TiSe2_effm_eq5}
\end{align}
with the electron-phonon coupling constant 
\begin{align}
g_{\mu \ell, \nu m} (\bm{k},\bm{q}) &= \sum_{\bm{R}_n} \left[ \bm{\nabla} t_{\mu \ell, \nu m} (\bm{R}_n) \right] 
 \label{TiSe2_effm_eq6} \\
&\cdot \left[ \frac{\bm{\varepsilon}(\bm{q},\mu)}{\sqrt{M_{\mu}}}e^{-i(\bm{k}-\bm{q})\cdot \bm{R}_n}
- \frac{\bm{\varepsilon}(\bm{q},\nu)}{\sqrt{M_{\nu}}} e^{-i\bm{k}\cdot \bm{R}_n} \right],  
\notag 
\end{align}
where $\bm{\nabla} t_{\mu \ell, \nu m} (\bm{R}_n)$ is the first derivative of the transfer integral 
with respect to $\bm{R}_n$.  

%%  Fig. 4  %%
\begin{figure}[tb]
\begin{center}
\includegraphics[width=0.9\columnwidth]{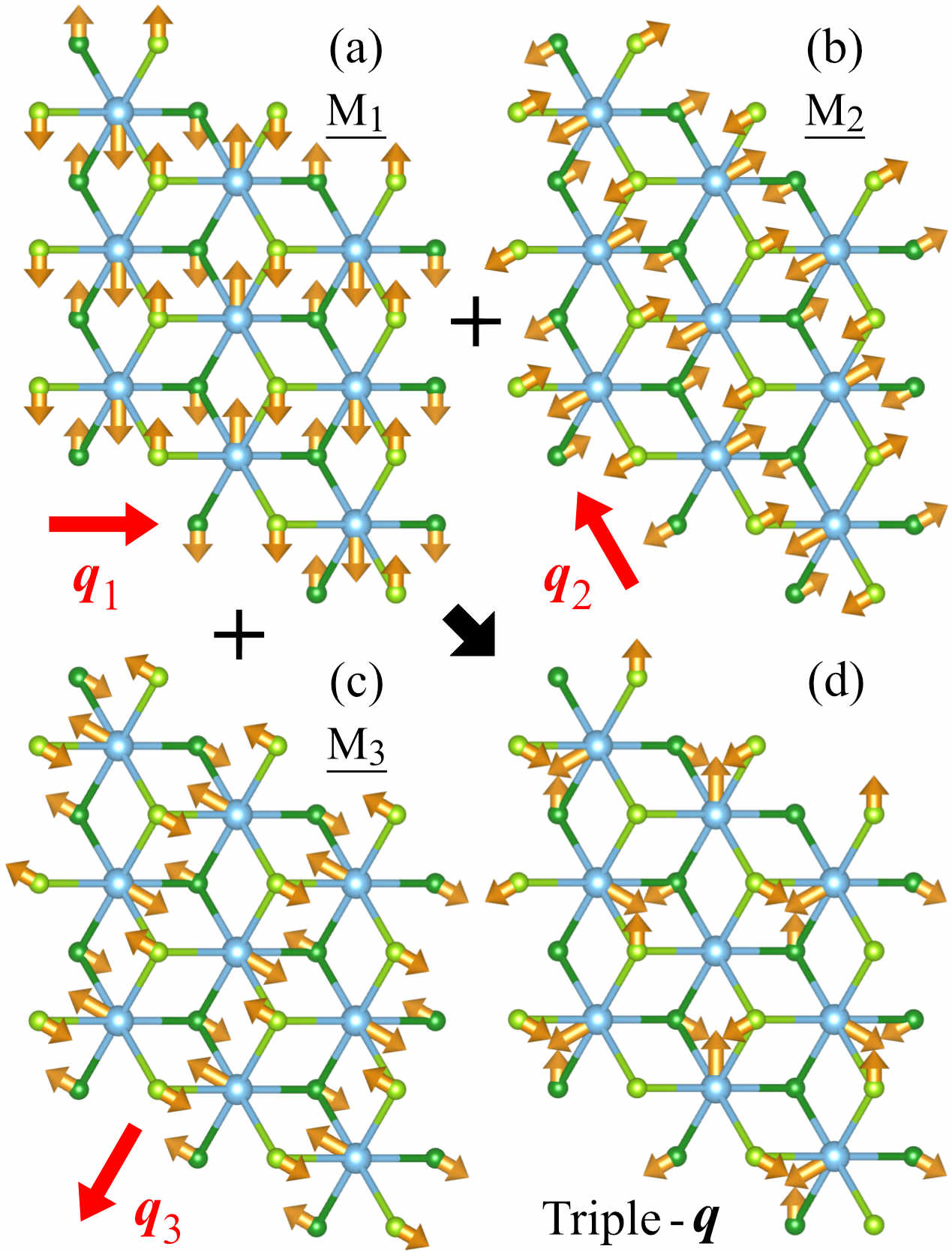}
\caption{
Schematic representation of the transverse phonon modes at the (a) M$_1$, (b) M$_2$ and (c) M$_3$ points 
of the BZ.  (d) Periodic lattice displacement in the triple-$\bm{q}$ state. 
}
\label{TiSe2_cs_fig3}
\end{center}
\end{figure}
%%%%%%%%

In Fig.~\ref{TiSe2_cs_fig3}(d), we show the schematic picture of the periodic lattice displacement (PLD) 
observed experimentally in TiSe$_2$~\cite{DMW76}.  
Realization of this PLD has been explained theoretically by Motizuki \textit{et al.}~\cite{SYM85,Mo86}.  
A first-principles calculation for this PLD has also been performed by Bianco $et$ $al$.~\cite{BCM15}.  
Accordingly, the 2$\times$2 PLD shown in Fig.~\ref{TiSe2_cs_fig3}(d) is realized by the sum of the 
transverse phonon modes at the M$_1$, M$_2$, and M$_3$ points of the BZ illustrated in 
Figs.~\ref{TiSe2_cs_fig3}(a)--\ref{TiSe2_cs_fig3}(c).  
Therefore the 2$\times$2 PLD is the triple-$\bm{q}$ structure characterized by the wave vectors 
$\bm{q}_{1} =  \bm{b}_1/2$, $\bm{q}_{2} = (\bm{b}_2 - \bm{b}_1)/2$, and $\bm{q}_{3} = - \bm{b}_2/2$ 
shown in Fig.~\ref{TiSe2_cs_fig2}(b).  
In this paper, we consider the transverse phonon modes shown in 
Figs.~\ref{TiSe2_cs_fig3}(a)--\ref{TiSe2_cs_fig3}(c) as 
the specific phonon modes in Eqs.~(\ref{TiSe2_effm_eq4})--(\ref{TiSe2_effm_eq6}). 

To estimate the coupling constants $g_{\mu \ell, \nu m} (\bm{k},\bm{q})$, we need the polarization 
vector $\bm{\varepsilon}(\bm{q},\mu)$ characterized by the eigenstate of the transverse phonon mode. 
$\bm{\varepsilon}(\bm{q},\mu)$ for the PLD in TiSe$_2$ has been provided by Motizuki {\it et al}.~\cite{SYM85,Mo86} (see also Appendix~\ref{app_tri}).  
When the ratio between the lattice displacements of Ti and Se ions is given as 
$\xi = |\bm{u}_{\bm{q}_{j},{\rm Se}}| / |\bm{u}_{\bm{q}_{j},{\rm Ti}}|$, 
the polarization vectors for the transverse phonon mode at the M$_1$ point, which are perpendicular 
to the vector $\bm{q}_1$, are given by
\begin{align}
&\bm{\varepsilon}(\bm{q}_1,{\rm Ti }) = \sqrt{M_{\rm Ti}/M^{*}}\bm{e}_Y, 
\label{TiSe2_effm_eq7}
 \\
&\bm{\varepsilon}(\bm{q}_1,{\rm Se1}) = \bm{\varepsilon}(\bm{q}_1,{\rm Se2}) = - \xi  \sqrt{M_{\rm Se}/M^{*}}\bm{e}_Y,
\label{TiSe2_effm_eq8}
\end{align}
where $M^{*} = M_{\rm Ti} + 2 \xi^2 M_{\rm Se}$ is the effective mass of the transverse mode 
at the M point.  
Similarly, the polarization vectors $\bm{\varepsilon}(\bm{q}_2,\mu)$ and $\bm{\varepsilon}(\bm{q}_3,\mu)$ 
are perpendicular to their respective wave vectors (see also Appendix~\ref{app_tri}).  
The ratio $\xi$ was estimated as $\xi \simeq 1/3$ in previous experimental~\cite{DMW76,FHCetal17} 
and theoretical~\cite{MYT81,BCM15} studies.  We therefore assume $\xi = 1/3$ 
 and $M^{*}=M_{\rm Ti} + (2/9)M_{\rm Se}$~($= 65.416$~u)~\cite{SYM85,Mo86} throughout this paper.  

\begin{figure*}[tb]
\begin{center}
\includegraphics[width=2\columnwidth]{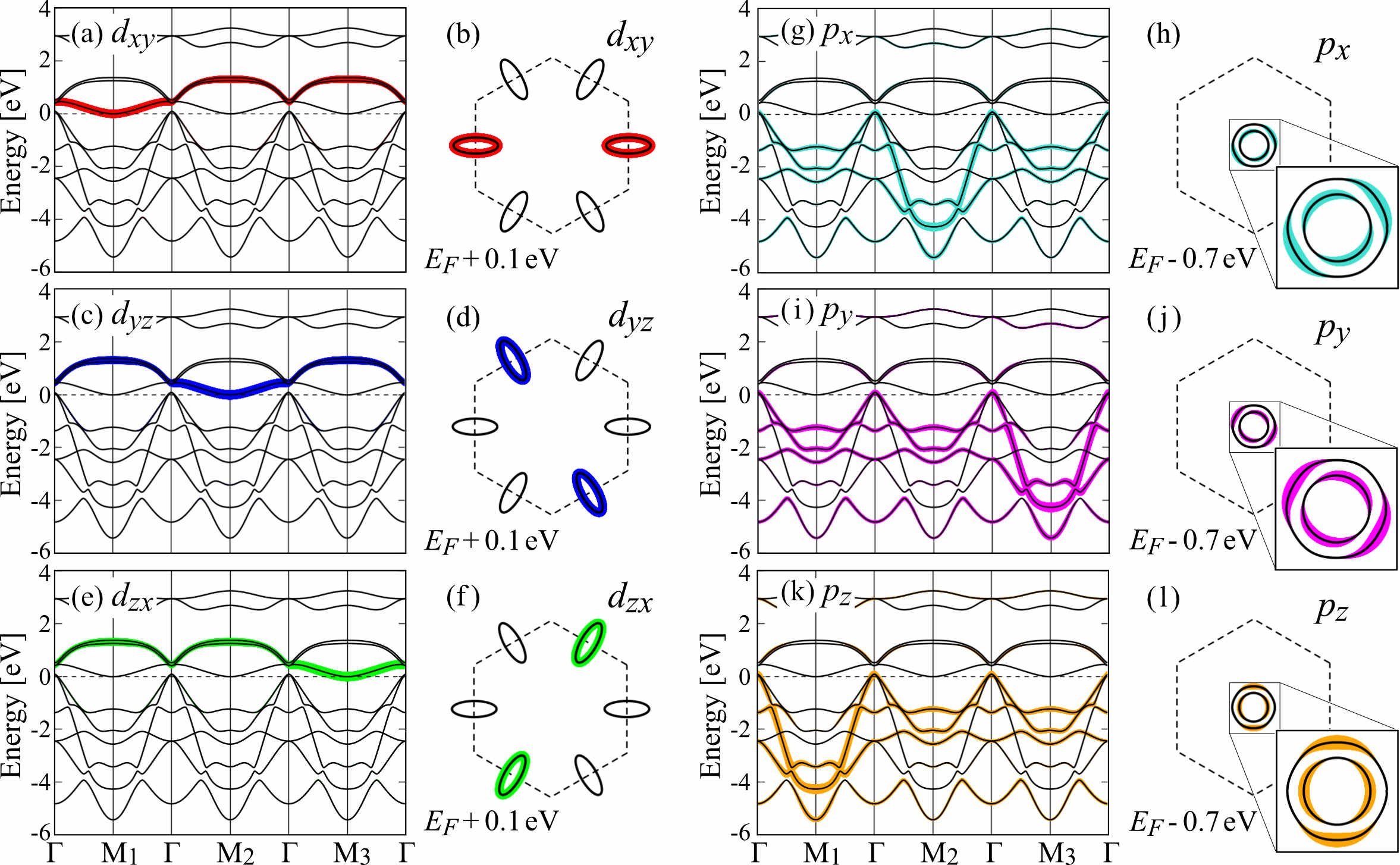}
\caption{
Weighted band dispersions and equal energy surfaces (lines) of the undistorted TB band structure. 
The width of the curves is in proportion to the weight of the (a,~b) Ti $d_{xy}$, (c,~d) Ti $d_{yz}$, 
(e,~f) Ti $d_{zx}$, (g,~h) Se $p_{x}$, (i,~j) Se $p_{y}$, and (k,~l) Se $p_{z}$ orbitals.  
In (b), (d), and (f), we plot the equal energy surfaces above the Fermi energy, $E_F + 0.1$~eV, 
and in (h), (j), and (l), we plot the equal energy surfaces below the Fermi energy, $E_F - 0.7$~eV.  
Dashed line indicates the original BZ without distortion. }  
\label{TiSe2_tb_fig2}
\end{center}
\end{figure*}

In addition to the polarization vector $\bm{\varepsilon}(\bm{q},\mu)$, 
the first derivative of the transfer integrals $\bm{\nabla} t_{\mu \ell, \nu m} (\bm{R}_n)$ is 
required to estimate the coupling constants $g_{\mu \ell, \nu m} (\bm{k},\bm{q})$. 
Here we briefly describe the estimation of this quantity and the details are found in Appendix~\ref{app_eelc}.  
We follow the approximation introduced 
by Motizuki \textit{et al.}~\cite{YM80,Mo86}: 
\begin{align}
\frac{t'(pd\sigma)}{t(pd\sigma)} = \alpha_{\rm c} \frac{s'(pd\sigma)}{s(pd\sigma)},\;\; {\rm etc.}, 
\label{TiSe2_effm_eq9}
\end{align}
where $t'(pd\sigma)$ is the first derivative of the transfer integral $t(pd\sigma)$ with respect to 
the interatomic distance, and $s(pd\sigma)$ and $s'(pd\sigma)$ indicate the overlap integral and 
its derivative, respectively.  
$\alpha_{\rm c}$ is the coupling constant that determines the strength of the electron-phonon coupling.  
In this paper, we treat $\alpha_{\rm c}$ as a tunable parameter; the value of $\alpha_{\rm c}$ 
is determined such that the calculated results are in good agreement with experiment.  
Note that $\alpha_{\rm c}=0$ does not indicate $g_{\mu \ell, \nu m} (\bm{k},\bm{q})=0$ since 
$\bm{\nabla} t_{\mu \ell, \nu m} (\bm{R}_n)$ also includes the terms given by the transfer integrals 
$t(pd\sigma)$ (see Appendix~\ref{app_eelc}).  
In the estimation of the overlap integrals and their derivatives, we use the Slater-type orbital~\cite{Sl30,CR63,CRR67}.  
We can thus calculate the values analytically (see Appendix~\ref{app_eelc}).  
The Slater-type orbital is characterized by the orbital exponents, which are estimated by 
Clementi {\it et al.}~\cite{CR63,CRR67}: 
$\zeta_{3d} = 2.7138$ and $\zeta_{4p} = 2.0718$ for the Ti 3$d$ and Se 4$p$ orbitals, respectively.  
As shown in Fig.~\ref{TiSe2_tb_fig1}(b), the valence and conduction bands near the Fermi level are 
composed of Se $p$ ($p_{x}$, $p_{y}$, $p_{z}$) and Ti $d\varepsilon$ ($d_{xy}$, $d_{yz}$, $d_{zx}$) 
orbitals, respectively (see also Fig.~\ref{TiSe2_tb_fig2}).  We therefore consider the electron-lattice 
coupling between the nearest-neighbor Ti $d\varepsilon$ ($d_{xy}$, $d_{yz}$, $d_{zx}$) and Se 
$p$ ($p_{x}$, $p_{y}$, $p_{z}$) orbitals only.  
In this approximation, we need the ratio between the overlap integral and its first derivative for both $pd\sigma$ 
and $pd\pi$; the estimated values given by the Slater-type orbitals are listed in Table~\ref{TiSe2_table2}. 

%% Table 2 %%
\begin{table}[t]
\caption{
Ratio between the Slater-Koster overlap integral and its first derivative estimated by the Slater-type orbital. 
$s(pd\sigma)$ and $s(pd\pi)$ are the overlap integrals between nearest-neighbor Ti $d$ and Se $p$ orbitals.
$s'(pd\sigma)$ and $s'(pd\pi)$ are the first derivative of the overlap integrals. 
In the Slater-type orbitals, we use the orbital exponents $\zeta_{3d} = 2.7138$ and $\zeta_{4p} = 2.0718$, for the Ti 3$d$ and Se 4$p$ orbitals, respectively~\cite{CR63,CRR67}. 
$R$ is the distance between Ti and Se ions: $R_{\rm Ti-Se} = 2.566$~\AA. }
\begin{center}
{\renewcommand\arraystretch{1.4}
\begin{tabular}{l c r c l c r }
\hline \hline 
\multicolumn{3}{c}{$R\times s'(pd)/s(pd)$ } & \multicolumn{1}{c}{} & \multicolumn{3}{c}{$s'(pd)/s(pd)$  [1/\AA]} \\
\hline 
\; $R \! \times \! s'(pd\sigma)/s(pd\sigma)$ &\!=\!&$\!-$3.860        &&\; $s'(pd\sigma)/s(pd\sigma)$ &\!=\!&$\!-$1.504  \;\\
\: $R \! \times \! s'(pd\pi)/s(pd\pi)              $ &\!=\!&$\!-$5.933           &&\; $s'(pd\pi)/s(pd\pi)$ &\!=\!&$\!-$2.312  \; \\
\hline\hline 
\end{tabular}
}
\end{center}
\label{TiSe2_table2}
\end{table}
%%%%%%%%

When the ions are displaced from their equilibrium position, the lattice system increases the elastic energy. 
The Hamiltonian of the elastic term is given by 
\begin{align}
\mathcal{H}_{p} = \frac{1}{2} \sum_{\bm{q}}  \omega_0^2(\bm{q}) |Q_{\bm{q}}|^2, 
\label{TiSe2_effm_eq10}
\end{align}
where $\omega_0(\bm{q})$ is the bare phonon frequency of the transverse mode at momentum $\bm{q}$.  
A bare phonon frequency $\omega_0(\bm{q})$ has been estimated by 
Motizuki \textit{et al.} 
in comparison with the experimentally observed phonon dispersions~\cite{MSYetal81,SYM85,Mo86}.  
Monney \textit{et al.} have also assumed the value close to it~\cite{MBCetal11}. 
In this paper, we use a similar value $M^{*}\omega^2_0(\bm{q}_{\rm M}) = 10$~eV/\AA$^2$ 
[$\omega_0(\bm{q}_{\rm M}) \simeq  6.11$~THz] at the M point ($\bm{q}=\bm{q}_{\rm M}$).  
We may also treat $\omega_0(\bm{q}_{\rm M})$ as a tunable parameter.  

%%  Fig. 5  %%
%%%%%%%%

%%%%%%%%%%%%%%%%%%%%
\subsection{Coulomb Interaction}
To treat the excitonic mechanism of the CDW formation, we also consider the intersite Coulomb 
interactions.  In general, the excitonic order (or excitonic insulator state) should be induced by 
the interband Coulomb interactions.  In TiSe$_2$, the interband Coulomb (or excitonic) interactions 
are given by the interactions between the valence Se 4$p$ and the conduction Ti 3$d$ bands.  
In real space, the interband Coulomb (or excitonic) interaction in TiSe$_2$ is essentially given 
by the intersite Coulomb interaction between the nearest-neighbor Ti and Se sites.  
Therefore, as the excitonic interactions, we consider the intersite Coulomb interaction between 
the nearest-neighbor Ti and Se sites given by 
\begin{align}
&\mathcal{H}_{ee} 
= \sum_{\bm{R}_j,\bm{R}_n} \sum_{\ell, \nu m}  V^{dp}_{ \ell,\nu m} (\bm{R}_n) n^{d}_{\ell}(\bm{R}_j) n^{p}_{\nu m } (\bm{R}_j+\bm{R}_n)
\notag \\
&=\! \frac{1}{N}\! \sum_{\bm{k},\bm{k}',\bm{q}} \sum_{\ell, \nu m}
V^{dp}_{\ell,\nu m}(\bm{k}-\bm{k}')  
d^{\dag}_{\bm{k},\ell} d_{\bm{k}',\ell} p^{\dag}_{\bm{k}'-\bm{q},\nu m} p_{\bm{k}-\bm{q},\nu m},
\label{TiSe2_effm_eq11} 
\end{align}
where $V^{dp}_{ \ell,\nu m} (\bm{R}_n)$ is the intersite Coulomb interaction between the 
nearest-neighbor Ti $d_\ell$ and Se($\nu$) $p_m$ orbitals, and $n^{d}_{\ell}(\bm{R}_j)$ and 
$n^{p}_{\nu m } (\bm{R}_j)$ are the number operators of the electron of the Ti $d_\ell$ and 
Se($\nu$) $p_m$ orbitals, respectively, in the unit cell at $\bm{R}_j$.  
Second line of Eq.~(\ref{TiSe2_effm_eq11}) indicates the Fourier transformed Coulomb interaction, where 
\begin{align}
V^{dp}_{\ell,\nu m}(\bm{k}-\bm{k}') 
= \sum_{\bm{R}_n}  V^{dp}_{\ell,\nu m}(\bm{R}_n) e^{-i(\bm{k}-\bm{k}')\cdot\bm{R}_n} ,
\label{TiSe2_effm_eq12}
\end{align}
and $d^{(\dag)}_{\bm{k},\ell}$ and $p^{(\dag)}_{\bm{k},\nu m}$ are the annihilation (creation) 
operators of an electron in the Ti $d_\ell$ and Se($\nu$) $p_m$ orbitals, respectively, at 
momentum $\bm{k}$.  In this paper, we assume the orbital independent interaction, 
$V^{dp}_{ \ell,\nu m} (\bm{R}_n) = V$, for simplicity, and we treat $V$ as a tunable parameter.  

%%  Fig. 6  %%
\begin{figure}[tb]
\begin{center}
\includegraphics[width=0.9\columnwidth]{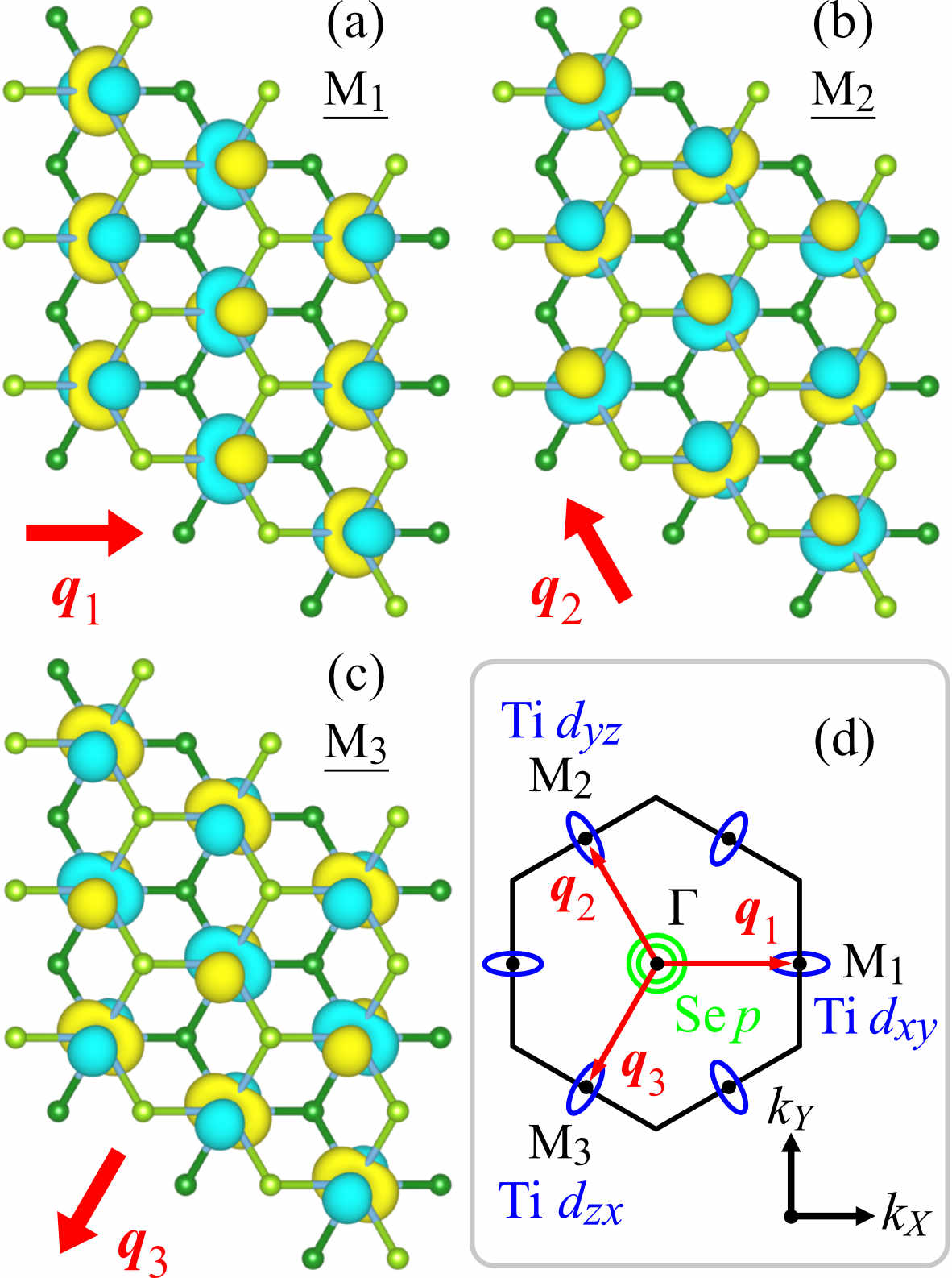}
\caption{
Bloch wave functions $\psi^{(0)}_{\bm{k},a}(\bm{r})$ of the conduction-band bottom at 
the (a) M$_1$, (b) M$_2$, and (c) M$_3$ points, where the Slater-type orbitals are employed as the 
atomic orbitals.  (d) Schematic Fermi surfaces of the monolayer TiSe$_2$.  Note that areas of the 
hole and electron pockets are slightly larger than those of our TB bands.}
\label{TiSe2_tb_fig3}
\end{center}
\end{figure}
%%%%%%%%

%%%%%%%%%%%%%%%%%%%%
%%%%%%%%%%%%%%%%%%%%
\section{Undistorted Band Structure}
Before discussing the CDW state caused by the electron-phonon and excitonic interactions, 
we overview the characters of undistorted band structure given by diagonalizing 
the TB Hamiltonian $\mathcal{H}_e$.  

Figure~\ref{TiSe2_tb_fig2} shows the calculated band dispersions along the $\bm{k}$-path 
through the M$_1$, M$_2$, and M$_3$ points of the BZ defined in Fig.~\ref{TiSe2_cs_fig2}(b).  
Here, we also plot the weight of orbitals on each band given by $|u^{(0)}_{\mu\ell,a}(\bm{k})|^2$, 
where $u^{(0)}_{\mu\ell,a}(\bm{k})$ is the $\mu\ell$ component of the eigenvector for the 
band $a$.  
Figures~\ref{TiSe2_tb_fig2}(a), \ref{TiSe2_tb_fig2}(c), and \ref{TiSe2_tb_fig2}(e) show 
the weighted band dispersions of the 
Ti $d_{xy}$, $d_{yz}$, and $d_{zx}$ orbitals, respectively. We find that the Ti $d_{xy}$, $d_{yz}$, 
and $d_{zx}$ orbital characters appear in the conduction-band bottom at the M$_1$, M$_2$, 
and M$_3$ points of the BZ, respectively.  
To show the corresponding characters in $\bm{k}$-space, 
we also plot in Figs.~\ref{TiSe2_tb_fig2}(b), ~\ref{TiSe2_tb_fig2}(d), and ~\ref{TiSe2_tb_fig2}(f) the equal energy 
surfaces (lines) above the Fermi level $E_F$ with the weight of the Ti $d$ orbitals.  
We find that the equal energy 
surface around the M$_1$ point is almost completely composed of the Ti $d_{xy}$ orbital.  Similarly, 
the equal energy surfaces around the M$_2$ and M$_3$ points are given by $d_{yz}$ and $d_{zx}$ 
orbitals, respectively.  
Thus the characters of the Ti $d_{xy}$, $d_{yz}$, and $d_{zx}$ orbitals are related to each other 
by the $2\pi/3$ rotation around the $\Gamma$ point.  These results are consistent with the 
orbital characters in the bulk TiSe$_2$, which was pointed out by van Wezel~\cite{vWe11}.  
Figures~\ref{TiSe2_tb_fig2}(g)--\ref{TiSe2_tb_fig2}(l) show the weighted band dispersions and the equal energy 
surfaces below $E_F$ for the Se $p_x$, $p_y$, and $p_z$ orbitals.  We find that the two valence bands 
around the $\Gamma$ point are composed of the Se $p$ orbitals but that the inequivalence 
in the weight of the $p_x$, $p_y$, and $p_z$ orbitals appears along the different $\bm{k}$-directions.  
The equal energy surfaces in the valence bands show the similar $2\pi/3$ rotational property of 
Se $p_x$, $p_y$, and $p_z$ orbitals around the $\Gamma$ point. 

Figures~\ref{TiSe2_tb_fig3}(a)--\ref{TiSe2_tb_fig3}(c) show the Bloch wave functions $\psi^{(0)}_{\bm{k},a}(\bm{r})$ 
of the conduction-band bottoms at the M$_1$, M$_2$, and M$_3$ points.  When the Hamiltonian 
in the TB approximation is diagonalized, the Bloch wave function of band $a$ is given by 
$\psi^{(0)}_{\bm{k},a}(\bm{r})=\sum_{\mu\ell}u^{(0)*}_{\mu\ell,a}(\bm{k})\phi^{(0)}_{\bm{k},\mu\ell}(\bm{r})$, 
where $\phi^{(0)}_{\bm{k},\mu\ell}(\bm{r})$ is the Bloch sum of the atomic orbitals 
$\phi^{(0)}_{\bm{k},\mu\ell}(\bm{r})=(1/\sqrt{N})\sum_{\bm{R}_i}\phi_{\ell}(\bm{r}-\bm{R}_{i\mu})e^{i\bm{k}\cdot\bm{R}_i}$ 
and we use the Slater-type orbital as the atomic orbital $\phi_{\ell}(\bm{r})$, as in the estimation 
of the overlap integrals discussed in Sec.~\ref{TiSe2_effm_ep}.  
We find in Fig.~\ref{TiSe2_tb_fig3}(a) that the Bloch wave function $\psi^{(0)}_{\bm{k},a}(\bm{r})$ 
at the M$_1$ point clearly shows the shape nearly consistent with the $d_{xy}$ orbital around Ti atoms.  
Note that, due to $e^{i\bm{k}\cdot\bm{R}_i} =e^{i\bm{q}_1\cdot\bm{R}_i}$ in the Bloch function 
at the M$_1$ point, the wave functions on Ti atoms change signs along the direction of $\bm{q}_1$.  
Similarly, the shapes of the $d_{yz}$ and $d_{zx}$ orbitals appear in the Bloch functions at the 
M$_2$ and M$_3$ points, respectively.  Ti $d_{xy}$, $d_{yz}$, and $d_{zx}$ orbitals, which appear in 
the Bloch functions at the M$_1$, M$_2$, and M$_3$ points, respectively, are rotated by $2\pi/3$ 
around the $Z$ axis in the global coordinates due to the three-fold rotational symmetry of the 
crystal structure.  
We do not show the Bloch functions of the valence-band top at the $\Gamma$ 
point here, but we have confirmed that they clearly show the shapes of the $p$ orbitals around Se 
atoms.  

Figure~\ref{TiSe2_tb_fig3}(d) summarizes the Fermi surfaces of the undistorted band structure 
of the monolayer TiSe$_2$ schematically.  The hole pockets (i.e., valence-band top) at the 
$\Gamma$ point are characterized by the Se $p$ orbitals and the electron pockets (i.e., 
conduction-band bottom) at the M$_1$, M$_2$, and M$_3$ points are characterized by the 
Ti $d_{xy}$, $d_{yz}$, and $d_{zx}$ orbitals, respectively.  The CDW state in TiSe$_2$ may 
therefore be given by the mixture of the Se $p$ orbitals at the $\Gamma$ point and 
Ti $d_{xy}$, $d_{yz}$, $d_{zx}$ orbitals at the M$_1$, M$_2$, and M$_3$ points.

%%%%%%%%%%%%%%%%%%%%
%%%%%%%%%%%%%%%%%%%%
\section{Phonon Softening and CDW}  \label{TiSe2_ep}

In this section, we discuss the realization of the CDW without introducing the excitonic interaction.  
We first discuss the softening of the transverse modes at the M points shown in 
Figs.~\ref{TiSe2_cs_fig3}(a)--\ref{TiSe2_cs_fig3}(c). 
We then examine the stability of the static triple-$\bm{q}$ CDW 
state, where the transverse phonon modes at the M$_1$, M$_2$, 
and M$_3$ points are frozen simultaneously, as shown in Fig.~\ref{TiSe2_cs_fig3}(d).

%%%%%%%%%%%%%%%%%%%%
\subsection{Phonon Softening} \label{TiSe2_epsus}

To discuss the structural instability in TiSe$_2$, we evaluate the effective phonon frequency 
$\omega(\bm{q})$ given as~\cite{Mo86} 
\begin{align}
\omega^2(\bm{q})  =  \omega^2_0(\bm{q}) - \chi(\bm{q}), 
\label{TiSe2_ep_eq1}
\end{align}
where $\omega_0(\bm{q})$ is the bare phonon frequency of the transverse mode and $\chi(\bm{q})$ 
is the susceptibility including the electron-phonon coupling $g_{\mu\ell,\nu m}(\bm{k},\bm{q})$ 
(see Appendix~\ref{app_epsus}).  Specifically, the susceptibility $\chi(\bm{q})$ is given by
%\begin{align}
%&
\begin{equation}
\chi(\bm{q})  
=\! -\frac{2}{N} \sum_{\bm{k}} \sum_{a,b}
\left| V_{ep}(a \bm{k}, b \bm{k} \! - \! \bm{q}) \right|^2 \!
\frac{f(\varepsilon^{(0)}_{\bm{k},a}) \! - \! f(\varepsilon^{(0)}_{\bm{k}-\bm{q},b}) }{\varepsilon^{(0)}_{\bm{k},a} \! - \! \varepsilon^{(0)}_{\bm{k}-\bm{q},b}} 
\label{TiSe2_ep_eq2}  
%\\
\end{equation}
with 
\begin{equation}
%&
V_{ep}(a \bm{k}, b \bm{k} \! - \! \bm{q}) 
= \! 
\sum_{\mu\ell,\nu m} u^{(0) *}_{\mu\ell,a}(\bm{k})g_{\mu\ell,\nu m}(\bm{k},\bm{q}) u^{(0)}_{\nu m,b}(\bm{k} \! -\! \bm{q}) ,
\label{TiSe2_ep_eq3}
\end{equation}
%\end{align}
where $\varepsilon^{(0)}_{\bm{k},a}$ is the undistorted energy band, 
$u^{(0)}_{\mu\ell,a}(\bm{k})$ is the $\mu\ell$ component of the eigenvector for the band $a$, 
and $f(\varepsilon^{(0)}_{\bm{k},a})$ is the Fermi distribution function (see Appendix~\ref{app_epsus}).  
In the calculations of the susceptibility, we use $500\times 500$ $\bm{k}$ points for summation.  
In Eq.~(\ref{TiSe2_effm_eq9}), we assume $\alpha_{\rm c} = 0.1$, which provides results in good 
agreement with the observed lattice displacement~\cite{DMW76} (see next subsection) if we 
use $M^{*}\omega^2_0(\bm{q}_{\rm M}) = 10$~eV/\AA$^2$ as the bare phonon frequency 
$\omega_0(\bm{q})$ at the M point.  
In this section, we assume $\alpha_{\rm c}=0.1$ unless otherwise stated.  
The $\alpha_{\rm c}$ dependence will be discussed in the next section. 

%%  Fig. 7  %%
\begin{figure}[tb]
\begin{center}
\includegraphics[width=0.9\columnwidth]{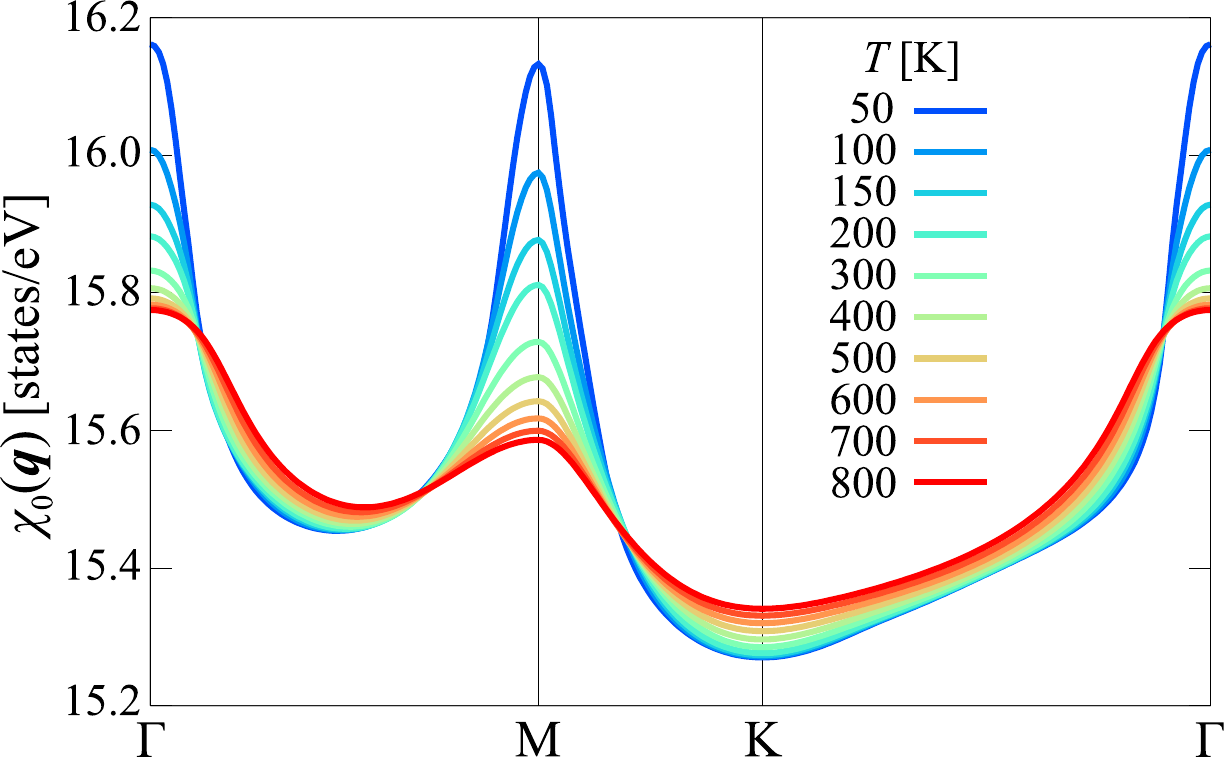}
\caption{Calculated temperature dependence of the bare electronic susceptibility 
$\chi_0(\bm{q})$ as a function of $\bm{q}$.}  
\label{TiSe2_ep_fig1}
\end{center}
\end{figure}
%%%%%%%%

%%  Fig. 8  %%
\begin{figure}[tb]
\begin{center}
\includegraphics[width=0.9\columnwidth]{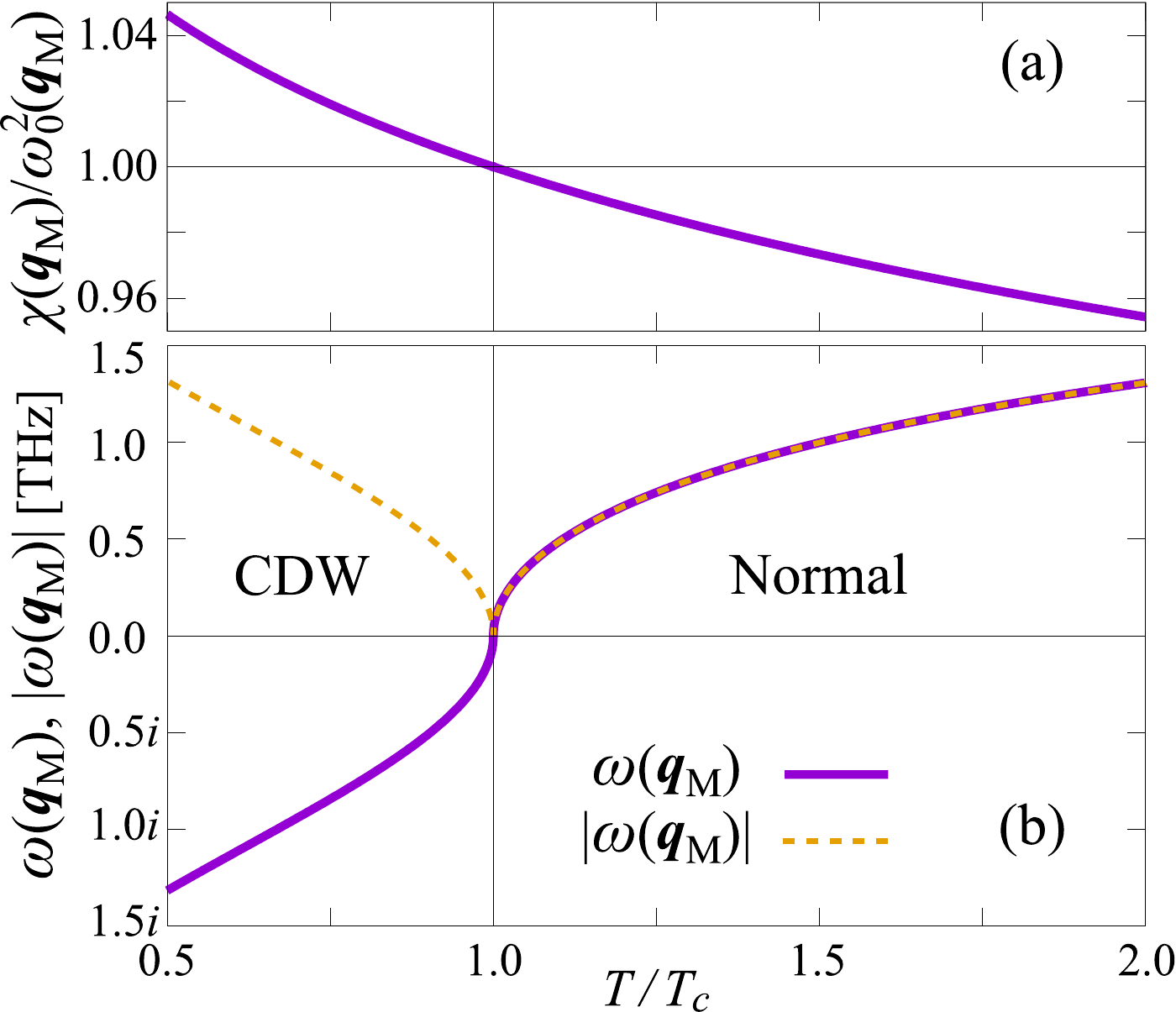}
\caption{Calculated temperature dependence of the (a) susceptibility 
$\chi(\bm{q}_{\rm M})$ [$\chi(\bm{q}_{\rm M})/\omega^2_0(\bm{q}_{\rm M})$] and 
(b) effective phonon frequency $\omega(\bm{q}_{\rm M})$ at the M point.  
We assume $M^{*}\omega^2_0(\bm{q}_{\rm M}) = 10$~eV/\AA$^2$ 
[$\omega_0(\bm{q}_{\rm M}) \simeq 6.11$~THz], which gives the transition 
temperature $T_c\simeq 443$~K. }  
\label{TiSe2_ep_fig2}
\end{center}
\end{figure}
%%%%%%%%

Before discussing the phonon softening, we show the character of the bare electronic 
susceptibility~\cite{YM80,Mo86,CM11} given as 
\begin{align}
\chi_0(\bm{q})  
= -\frac{1}{N} \sum_{\bm{k}} \sum_{a,b}
\frac{f(\varepsilon^{(0)}_{\bm{k},a}) -f(\varepsilon^{(0)}_{\bm{k}-\bm{q},b}) }
{\varepsilon^{(0)}_{\bm{k},a}-\varepsilon^{(0)}_{\bm{k}-\bm{q},b}}. 
\label{TiSe2_ep_eq4}
\end{align}
Note that, if $\bm{k}$- and $\bm{q}$-dependences of $V_{ep}(a \bm{k}, b \bm{k} \! - \! \bm{q})$ 
in Eq.~(\ref{TiSe2_ep_eq2}) are negligible, $\chi_0(\bm{q})$ corresponds to $\chi(\bm{q})$. 
In Fig.~\ref{TiSe2_ep_fig1}, we show the calculated bare electronic susceptibility $\chi_0(\bm{q})$ 
at different temperatures $T$.  The behavior of the $\bm{q}$ dependence of $\chi_0(\bm{q})$ 
reflects the band structure near $E_F$ [see Fig.~\ref{TiSe2_tb_fig1}(a)], which is in good 
agreement with previous theoretical estimates~\cite{YM80,CM11}.  
We find the temperature sensitive peak in $\chi_0(\bm{q})$ at the M point ($\bm{q}=\bm{q}_{\rm M}$), 
which corresponds to the wave vector of the CDW in monolayer TiSe$_2$.  
An enhancement of $\chi_0(\bm{q}_{\rm M})$ with decreasing temperature induces softening 
of the phonon mode at the M point.  
We note that $\chi_0(\bm{q})$ has a peak also at the $\Gamma$ point. 
However, previous studies have found that the phonon mode at the $\Gamma$ point does not 
show softening~\cite{MSYetal81,CM11,BCM15,DBS15,FHYetal16,SHTetal17,HBBetal17} 
because the phonon frequencies of the optical modes at the $\Gamma$ point is higher than the 
frequency of the softened transverse mode at the M point.  We therefore consider the susceptibility 
$\chi(\bm{q})$ only at the M point in the following discussion.  

Figure~\ref{TiSe2_ep_fig2} shows the temperature dependence of the susceptibility $\chi(\bm{q})$ 
and effective phonon frequency $\omega(\bm{q})$ at the M point ($\bm{q} =\bm{q}_{\rm M}$), 
where we assume $M^{*}\omega^2_0(\bm{q}_{\rm M}) = 10$~eV/\AA$^2$ 
[$\omega_0(\bm{q}_{\rm M}) \simeq  6.11$~THz] as a bare phonon frequency. 
At this  frequency $\omega_0(\bm{q}_{\rm M})$, we find that the susceptibility 
$\chi(\bm{q}_{\rm M})$ becomes larger than $\omega^2_0(\bm{q}_{\rm M})$ 
at $T \simeq 443$~K.  To show this character clearly, we plot 
$\chi(\bm{q}_{\rm M})/\omega^2_0(\bm{q}_{\rm M})$ as a function of $T$ in 
Fig.~\ref{TiSe2_ep_fig2}(a).  
When the susceptibility reaches $\chi(\bm{q}_{\rm M})/\omega^2_0(\bm{q}_{\rm M})=1$, 
the effective phonon frequency $\omega(\bm{q}_{\rm M})$ 
[$=\omega_0(\bm{q}_{\rm M})\sqrt{1-\chi(\bm{q}_{\rm M})/\omega_0^2(\bm{q}_{\rm M})}$] 
vanishes, resulting in the structural phase transition.  
The transition temperature $T_c$ is given by $\omega(\bm{q}_{\rm M})=0$, or 
$\chi(\bm{q}_{\rm M})/\omega^2_0(\bm{q}_{\rm M})=1$, in this estimation.  
Although $T_c$ is higher than the experimental value in this parameter setting, 
the temperature dependent curve of 
$\omega(\bm{q}_{\rm M})$ is in good agreement with experimental result obtained by 
the x-ray diffuse scattering~\cite{HZHetal01}. 

%%  Fig. 9  %%
\begin{figure}[tb]
\begin{center}
\includegraphics[width=\columnwidth]{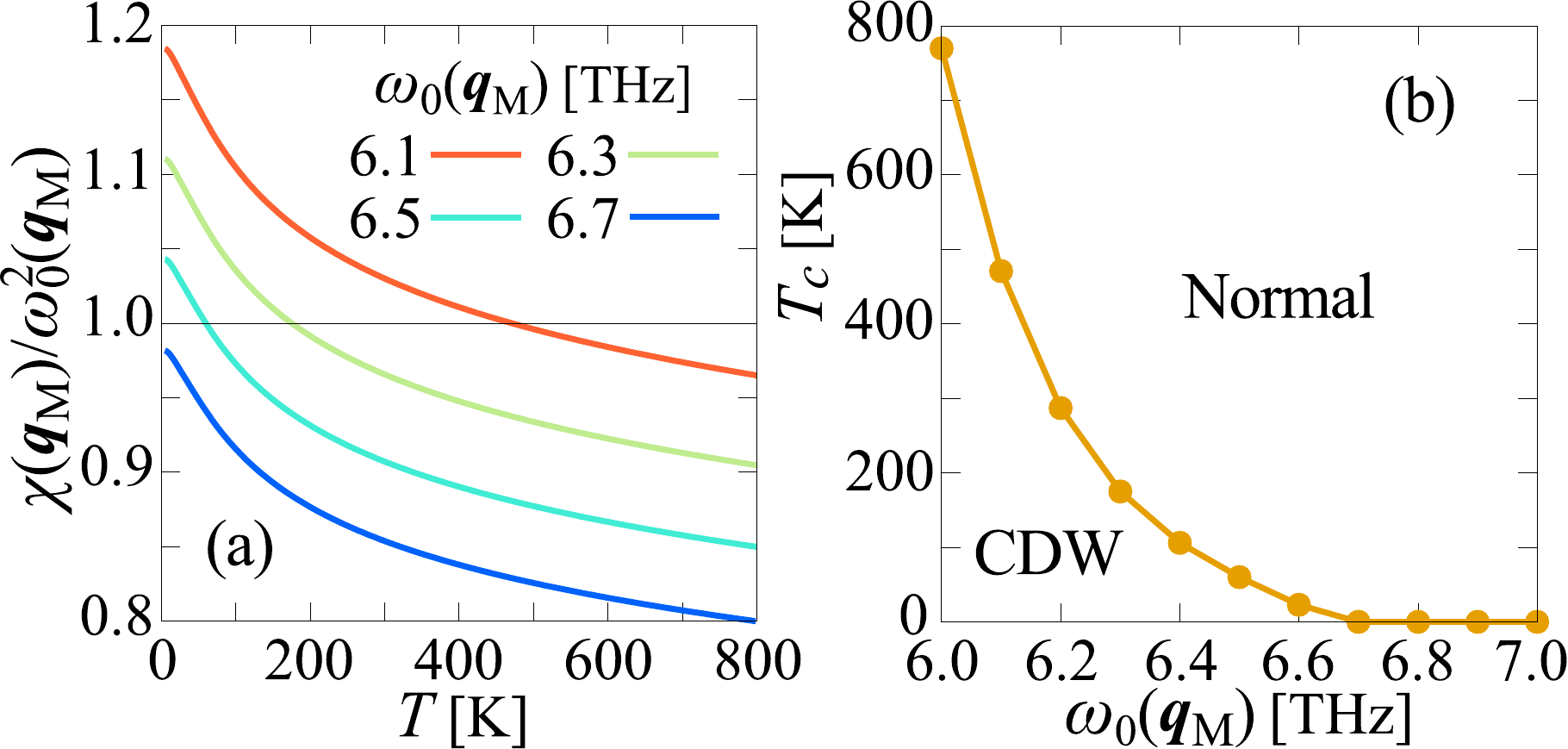}
\caption{
(a) Calculated temperature dependence of $\chi(\bm{q}_{\rm M})/\omega^2_0(\bm{q}_{\rm M})$ 
at different values of $\omega_0(\bm{q}_{\rm M})$.  
(b) Calculated transition temperature $T_c$ as a function of $\omega_0(\bm{q}_{\rm M})$. }
\label{TiSe2_ep_fig3}
\end{center}
\end{figure}
%%%%%%%%

To investigate $\omega_0(\bm{q}_{\rm M})$-dependence of the  critical temperature $T_c$, 
we also show the temperature dependence of $\chi(\bm{q}_{\rm M})/\omega^2_0(\bm{q}_{\rm M})$ 
in Fig.~\ref{TiSe2_ep_fig3}(a) for different values of $\omega_0(\bm{q}_{\rm M})$.  With increasing 
$\omega_0(\bm{q}_{\rm M})$ and thus decreasing $\chi(\bm{q}_{\rm M})/\omega^2_0(\bm{q}_{\rm M})$, 
$T_c$ is suppressed and vanishes at $\omega_0(\bm{q}_{\rm M}) = 6.7$ THz.  
Figure~\ref{TiSe2_ep_fig3}(b) shows the transition temperature $T_c$ as a function of 
$\omega_0(\bm{q}_{\rm M})$.  We find that the calculated $T_c$ is in good agreement 
with the experimental value $T_c\simeq200$ K when $\omega_0(\bm{q}_{\rm M}) = 6.2-6.3$ THz.  
Note that the estimation of $\chi({\bm{q}})$ in Eq.~(\ref{TiSe2_ep_eq2}) corresponds to a random phase approximation~\cite{Mo86} and overestimations of $T_c$ may be due to this approximation.

%%%%%%%%%%%%%%%%%%%%
\subsection{Triple-$\bm{q}$ CDW}

In this subsection, we discuss the stability of the static triple-$\bm{q}$ CDW state induced by 
the electron-phonon coupling $g_{\mu \ell, \nu m} (\bm{k},\bm{q})$.  Here, we estimate the 
change in the total energy when the static triple-$\bm{q}$ crystal structure shown in 
Fig.~\ref{TiSe2_cs_fig3}(d) is realized.  

When the transverse phonon modes at $\bm{q}_{1}$, $\bm{q}_{2}$, and $\bm{q}_{3}$ are frozen 
simultaneously, the corresponding expectation value is $\braket{ Q_{\bm{q}_j}  } = \sqrt{N M^{*}} u$ 
%\cite{SYM85,Mo86} 
and the electron-phonon coupling in the static triple-$\bm{q}$ structure is 
given by
\begin{align}
\mathcal{H}_{ep} = \sum_{\bm{k},\bm{q}_j} \sum_{\mu \ell, \nu m}  \bar{g}_{\mu \ell, \nu m} (\bm{k},\bm{q}_j) u c^{\dag}_{\bm{k},\mu \ell} c_{\bm{k}-\bm{q}_j,\nu m}, 
\label{TiSe2_ep_eq5}
\end{align}
where $ \bar{g}_{\mu \ell, \nu m} (\bm{k},\bm{q}_j) \equiv  \sqrt{M^{*}}g_{\mu \ell, \nu m} (\bm{k},\bm{q}_j)$ 
for $\bm{q}_j=\bm{q}_1$, $ \bm{q}_2$, $\bm{q}_3$ and $u$ corresponds to the magnitude of the 
displacement of the Ti atoms~\cite{SYM85,Mo86}.  
Since Eq.~(\ref{TiSe2_ep_eq5}) is not diagonal for $\bm{k}$ in the original BZ without distortion, we must introduce 
the reduced BZ (RBZ) shown in Fig.~\ref{TiSe2_cs_fig2}(b). 
In order to write the Hamiltonian simply 
in the matrix notation, here we introduce the 11$\times$11 matrices of the transfer integral $\hat{t} (\bm{k})$ 
and electron-phonon coupling $\hat{\bar{g}} (\bm{k},\bm{q})$, and the eleven dimensional vector 
of the annihilation (creation) operator $\bm{c}^{(\dag)}_{\bm{k}}$.  
When we define the row vector 
$\bar{\bm{c}}^{\dag}_{\bm{k}} = (\, \bm{c}^{\dag}_{\bm{k}_0} \; \bm{c}^{\dag}_{\bm{k}_1} \; 
\bm{c}^{\dag}_{\bm{k}_2} \; \bm{c}^{\dag}_{\bm{k}_3} \,)$ with $\bm{k}_j = \bm{k} -\bm{q}_j$ 
and $\bm{q}_0=\bm{0}$, the Hamiltonian of the tight-binding band and electron-phonon coupling 
$\mathcal{H}_{cdw}^{ep} = \mathcal{H}_e + \mathcal{H}_{ep}$ may be written as 
\begin{align}
\mathcal{H}_{cdw}^{ep}
= \sum_{\bm{k} \in {\rm RBZ}} \bar{\bm{c}}^{\dag}_{\bm{k}} \hat{\mathcal{H}}^{ep}_{\bm{k}} \bar{\bm{c}}_{\bm{k}}
= \sum_{\bm{k} \in {\rm RBZ}} \sum_{i,j} \bm{c}^{\dag}_{\bm{k}_i} \hat{\mathcal{H}}^{ep}_{\bm{k}_i,\bm{k}_j} \bm{c}_{\bm{k}_j}, 
\label{TiSe2_ep_eq6}
\end{align}
where $\hat{\mathcal{H}}^{ep}_{\bm{k}_i,\bm{k}_j}$ is the 11$\times$11 block matrix of 
$(\bm{k}_i,\bm{k}_j)$ component of $\hat{\mathcal{H}}^{ep}_{\bm{k}}$ and is given as 
\begin{align}
\hat{\mathcal{H}}^{ep}_{\bm{k}_i,\bm{k}_j}= 
\begin{cases}
\hat{t}({\bm{k}_i}) & (\bm{k}_i=\bm{k}_j) \\
\hat{\bar{g}} (\bm{k}_i,\bm{q}_i+\bm{q}_j) u & (\bm{k}_i \ne \bm{k}_j)
\end{cases}.
\label{TiSe2_ep_eq7}
\end{align}
We estimate the distorted energy band $\varepsilon_{\bm{k},a}$ in the static triple-$\bm{q}$ 
structure by diagonalizing the 44$\times$44 matrix $\hat{\mathcal{H}}^{ep}_{\bm{k}}$ 
in the RBZ.  See Appendix \ref{app_tri} for details.  

Figure~\ref{TiSe2_ep_fig4}(a) shows the calculated energy bands $\varepsilon_{\bm{k},a}$ 
of the undistorted ($u/a=0$) and triple-$\bm{q}$ superlattice ($u/a=0.02$) structures in the 
RBZ.  In the normal state, the conduction-band bottoms at the M points are folded to the 
$\Gamma$ point of the RBZ and the semi-metallic state is realized with the small band 
overlap.  When the electron-phonon coupling induces the lattice displacement with $u\ne0$, 
the band hybridization occurs to open the band gap 
around the $\Gamma$ point in the RBZ. 

%%  Fig. 10  %%
\begin{figure}[tb]
\begin{center}
\includegraphics[width=\columnwidth]{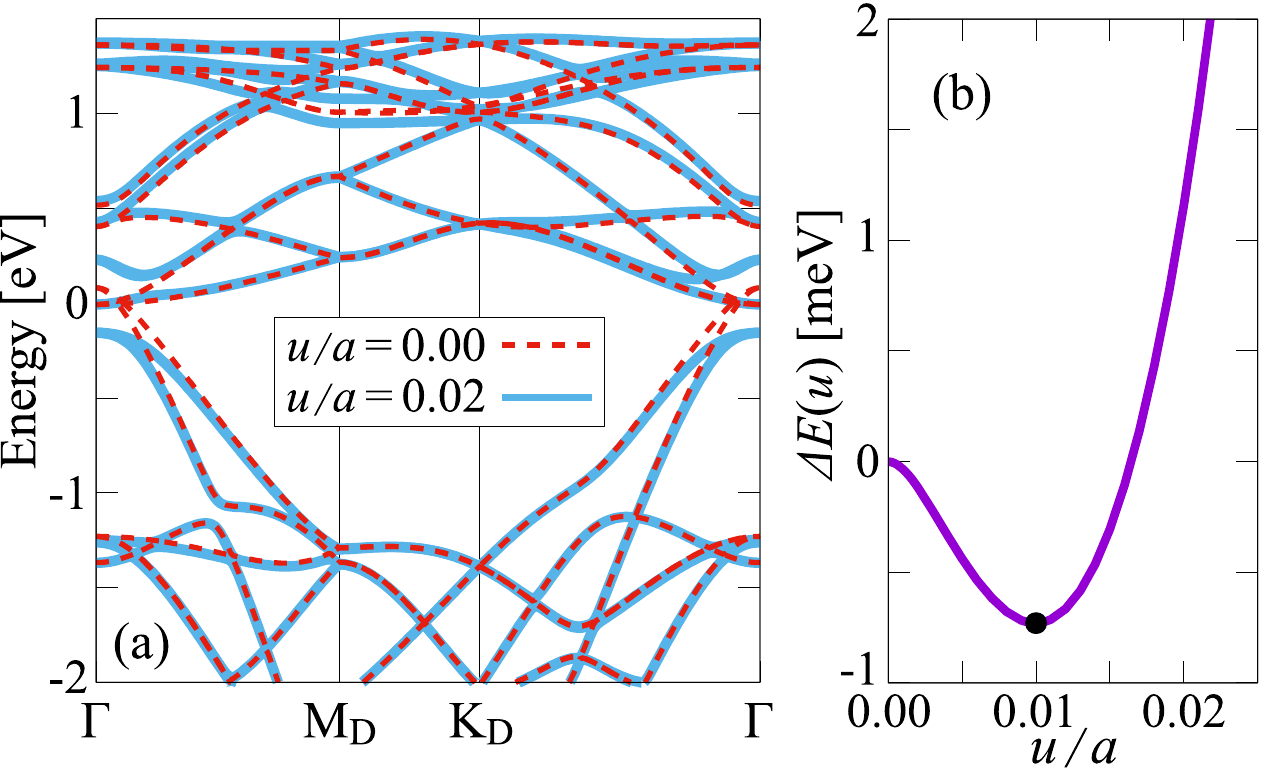}
\caption{
(a) Calculated band dispersions of TiSe$_2$ in the undistorted structure ($u/a=0$) and 
triple-$\bm{q}$ superlattice structure ($u/a=0.02$) displayed in the RBZ.  
The energy 0 corresponds to the Fermi energy in the undistorted structure ($u/a=0$).  
(b) Calculated energy $\Delta E(u)$ as a function of $u$.  We assume $\alpha_{\rm c} = 0.1$ 
and $M^{*}\omega^2_0(\bm{q}_{\rm M}) = 10$~eV/\AA$^2$ 
[$\omega_0(\bm{q}_{\rm M}) \simeq 6.11$~THz], and $a$~($=3.54$~\AA) is the lattice constant.
A solid dot indicates the stationary point in (b).  
}
\label{TiSe2_ep_fig4}
\end{center}
\end{figure}
%%%%%%%%

By the gap opening at the Fermi level, the electronic energy in the triple-$\bm{q}$ 
structure is lowered. The energy difference at zero temperature is simply given as 
\begin{align}
\Delta E_{\rm elec} (u) = \frac{2}{N} \left[ \sum_{\bm{k},a}^{{\rm occ.}} \varepsilon_{\bm{k},a}(u) 
- \sum_{\bm{k},a}^{{\rm occ.}}\varepsilon_{\bm{k},a}(0)  \right], 
\label{TiSe2_ep_eq8}
\end{align}
where $\varepsilon_{\bm{k},a}(u)$ and $\varepsilon_{\bm{k},a}(0)$ are the band energies in 
the triple-$\bm{q}$ and undistorted structures, respectively, and ${\rm occ.}$ indicates the 
sum over the occupied $\bm{k}$ points in the RBZ.  $N$ and 2 in Eq.~(\ref{TiSe2_ep_eq8}) 
correspond to the number of the unit cells in the normal phase and spin degrees of freedom, 
respectively.  
When the atoms are displaced from their equilibrium positions, the energy of the lattice 
system increases as 
\begin{align}
\Delta E_{\rm elas} (u) = \frac{1}{2} \sum_{\bm{q}_j} M^{*} \omega^2_0(\bm{q}_j) u^{2}, 
\label{TiSe2_ep_eq9}
\end{align}
where $\omega_0(\bm{q}_j)$ [$=\omega_0(\bm{q}_{\rm M})$] is the bare phonon frequency 
for $\bm{q}_j = \bm{q}_1$, $\bm{q}_2$, and $\bm{q}_3$.  
The sum of the electronic and elastic terms in Eqs.~(\ref{TiSe2_ep_eq8}) and (\ref{TiSe2_ep_eq9}) 
gives the change in the total energy in the triple-$\bm{q}$ structure, 
\begin{align}
\Delta E (u) = \Delta E_{\rm elec} (u) + \Delta E_{\rm elas} (u). 
\label{TiSe2_ep_eq10}
\end{align}

Figure~\ref{TiSe2_ep_fig4}(b) shows the calculated $\Delta E (u)$ as a function of $u$, 
where we assume $M^{*}\omega^2_0(\bm{q}_{\rm M})\! =\! 10$~eV/\AA$^2$ 
[$\omega_0(\bm{q}_{\rm M}) \! \simeq \! 6.11$~THz] and $\alpha_{\rm c}=0.1$ in the 
electron-phonon coupling $g_{\mu \ell, \nu m} (\bm{k},\bm{q}_j)$.  
The sum over $\bm{k}$ in Eq.~(\ref{TiSe2_ep_eq8}) is evaluated by the tetrahedron 
method~\cite{BJA94} with a sampling of 100$\times$100 $\bm{k}$ points in the RBZ.  
In this parameter setting, the energy curve of $\Delta E (u)$ has a stationary point 
at a finite value of $u$, indicating the realization of the stable triple-$\bm{q}$ CDW state.  
The calculated lattice displacement $u/a=0.010$ at the stationary point is consistent 
with the experimental value $u/a = 0.012$ estimated by the neutron diffraction~\cite{DMW76}. 
Recent x-ray study for monolayer TiSe$_2$ also observed a consistent value~\cite{FHCetal17}. 

%%  Fig. 11  %%
\begin{figure}[tb]
\begin{center}
\includegraphics[width=\columnwidth]{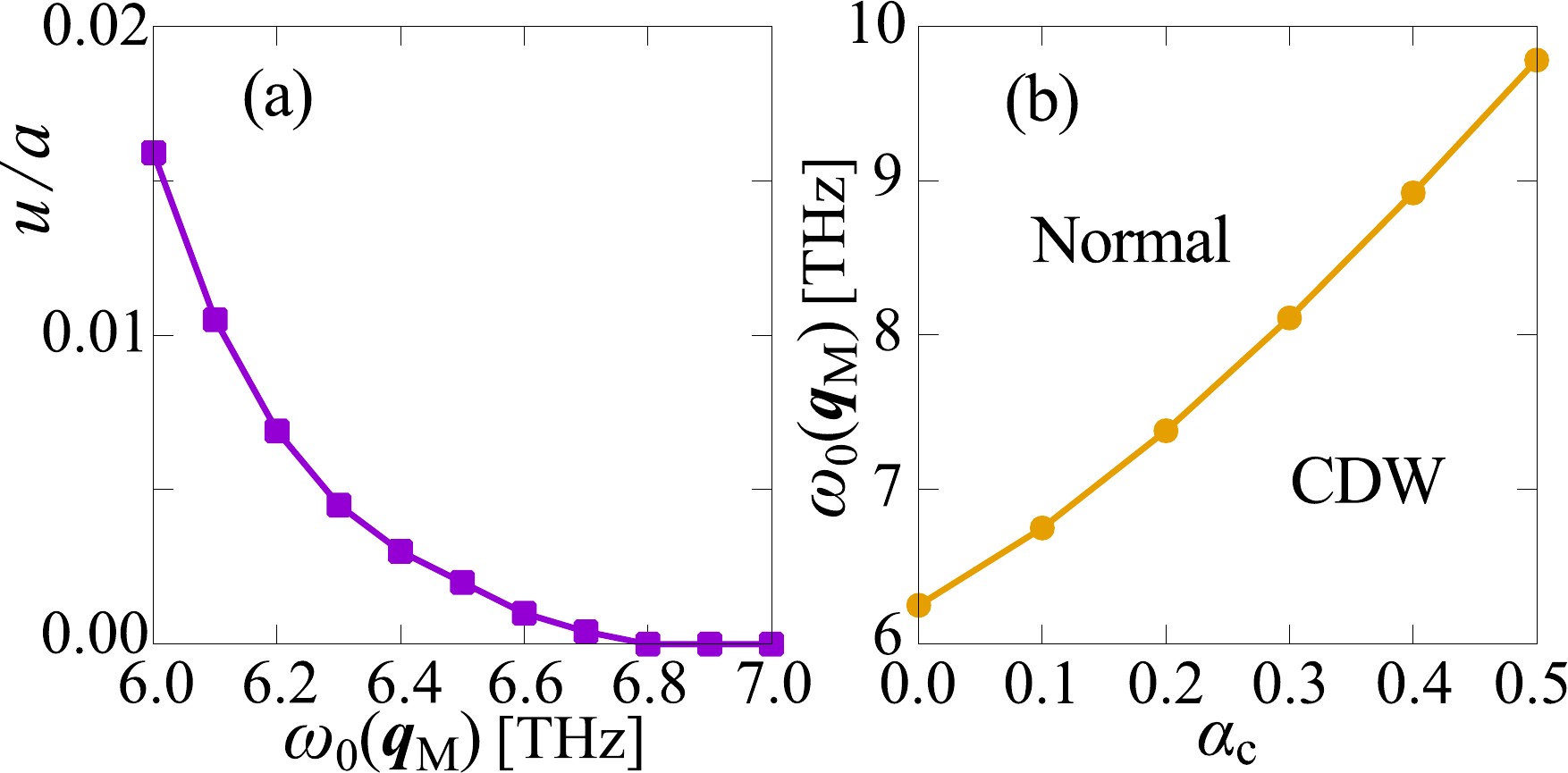}
\caption{
(a) Calculated lattice displacement $u$ in the triple-$\bm{q}$ CDW state as a function 
of $\omega_0(\bm{q}_{\rm M})$ at $\alpha_{\rm c} = 0.1$.  $a$ ($=3.54$~\AA) is the lattice constant. 
(b) Calculated phase boundary between the normal ($u=0$) and triple-$\bm{q}$ CDW ($u\ne0$) states. }
\label{TiSe2_ep_fig5}
\end{center}
\end{figure}
%%%%%%%%

We also check the stability of the triple-$\bm{q}$ CDW state for different values 
of $\omega_0(\bm{q}_{\rm M})$ and $\alpha_{\rm c}$ at $T=0$.  
In Fig.~\ref{TiSe2_ep_fig5}(a), we show the stationary $u$ point in $\Delta E (u)$ 
as a function of $\omega_0(\bm{q}_{\rm M})$ at $\alpha_{\rm c} = 0.1$.  
The lattice displacement $u$ is suppressed with increasing $\omega_0(\bm{q}_{\rm M})$ 
and vanishes at $\omega_0(\bm{q}_{\rm M})=6.8$~THz.  
Note that the phase boundary of $\omega_0(\bm{q}_{\rm M})$ shown in Fig.~\ref{TiSe2_ep_fig5}(a) 
is slightly larger than the boundary shown in Fig.~\ref{TiSe2_ep_fig3}(b) estimated 
from the phonon softening $\chi(\bm{q}_{\rm M})/\omega^2_0(\bm{q}_{\rm M})$.  
This is because the susceptibility  $\chi(\bm{q})$ is derived from the perturbation for a single-$\bm{q}$ phonon mode (see Appendix.~\ref{app_epsus}) and the triple-$\bm{q}$ CDW state including the couplings among different $\bm{q}$ phonon modes is more stable than a single-$\bm{q}$ CDW state, which is also discussed in Refs.~\cite{SYM85} and \cite{BCM15}.
As shown in Fig.~\ref{TiSe2_ep_fig5}(a), the lattice displacement $u$ is in good 
agreement with the experimental value when $\omega_0(\bm{q}_{\rm M}) = 6.0$ -- $6.1$~THz.  
Moreover, we estimate in Fig.~\ref{TiSe2_ep_fig5}(b) the phase boundary between the normal ($u=0$) 
and triple-$\bm{q}$ CDW ($u\ne0$) state in the parameter space of $\alpha_{\rm c}$ 
and $\omega_0(\bm{q}_{\rm M})$ at $T=0$.  We find that, with increasing $\alpha_{\rm c}$, the 
triple-$\bm{q}$ CDW state becomes more stable due to the enhancement of the electron-phonon 
coupling $g_{\mu\ell,\nu m}(\bm{k},\bm{q})$, despite the fact that the bare phonon frequency 
$\omega_0(\bm{q}_{\rm M})$ becomes larger.

%%%%%%%%%%%%%%%%%%%%
%%%%%%%%%%%%%%%%%%%%
\section{Roles of Excitonic Interaction} \label{TiSe2_ex}

In this section, we treat the intersite Coulomb interaction term $\mathcal{H}_{ee}$ 
in the mean-field approximation and discuss roles of the excitonic interaction for 
the triple-$\bm{q}$ CDW state shown in the previous section.  
Hereafter, we assume $\alpha_{\rm c}=0.1$ in the electron-phonon coupling 
$g_{\mu \ell, \nu m} (\bm{k},\bm{q})$ unless otherwise indicated.

%%%%%%%%%%%%%%%%%%%%
\subsection{Excitonic Order}
Let us briefly discuss the mean-field approximation for 
the intersite Coulomb interaction term $\mathcal{H}_{ee}$.  
Details of the calculations are given in 
Appendix~\ref{app_icmf}.  
In TiSe$_2$, the locations of the top of the valence Se $p$ bands and the bottom of 
the conduction Ti $d$ bands are separated in momentum space by 
$\bm{q}_{j}=\bm{q}_1$, $\bm{q}_2$, and $\bm{q}_3$.  
We therefore introduce the excitonic order parameters defined by 
\begin{align}
\Delta^{dp}_{\ell, \nu m}(\bm{k},\bm{q}_j)
\! \equiv
\! - \frac{1}{N}  \sum_{\bm{k}'}   V^{dp}_{\ell, \nu m}(\bm{k} \! - \! \bm{k}') \braket{p^{\dag}_{\bm{k}'-\bm{q}_j,\nu m} d_{\bm{k}',\ell}}, 
\label{TiSe2_mf_eq1} 
\end{align}
for $\bm{q}_{j}=\bm{q}_1$, $\bm{q}_2$, $\bm{q}_3$.
The order parameters thus defined indicate the spontaneous hybridization between 
the Se $p$ and Ti $d$ bands due to the Coulomb interaction $V^{dp}_{\ell, \nu m}(\bm{k}\!-\!\bm{k}')$, 
which results in the excitonic CDW state.  The driving force of the CDW state is 
hence the interband Coulomb interaction.  
The mean-field Hamiltonian may then be written as 
$\mathcal{H}_{ee} \sim \mathcal{H}^{\rm MF}_{ee} = \mathcal{H}^{ex}_{cdw}+ E^{ex}_{0}$ with 
\begin{align}
\mathcal{H}^{ex}_{cdw}
&=  \sum_{\bm{k},\bm{q}_j} \sum_{\ell, \nu m}  
\Delta^{dp}_{\ell, \nu m}(\bm{k},\bm{q}_j) d^{\dag}_{\bm{k},\ell} p_{\bm{k} -\bm{q}_j,\nu m}
+  {\mathrm {H.c.}},
\label{TiSe2_mf_eq2}  \\
E^{ex}_{0}&=- \sum_{\bm{k},\bm{q}_j} \sum_{\ell, \nu m} \Delta^{dp}_{\ell, \nu m}(\bm{k},\bm{q}_j)\braket{d^{\dag}_{\bm{k},\ell} p_{\bm{k} -\bm{q}_j,\nu m}}. 
\label{TiSe2_mf_eq3}
\end{align}
We may write the Hamiltonian in the matrix form in the RBZ, using the 5$\times$6 matrix 
of the order parameter $\hat{\Delta}(\bm{k},\bm{q}_j)$, the five-dimensional vector 
of the annihilation (creation) operator of Ti $d$ orbitals $\bm{d}^{(\dag)}_{\bm{k}}$, 
and six-dimensional vector of the annihilation (creation) operator of Se($\nu$) $p$ 
orbitals $\bm{p}^{(\dag)}_{\bm{k}}$.  Thus we may rewrite Eq.~(\ref{TiSe2_mf_eq2}) as 
\begin{align}
\mathcal{H}^{ex}_{cdw}
= \! \sum_{\bm{k} \in {\rm RBZ}} \bar{\bm{c}}^{\dag}_{\bm{k}} \hat{\mathcal{H}}^{ex}_{\bm{k}} \bar{\bm{c}}_{\bm{k}}
= \! \sum_{\bm{k} \in {\rm RBZ}} \sum_{i\ne j} \bm{c}^{\dag}_{\bm{k}_i} \hat{\mathcal{H}}^{ex}_{\bm{k}_i,\bm{k}_j} \bm{c}_{\bm{k}_j},
\label{TiSe2_mf_eq4}
\end{align}
where $\bm{c}^{\dag}_{\bm{k}_i} = (\, \bm{d}^{\dag}_{\bm{k}_i} \; \bm{p}^{\dag}_{\bm{k}_i} \,)$ 
and $\hat{\mathcal{H}}^{ex}_{\bm{k}_i,\bm{k}_j}$ is the 11$\times$11 block matrix consisting 
of the $(\bm{k}_i,\bm{k}_j)$ components of $\hat{\mathcal{H}}^{ex}_{\bm{k}}$, i.e.,  
\begin{align}
\hat{\mathcal{H}}^{ex}_{\bm{k}_i,\bm{k}_j} 
=
\left[
\begin{array}{c c }
\hat{0}   &   \hat{\Delta}(\bm{k}_i,\bm{q}_{i}+\bm{q}_j)  \\
\hat{\Delta}^{\dag}(\bm{k}_j,\bm{q}_{i}+\bm{q}_j)   &   \hat{0}  \\
\end{array}
\right].
\label{TiSe2_mf_eq5}
\end{align}

In the calculation, we assume the excitonic order parameters defined between the nearest-neighbor Ti 
$d\varepsilon$ ($d_{xy}$, $d_{yz}$, $d_{zx}$)  and Se $p$ ($p_{x}$, $p_{y}$, $p_{z}$) orbitals only.  
We diagonalize the mean-field Hamiltonian, $\hat{\mathcal{H}}^{ex}_{\bm{k}}$ defined above 
plus $\hat{\mathcal{H}}^{ep}_{\bm{k}}$ defined in Eq.~(\ref{TiSe2_ep_eq6}), and optimize 
the order parameter $\hat{\Delta}(\bm{k}_i,\bm{q}_j)$ self-consistently at each value 
of the lattice displacement $u$.  
Using the band dispersion with the optimized order parameters,  we evaluate 
$\Delta E (u) = \Delta E_{\rm elec} (u) + \Delta E_{\rm elas} (u) + E^{ex}_{0}/N$ 
and find the stationary point of $\Delta E (u)$.  
In the self-consistent calculation, we use a sampling of 50$\times$50 $\bm{k}$ 
points in the RBZ.  

%%  Fig. 12  %%
\begin{figure}[t]
\begin{center}
\includegraphics[width=\columnwidth]{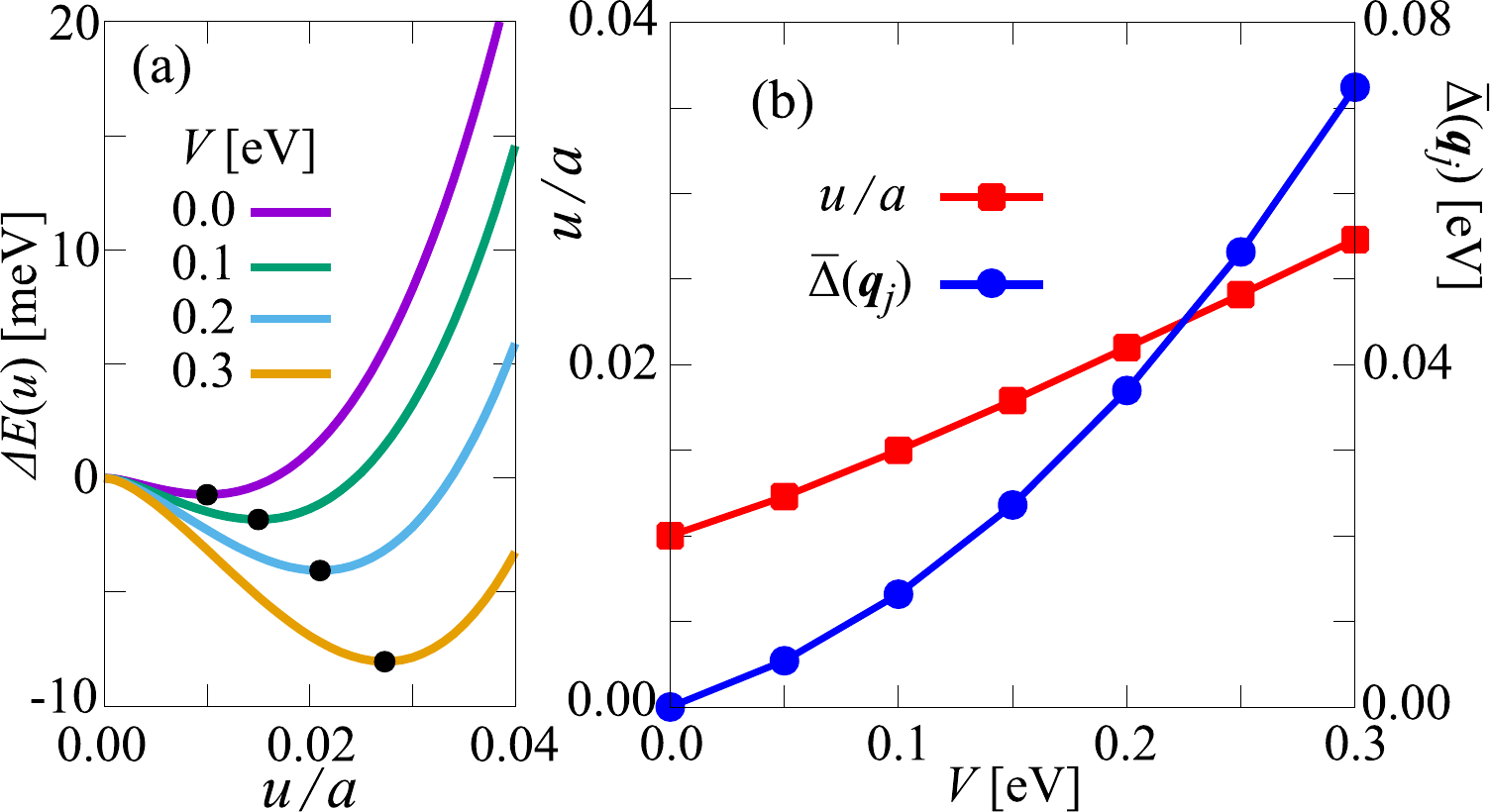}
\caption{
(a) Calculated energy $\Delta E(u)$ as a function of $u$ with different values of $V$.  
We assume $M^{*}\omega^2_0(\bm{q}_{\rm M}) = 10$~eV/\AA$^2$ and $a=3.54$ \AA.  
Solid dots indicate the stationary points. 
(b) Calculated lattice displacement $u$ at the stationary point and the corresponding 
order parameter $\bar{\Delta}(\bm{q}_j)$ as a function of $V$.}  
\label{TiSe2_mf_fig1}
\end{center}
\end{figure}
%%%%%%%%

%%%%%%%%%%%%%%%%%%%%
\subsection{Enhancement of CDW}
 
Figure~\ref{TiSe2_mf_fig1}(a) shows the calculated $u$ dependence of the energy 
$\Delta E (u)$ at $M^{*}\omega^2_0(\bm{q}_{\rm M}) = 10$~eV/\AA$^2$ for different 
values of the Coulomb interaction $V$ [$=V^{dp}_{\ell,\nu m}(\bm{R}_n)$].  
We find that, with increasing $V$, the energy of the triple-$\bm{q}$ CDW state 
becomes more stable and the lattice displacement $u$ at the stationary point is 
enhanced.  The stationary values of $u$ are shown in Fig.~\ref{TiSe2_mf_fig1}(b) 
as a function of $V$, which clearly indicates that the excitonic (intersite Coulomb) 
interactions stabilize the triple-$\bm{q}$ CDW state in TiSe$_2$, working 
cooperatively with the electron-phonon coupling.  

%%  Fig. 13  %%
\begin{figure}[t]
\begin{center}
\includegraphics[width=0.92\columnwidth]{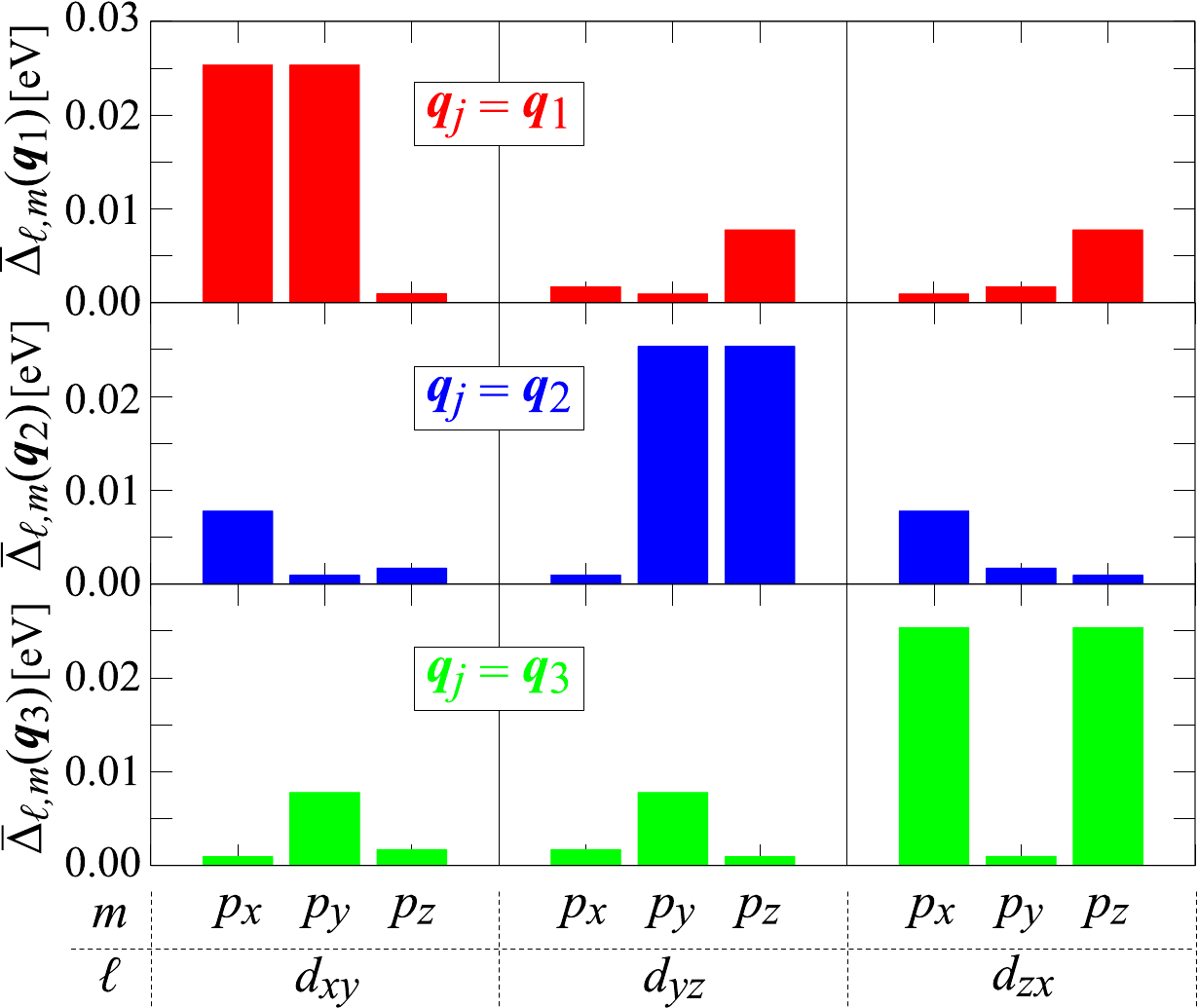}
\caption{Orbital dependence of the averaged order parameters: 
$\bar{\Delta}_{\ell,m}(\bm{q}_1)$ (upper panels), 
$\bar{\Delta}_{\ell,m}(\bm{q}_2)$ (middle panels), and 
$\bar{\Delta}_{\ell,m}(\bm{q}_3)$ (lower panels) 
at $V = 0.3$~eV.  Also see Fig.~\ref{TiSe2_mf_fig1}.}
\label{TiSe2_mf_fig2}
\end{center}
\end{figure}
%%%%%%%%

To study the character of the excitonic ordering, we calculate the average of the 
absolute values of the order parameters defined by
\begin{align}
\bar{\Delta}_{\ell,m} (\bm{q}_j) \equiv  \frac{1}{N}\sum_{\bm{k}\in {\rm RBZ}} 
\sum_{i=0}^{3}\sum_{\nu=1,2} | \Delta^{dp}_{\ell, \nu m}(\bm{k}_i,\bm{q}_j) |.
\label{TiSe2_mf_eq6} 
\end{align}
As an indicator of the excitonic ordering, we also define the total value of the 
averaged order parameters $\bar{\Delta}_{\ell,m} (\bm{q}_j)$, 
\begin{align}
\bar{\Delta} (\bm{q}_j) \equiv \sum_{\ell, m}   \bar{\Delta}_{\ell,m} (\bm{q}_j) . 
\label{TiSe2_mf_eq7} 
\end{align}
As shown in Fig.~\ref{TiSe2_mf_fig1}(b), 
the calculated total order parameter $\bar{\Delta} (\bm{q}_j)$ satisfies 
the relation 
$\bar{\Delta} (\bm{q}_1)=\bar{\Delta} (\bm{q}_2)=\bar{\Delta} (\bm{q}_3)$ 
due to the three-fold rotational symmetry.  
With increasing $V$, $\bar{\Delta} (\bm{q}_j)$ increases monotonically from 
$\bar{\Delta} (\bm{q}_j)=0$ at $V=0$, which indicates that the excitonic order 
coexists with the phononic triple-$\bm{q}$ CDW order and enhances the $d$-$p$ 
hybridizations, supporting the realization of the stable triple-$\bm{q}$ CDW state. 

Figure~\ref{TiSe2_mf_fig2} shows the orbital dependence of the averaged order 
parameters $\bar{\Delta}_{\ell,m} (\bm{q}_j)$ at $V=0.3$~eV.  
We find that the components between the Ti $d_{xy}$ orbital and Se $p_x$ and $p_y$ 
orbitals [$\bar{\Delta}_{xy,x} (\bm{q}_1)  =  \bar{\Delta}_{xy,y} (\bm{q}_1)$] are 
dominant in the order parameter with $\bm{q}_1$.  
This behavior is understood from the orbital character of the undistorted band 
structure shown in Figs.~\ref{TiSe2_tb_fig2} and ~\ref{TiSe2_tb_fig3}; 
the conduction band around the M$_1$ point is mostly given by the Ti $d_{xy}$ orbital 
and the valence bands around the $\Gamma$ point are mostly given by the Se $p$ 
orbitals.  We find that the components $\bar{\Delta}_{xy,x} (\bm{q}_1)$ and 
$\bar{\Delta}_{xy,y} (\bm{q}_1)$ are dominant 
but the component $\bar{\Delta}_{xy,z} (\bm{q}_1)$ is very small.  This is because 
the Se $p_z$ orbital are nearly perpendicular to the Ti $d_{xy}$ orbital but the Se 
$p_x$ and $p_y$ orbitals can enhance the $pd\pi$ bonding with the Ti $d_{xy}$ orbital.  
In the same way, the $d_{yz}$ components are dominant in the order parameter 
with $\bm{q}_2$ [$\bar{\Delta}_{yz,y} (\bm{q}_2)=\bar{\Delta}_{yz,z} (\bm{q}_2)$] 
and the $d_{zx}$ components are dominant in the order parameter with $\bm{q}_3$ 
[$\bar{\Delta}_{zx,z} (\bm{q}_3)=\bar{\Delta}_{zx,x} (\bm{q}_3)$], reflecting the 
orbital character in the undistorted band dispersions.  

%%  Fig. 14  %%
\begin{figure}[t]
\begin{center}
\includegraphics[width=0.95\columnwidth]{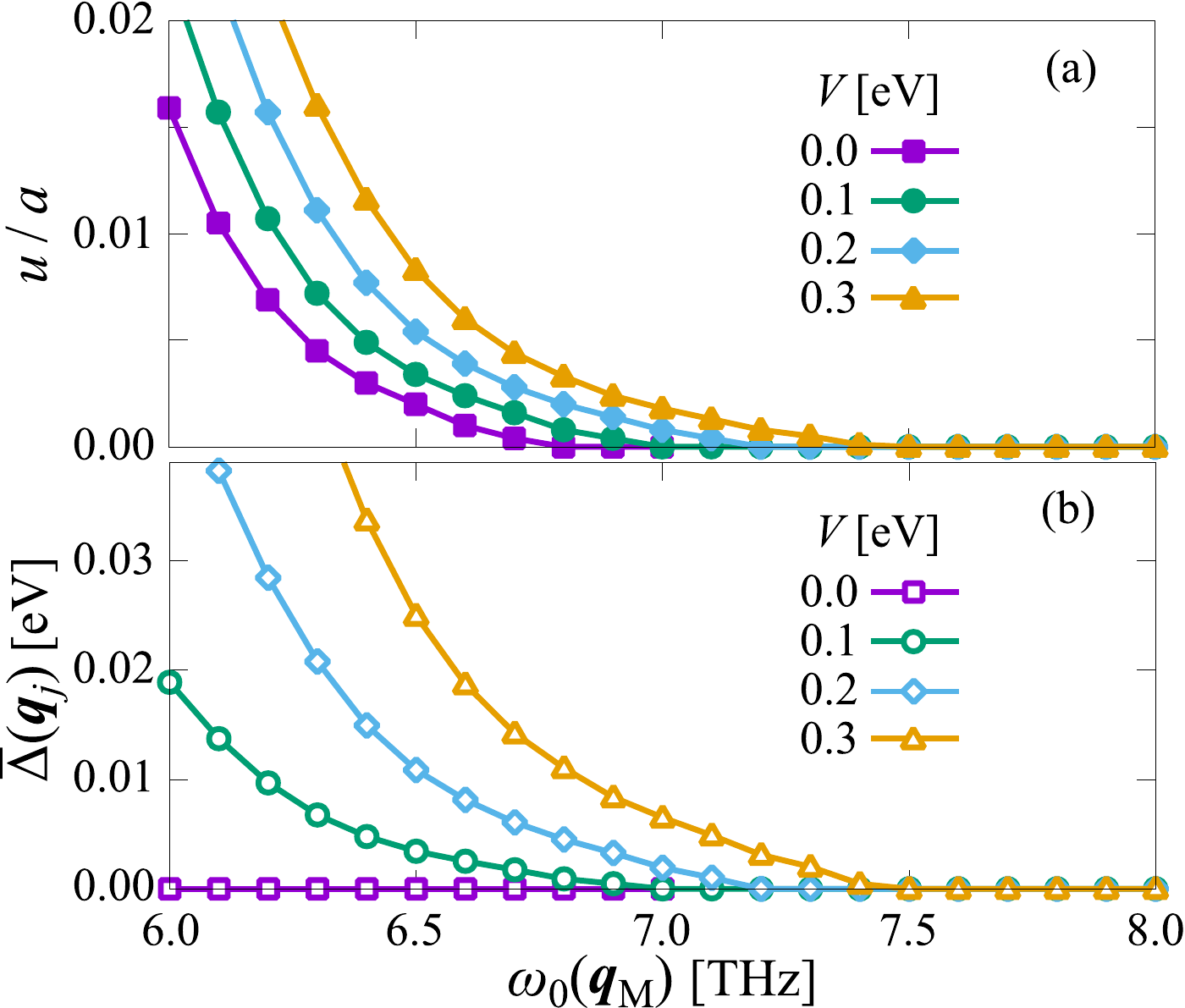}
\caption{
Calculated $\omega_0(\bm{q}_{\rm M})$ dependence of (a) the lattice displacement $u$ 
and (b) order parameter $\bar{\Delta}(\bm{q}_j)$ for different values of $V$.  
$a$~$(=3.54$ \AA) is the lattice constant.}  
\label{TiSe2_mf_fig3}
\end{center}
\end{figure}
%%%%%%%%

%%  Fig. 15  %%
\begin{figure}[tb]
\begin{center}
\includegraphics[width=0.6666\columnwidth]{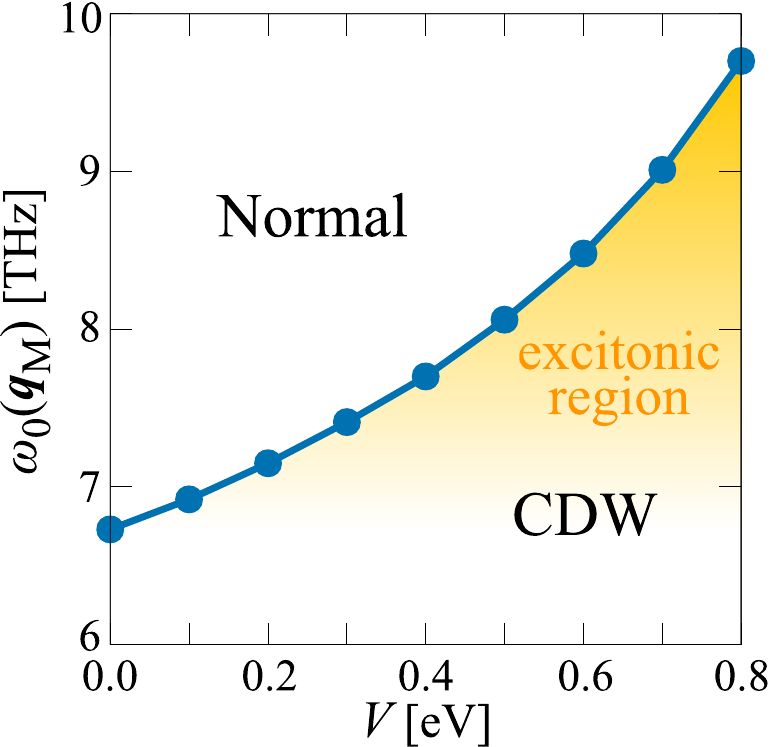}
\caption{Calculated ground-state phase diagram of the triple-$\bm{q}$ CDW state 
in the parameter space of $V$ and $\omega_0(\bm{q}_{\rm M})$.}  
\label{TiSe2_mf_fig4}
\end{center}
\end{figure}
%%%%%%%%

We also check the stability of the triple-$\bm{q}$ CDW state for different values 
of $\omega_0(\bm{q}_{\rm M})$ and $V$.  Figure~\ref{TiSe2_mf_fig3} shows 
$\omega_0(\bm{q}_{\rm M})$ dependence of the lattice displacement $u$ 
and order parameter $\bar{\Delta}(\bm{q}_j)$ for different values of $V$.   
With increasing $\omega_0(\bm{q}_{\rm M})$, the PLD in the triple-$\bm{q}$ CDW state 
is suppressed, but with increasing $V$, the lattice displacement $u$ is enhanced.  
Similarly, $\bar{\Delta}(\bm{q}_j)$ is suppressed with increasing $\omega_0(\bm{q}_{\rm M})$.  
We note that the pure excitonic state, where the triple-$\bm{q}$ CDW state occurs 
without lattice displacements, is not realized in our calculations, similar to the previous report in 
Refs.~\cite{ZFBetal13} and \cite{WSY15}.  
Figure~\ref{TiSe2_mf_fig4} shows the ground-state phase diagram in the 
parameter space of $V$ and $\omega_0(\bm{q}_{\rm M})$. 
Apparently, the area 
of the triple-$\bm{q}$ CDW phase is enlarged with increasing $V$.  
The excitonic interaction $V$ thus enhances the triple-$\bm{q}$ CDW state 
in TiSe$_2$.  We may therefore regard the CDW state in this enlarged region 
as the exciton-induced CDW state.  

%%  Fig. 16  %%
\begin{figure*}[t]
\begin{center}
\includegraphics[width=2.05\columnwidth]{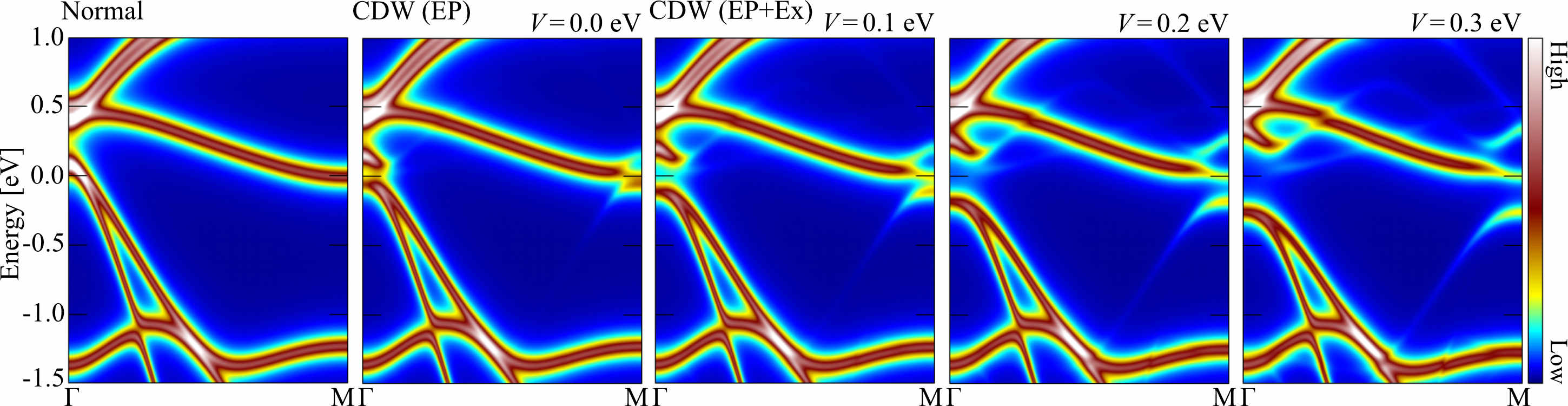}
\caption{
Calculated single-particle spectra $A(\bm{k},\omega)$ of the normal and triple-$\bm{q}$ 
CDW states of the monolayer TiSe$_2$, presented in the unfolded BZ.  We assume the 
self-consistent solutions for $V=0.0$, $0.1$, $0.2$, and $0.3$ eV shown in 
Fig.~\ref{TiSe2_mf_fig1}.  The spectra are broadened with $\eta=0.05$ eV.  
The energy zero is set to the lowest edge of the band above the Fermi level.
}  
\label{TiSe2_spec_fig1}
\end{center}
\end{figure*}
%%%%%%%%

%%%%%%%%%%%%%%%%%%%%
%%%%%%%%%%%%%%%%%%%%
\section{Electronic structure in CDW} \label{TiSe2_es}

In order to discuss the electronic structure of the triple-$\bm{q}$ CDW state, 
here we calculate the single-particle spectrum, simulating the angle-resolved photoemission 
spectroscopy (ARPES), and also the electronic charge density distribution in the TB 
approximation, discussing the local charge distribution in the CDW state of TiSe$_2$.

%%%%%%%%%%%%%%%%%%%%
\subsection{Single-particle Spectrum}

In our one-body approximation, the single-particle spectrum is given by 
\begin{align}
A(\bm{k},\omega) = \sum_{\mu\ell} \sum_{a} |u_{\bm{q}_0\mu\ell,a}(\bm{k})|^{2} \delta(\omega-\varepsilon_{\bm{k},a} ), 
\label{TiSe2_spec_eq1} 
\end{align}
where $u_{\bm{q}_0\mu\ell,a}(\bm{k})$ is the coefficient of the unitary  transformation 
$c_{\bm{k}-\bm{q}_j,\mu\ell} = \sum_{a} u_{\bm{q}_j\mu\ell,a}(\bm{k}) \gamma_{\bm{k},a}$ 
in the diagonalization of the 44$\times$44 Hamiltonian matrix 
$\hat{\mathcal{H}}^{ep}_{\bm{k}}+\hat{\mathcal{H}}^{ex}_{\bm{k}}$. 
Detailed derivation is given in Appendix~\ref{app_spec}.  
In the spectral calculation, each $\delta$-function in Eq.~(\ref{TiSe2_spec_eq1}) is 
represented by a Lorentzian function with a finite broadening factor $\eta$. 

Calculated results for $A(\bm{k},\omega)$ in the normal and triple-$\bm{q}$ CDW states 
are shown in Fig.~\ref{TiSe2_spec_fig1} along the line $\Gamma - \mathrm{M}$ $(\mathrm{M}_1$) 
of the unfolded BZ.  
We assume $V=0.0$, $0.1$, $0.2$, and $0.3$ eV to obtain the self-consistent solutions 
for the CDW states as in Fig.~\ref{TiSe2_mf_fig1}.  
In the normal state, the single-particle spectrum reproduces the semimetallic band structure 
with a small band overlap as shown in Fig.~\ref{TiSe2_tb_fig1} and Fig.~\ref{TiSe2_tb_fig2}.  
In the CDW state without the excitonic interaction ($V=0$ eV), the single-particle spectrum 
shows the small hybridization gaps both in the valence band around the $\Gamma$ point and 
in the conduction band around the M point.  The gaps open due to folding and splitting of the 
bands in the RBZ, caused by the lattice distortion, as shown in Fig.~\ref{TiSe2_ep_fig4}.  
With increasing $V$, the energy gap $E_g$ becomes larger due to the enhancement of 
the triple-$\bm{q}$ CDW state, where the calculated energy gaps are given by $E_g \sim$ 0.06, 0.11, 0.18, and 0.26 eV at $V=0.0$, $0.1$, $0.2$, and $0.3$ eV, respectively. 
In addition, with increasing $V$, the single-particle spectrum clearly indicates the band 
folding behavior, giving rise to the 2$\times$2 superlattice formation.  
The effect of the band folding has clearly been observed at the M point of the unfolded 
BZ in the ARPES experiments~\cite{CCFetal15,CCWetal16,SNSetal16}, which is consistent 
with our calculated results shown in Fig.~\ref{TiSe2_spec_fig1}.  The additional spectral weight can 
clearly be observed around the M point of the BZ, reflecting the bands around the $\Gamma$ point, 
which is caused by the spontaneous hybridization between the valence and conduction bands.  

%%  Fig. 17  %%
\begin{figure*}[tb]
\begin{center}
\includegraphics[width=1.4\columnwidth]{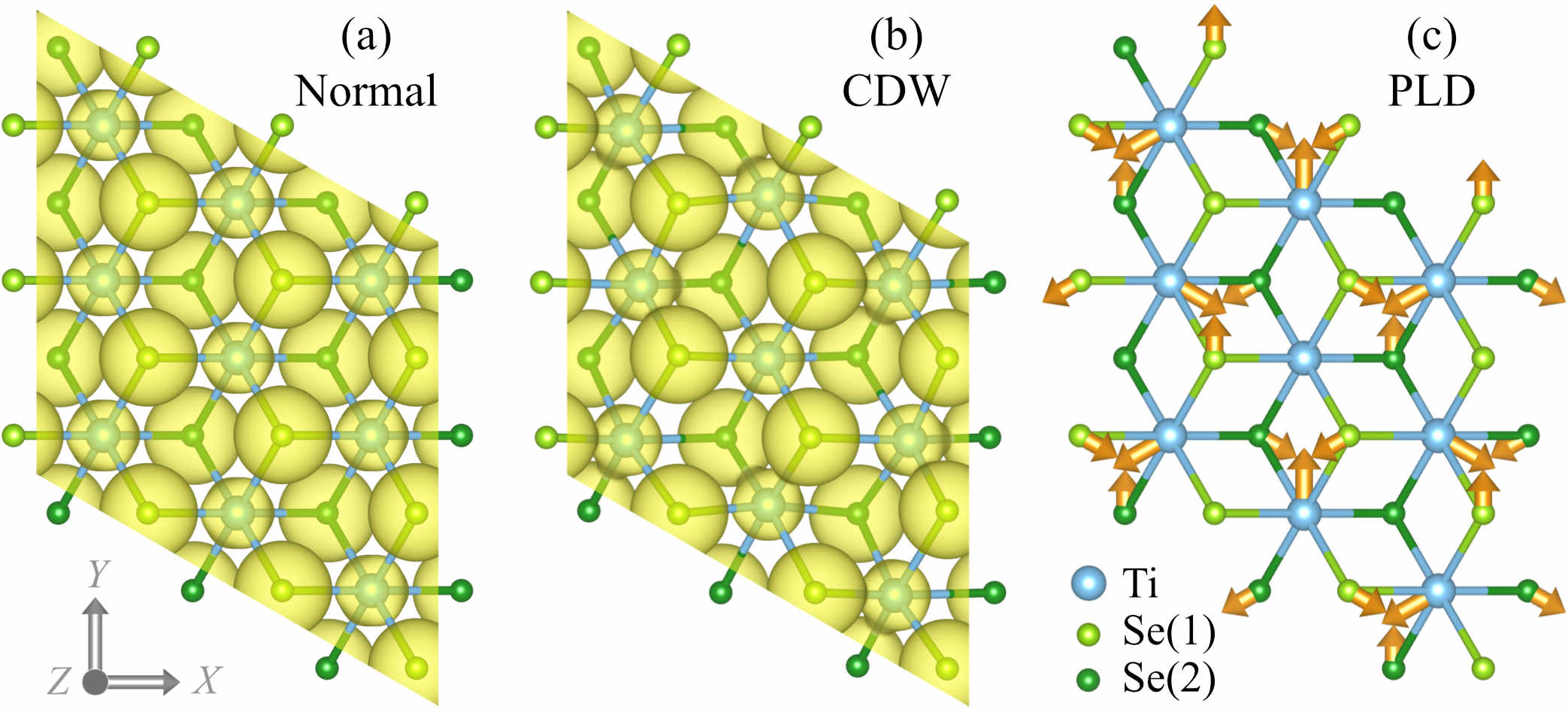}
\caption{
Calculated isosurface of the electronic charge density $\rho(\bm{r})$ in the (a) normal 
and (b) triple-$\bm{q}$ CDW states.  In (b), we assume the self-consistent solution 
for $V=0.3$ eV shown in Fig.~\ref{TiSe2_mf_fig1}.  We use the Slater-type orbitals as 
the atomic orbitals and plot the isosurfaces at an isovalue 0.025 in both (a) and (b).  
(c) Schematic representation of the periodic lattice displacement (PLD) in the 
triple-$\bm{q}$ CDW state.}
\label{TiSe2_cden_fig1}
\end{center}
\end{figure*}
%%%%%%%%

Two remarks are in order. 
First, the intersite Coulomb interaction $V$ is essential to reproduce the experimental ARPES 
spectrum in the monolayer TiSe$_2$~\cite{CCFetal15,CCWetal16,SNSetal16}. 
When the intersite Coulomb interaction is absent ($V=0$), the calculated band gap and 
effect of band folding are small and weak in comparison with the experiment.  
However, the band gap and folding spectrum are enhanced with increasing $V$ 
and the single-particle spectrum around $V\sim0.2$ eV may be in good agreement with 
the ARPES spectrum.  
Second, since our model omits the spin-orbit coupling, 
our calculations do not reproduce the 
spin-orbit splitting of the valence Se 4$p$ bands at the $\Gamma$ point~\cite{AMS85,GPJ17}. 
The spin-orbit interaction is required for more accurate comparison.

%%%%%%%%%%%%%%%%%%%%
\subsection{Charge Density Distribution}

To elucidate the local electronic structure in the triple-$\bm{q}$ CDW state, we calculate 
the charge density distribution in TiSe$_2$.  In general, the electronic charge density is 
given by 
\begin{align}
\rho(\bm{r}) =  \sum_{\bm{k}}\sum_{a}  |\psi_{\bm{k},a} (\bm{r})|^2 f(\varepsilon_{\bm{k},a}), 
\label{TiSe2_cden_eq1}
\end{align}
where $\psi_{\bm{k},a} (\bm{r})$ is the Bloch wave function of band $a$.  Here, we do not 
write the elementary charge $e$ explicitly.  Note that the charge density $\rho(\bm{r})$ 
in Eq.~(\ref{TiSe2_cden_eq1}) is a density of electrons; the charge distribution of atomic 
cores (ions) should be added in the evaluation of the total charge density or net electric 
polarization.  
In the TB approximation, the Bloch functions are given by the linear combinations of atomic 
orbitals.  The charge density $\rho(\bm{r})$ can then be rewritten as 
\begin{align}
\rho(\bm{r}) 
= \! \sum_{\bm{R}_i,\bm{R}_j}\sum_{\mu\ell,\nu m} \! \braket{c^{\dag}_{i,\mu\ell} c_{j,\nu m}} \phi_{\ell} (\bm{r} \! - \! \bm{R}_{i\mu}) \phi_{m} (\bm{r} \! -\! \bm{R}_{j\nu}), 
\label{TiSe2_cden_eq8}
\end{align}
where $c^{(\dag)}_{i,\mu\ell}$ is the annihilation (creation) operator of an electron on the 
atomic orbital $\mu\ell$ in the $i$-th unit cell~\cite{KO16}.  Here, we use the Slater-type orbitals as 
the atomic orbitals $\phi_{\ell} (\bm{r})$, which we have used in the estimation of the 
overlap integrals in Sec. II C.  
In the evaluation of Eq.~(\ref{TiSe2_cden_eq8}), we include the on-site expectation 
values $\braket{c^{\dag}_{i,\mu\ell} c_{i,\mu m}}$ for all the atoms and the $d$-$p$ 
bonding contributions $\braket{d^{\dag}_{i,\ell} p_{j,\nu m}}$ between the nearest-neighbor 
Ti and Se($\nu$) atoms.   
We omit other (more distant) expectation values because they 
are negligibly small.  In the triple-$\bm{q}$ CDW state, we extend the unit cell as shown 
in Fig.~\ref{TiSe2_cs_fig3}(d) and estimate the expectation values for the four TiSe$_2$ 
units in the extended 2$\times$2 unit cell.  

Figure~\ref{TiSe2_cden_fig1} shows the calculated charge density distributions in the normal and 
triple-$\bm{q}$ CDW states.  
Here, we assume as the CDW state the self-consistent solution for $V=0.3$~eV shown in Fig.~\ref{TiSe2_mf_fig1}. 
Note that the results are qualitatively the same even if we assume $V=0.2$~eV.  
In the normal state, we find that the isosurface surrounds each atom and the charge densities $\rho(\bm{r})$ 
around Se sites are larger than those around Ti sites, reflecting the occupation numbers of electrons.  
We also find that, in the CDW state, the radius of the isosurface surrounding each atom does not change 
drastically from the normal state, indicating that the CDW state in TiSe$_2$ is not a site-centered charge 
order~\cite{vdBK08} that should have an inequivalent deviation in the on-site electronic occupations. 
Instead, the deviation in $\rho(\bm{r})$ appears between the Ti and Se sites due to the formation of the bonding 
orbital (trimer) of the Ti $d$ and two Se $p$ orbitals in the distorted TiSe$_6$ octahedra.  
Therefore the trimerization of the Ti and two Se orbitals is the essence in the electronic structure, 
and the bond-centered CDW~\cite{vdBK08} is a suitable description of the CDW in TiSe$_2$. 

\begin{figure}[tb]
\begin{center}
\includegraphics[width=0.9\columnwidth]{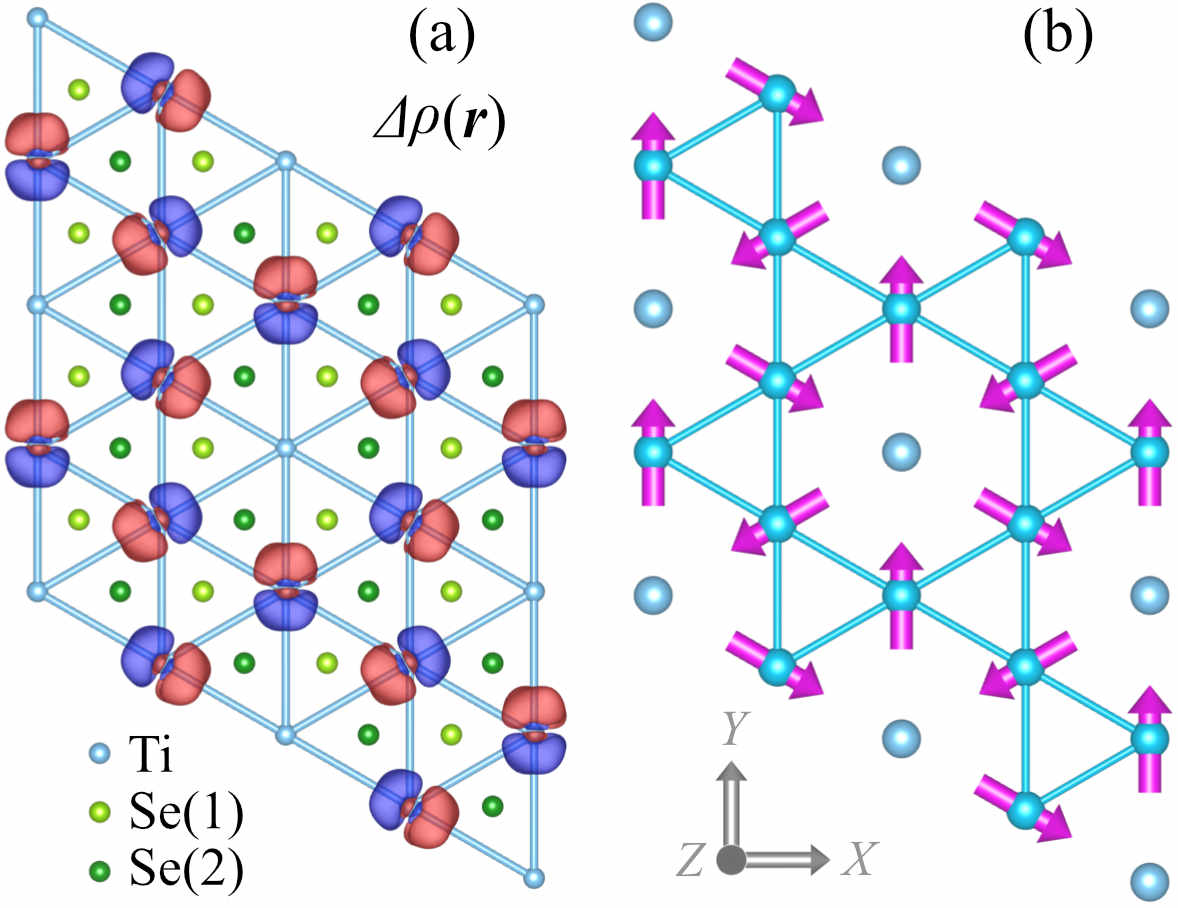}
\caption{
(a) Isosurface of the difference in the electronic charge densities between the triple-$\bm{q}$ CDW 
and normal states, $\Delta \rho(\bm{r}) =  \rho_{\rm CDW}(\bm{r}) -  \rho_{\rm N}(\bm{r})$, where 
we assume the self-consistent solutions for $V=0.3$ eV given in Fig.~\ref{TiSe2_mf_fig1}.   
We plot the isosurface at an isovalue $\pm$0.007 and the red and blue surfaces indicate the 
positive and negative part of $\Delta \rho(\bm{r})$, respectively.  
(b) Schematic representation of the electric polarization in the electronic density shown in (a).  
An arrow indicates the electric dipole at the Ti sites.}
\label{TiSe2_cden_fig2}
\end{center}
\end{figure}

To illustrate the deviation in the electronic density clearly, Fig.~\ref{TiSe2_cden_fig2}(a) shows the difference in the 
electron density distributions between the CDW and normal states 
$\Delta \rho(\bm{r}) =  \rho_{\rm CDW}(\bm{r}) -  \rho_{\rm N}(\bm{r})$.  
Clearly, $\Delta \rho(\bm{r})$ exhibits the electric dipole structure 
in the distorted TiSe$_6$ octahedra
due to the deviation in the electronic density.  The schematic representation 
of the dipole structure is shown in Fig.~\ref{TiSe2_cden_fig2}(b), where we only describe the Ti 
sites, omitting the Se sites, and an arrow indicates the electric polarization in  $\Delta \rho(\bm{r})$. 
We thus find in Fig.~\ref{TiSe2_cden_fig2}(b) that the polarization in the Ti sites 
forms a kagom\'e network and the 
dipoles show the structure of clockwise and anticlockwise 
vortices on the triangles in the kagom\'e network.  

The polarization structure in Fig.~\ref{TiSe2_cden_fig2} given by the electronic density $\rho(\bm{r})$ 
demonstrates the presence of a vortex-like antiferroelectric structure.  In this vortex-like structure shown 
in Fig.~\ref{TiSe2_cden_fig2}(b), we identify the local electric toroidal moment~\cite{DT90,PPKetal06,GLYetal12} 
along the $Z$-axis at the center of the vortex defined by the three 
dipoles on the triangle.  
The clockwise and anticlockwise vortices making opposite axial toroidal vectors can be regarded 
as the antiferroelectric toroidal network.  

%%  Fig. 18  %%
%%%%%%%%

%%%%%%%%%%%%%%%%%%%%
%%%%%%%%%%%%%%%%%%%%
\section{Discussion and Summary}

%Discussion
Here, we 
discuss the implications of our results in recent developments in studies of TiSe$_2$.
Recent experimental studies have pointed out the difficulty in the pure excitonic driving force in 
the formation of the CDW state in TiSe$_2$, where the PLD is survived even if the excitonic 
interactions are screened~\cite{HJDetal16,NSSetal17,PLMetal14,KdlPLetal17}.  
However, these studies have admitted possible contributions of the excitonic correlations 
to the development in the rigid CDW state of the electron-phonon coupled system~\cite{HJDetal16,NSSetal17,PLMetal14,KdlPLetal17}.  Moreover, other experimental results ~\cite{MPNetal16,KRVetal17}
 have rather supported the cooperative scenario for the CDW formation, 
as was suggested theoretically~\cite{vWNS10-2,ZFBetal13,WSY15,KZFetal15}.  
In this paper, we have shown that the electron-phonon and excitonic interactions work cooperatively 
with each other to enhance the stability of the triple-$\bm{q}$ CDW state.  
Our microscopic theory has thus advanced and strengthened the cooperative 
scenario suggested in simplified models~\cite{vWNS10-2,ZFBetal13,WSY15,KZFetal15}.  

However, the pure phononic mechanism may not be denied completely in our present theory 
since we have shown the stability of the CDW state without the excitonic interaction, as discussed 
in Sec.~\ref{TiSe2_ep}.  
To regard the CDW state in TiSe$_2$ as an excitonic insulator or excitonic condensation state 
assertively, we must elucidate the contribution of an excitonic interaction, namely an 
inter-band Coulomb interaction, in comparison with experiment.  
As one of the methods of verification, we may suggest the application of time-resolved 
experiments~\cite{MJBetal11,RHWetal11,HRKetal12,MEUetal16,MHGetal17,WTHetal18}, 
where we can make use of the difference in the time scales between the excitonic and 
phononic systems.  In particular, theoretical studies of the photo-induced dynamics for excitonic orders
have been investigated in the two-band excitonic insulator models~\cite{GWE16,MGEetal17}.   To understand the real 
materials, however, we need a quantitative microscopic models, for which our theoretical study 
for TiSe$_2$ will be proven to be useful.  
Besides the photo-induced dynamics, responses to other external fields~\cite{SKO16,SO16,MO16} 
should also be studied theoretically, for which our microscopic model for the CDW state of 
TiSe$_2$ will be valuable to elucidate the contributions of the excitonic interaction.  

In such studies, we also need to extend our monolayer model to the bulk 1$T$-TiSe$_2$ model.  
In the bulk structure, because the bottoms of the conduction band are located at the L points, which are $k_Z = \pi/c$ above the M points~\cite{ZF78,FGH97}, a triple-$\bm{q}$ CDW state with the modulation vector $\bm{q}_{\rm L}=\bm{q}_{\rm M} + (\pi/c) \bm{e}_{Z}$ is anticipated, where the TiSe$_2$ layers with antiparallel lattice displacements stack alternately along the $Z$-direction, keeping the in-plane structure to be the same as our monolayer triple-$\bm{q}$ structure.
Our monolayer studies have thus captured the essential characters of the in-plane structure of the bulk system.
However, to understand the bulk 1$T$-TiSe$_2$ in detail, it is necessary to investigate the roles of the inter-layer coupling carefully.

% summary
To conclude, we have investigated the electronic structure and microscopic mechanism 
of the triple-$\bm{q}$ CDW state in the monolayer TiSe$_2$ on the basis of the realistic 
multi-orbital $d$-$p$ model  with the electron-phonon coupling and intersite Coulomb 
(excitonic) interactions.  The phononic and excitonic mechanisms of the CDW transition 
have thus been considered.  
First, using the first-principles band-structure calculations, we have constructed the tight-binding 
bands made from the Ti 3$d$ and Se 4$p$ orbitals in the monolayer TiSe$_2$.  
From the undistorted band structure, we have shown that the valence-band top at the 
$\Gamma$ point is characterized by the Se $p$ orbitals and the conduction-band bottom 
at the M$_1$, M$_2$, and M$_3$ points are characterized by the Ti $d_{xy}$, $d_{yz}$, and 
$d_{zx}$ orbitals, respectively.  
Next, we have constructed the electron-phonon coupling in the tight-binding approximation 
for the transverse phonon modes, of which the softening has been observed experimentally~\cite{HZHetal01}.  
Taking into account the electron-phonon coupling only, we have shown that 
the transverse phonon mode softens at the M point of the BZ and that the instability toward 
the triple-$\bm{q}$ CDW state occurs when the transverse 
modes at the M$_1$, M$_2$, and M$_3$ points are frozen simultaneously 
(i.e., representing the phononic mechanism for the triple-$\bm{q}$ CDW state).   

Furthermore, we have introduced the intersite Coulomb interaction between the nearest-neighbor 
Ti and Se atoms, which induces the excitonic instability between the valence Se 4$p$ and 
conduction Ti 3$d$ bands.  
We have treated the intersite Coulomb (excitonic) interaction in the mean-field approximation 
and have shown that the excitonic interaction favors to further stabilize the triple-$\bm{q}$ CDW 
state caused by the phononic mechanism.  
We have thus demonstrated that the electron-phonon and excitonic interactions 
cooperatively stabilize the triple-$\bm{q}$ CDW state in the monolayer TiSe$_2$.  
Here, we have also shown the orbital characters of the excitonic order parameters explicitly in the triple-$\bm{q}$ CDW state.
Using the mean-field solution for the ground state of the proposed model, we have 
calculated the single-particle spectrum in the triple-$\bm{q}$ CDW state to reproduce the 
band folding spectrum observed in the ARPES experiments.  
To illustrate the electronic structure in the triple-$\bm{q}$ CDW state intuitively, we 
have also calculated the charge density distribution in real space and have shown that the 
the bond-type CDW occurs in the monolayer TiSe$_2$. 
In addition, we have found out a vortex-like antiferroelectric electron polarization in the kagom\'e network of Ti atoms.

\begin{acknowledgments}
The authors would like to thank Y.~\={O}no, K.~Sugimoto, T.~Toriyama, H.~Watanabe, and T.~Yamada 
for enlightening discussions on theoretical aspects and S.~Kito, A.~Nakano, and H.~Sawa for 
experimental aspects.  
This work was supported in part by Grants-in-Aid for Scientific Research from JSPS 
(Projects No.\,26400349, No.\,17K05530, No.\,18H01183, and No.\,18K13509) of Japan, 
and also in part by RIKEN iTHES Project.   
\end{acknowledgments}

%%%%%%%%%%%%%%%%%%%%
%%%%%%%%%%%%%%%%%%%%
%%%%%%%%%%%%%%%%%%%%
\begin{appendix}

%%%%%%%%%%%%%%%%%%%%
%%%%%%%%%%%%%%%%%%%%
\section{Electron-Phonon Coupling}

%%%%%%%%%%%%%%%%%%%%
\subsection{Derivation of electron-phonon coupling} \label{app_ep}
Here, following Motizuki {\it et al.}~\cite{Mo86}, we derive the electron-phonon coupling used 
in Sec.~\ref{TiSe2_effm_ep}. The electron-phonon coupling is derived from 
the change in energy when the ions are displaced from their 
equilibrium positions.  
Motizuki {\it et al.}~\cite{Mo86} adopted the Fr\"ohlich approach~\cite{Fr52} in the tight-binding 
approximation, where the atomic wave functions move rigidly with the ions.  

First, for the undistorted system, we write the Bloch wave function in the tight-binding approximation as 
\begin{align}
\phi^{(0)}_{\bm{k},\mu\ell}(\bm{r}) = \frac{1}{\sqrt{N}}\sum_{\bm{R}_i} e^{i\bm{k}\cdot\bm{R}_i}\phi_{\ell}(\bm{r}-\bm{R}_{i\mu}),
\label{app_ep_eq1}
\end{align}
where $\phi_{\ell}(\bm{r})$ is the atomic wave function of orbital $\ell$, 
$\bm{R}_{i\mu}=\bm{R}_i + \bm{\tau}_{\mu}$, $\bm{R}_i$ is the lattice vector, and 
$\bm{\tau}_{\mu}$ is the position of atom $\mu$ in the $i$-th unit cell.  
Using this wave function, we write the transfer integrals in the undistorted system as 
\begin{equation}
T^{(0)}_{\mu\ell,\nu m} (\bm{k}) 
= \sum_{\bm{R}_i-\bm{R}_j} e^{-i\bm{k}\cdot(\bm{R}_i-\bm{R}_{j})}T^{(0)}_{i\mu\ell,j\nu m} ,
\label{app_ep_eq2} 
\end{equation}
where 
\begin{equation}
T^{(0)}_{i\mu\ell,j\nu m}  \equiv \int d\bm{r} \phi^{*}_{\ell}(\bm{r}-\bm{R}_{i\mu}) H_e \phi_{m}(\bm{r}-\bm{R}_{j\nu}) 
\label{app_ep_eq3}
\end{equation}
and $H_e$ represents the one-electron Hamiltonian. 
The transfer integral $T^{(0)}_{i\mu\ell,j\nu m}$ is a function of $\bm{R}_n = \bm{R}_i-\bm{R}_j$ 
in the two-center  approximation~\cite{SK54}.  
When we write $T^{(0)}_{i\mu\ell,j\nu m} = t_{\mu\ell,\nu m} (\bm{R}_n)$ and 
$T^{(0)}_{\mu\ell,\nu m} (\bm{k}) = t_{\mu\ell,\nu m} (\bm{k}) $, 
the transfer integrals in Eq.~(\ref{app_ep_eq2}) become 
\begin{align}
t_{\mu\ell,\nu m} (\bm{k}) = \sum_{\bm{R}_n} e^{-i\bm{k}\cdot \bm{R}_n}t_{\mu\ell,\nu m} (\bm{R}_n),  
\label{app_ep_eq4}
\end{align}
which correspond to Eq.~(\ref{TiSe2_effm_eq2}) in the main text.

Next, to derive the electron-phonon coupling, we consider the Bloch functions 
when the ions are displaced from their equilibrium positions.  
The Bloch wave functions in the distorted system with a lattice displacement 
$\delta \bm{R}_{i\mu}$ are given by 
\begin{align}
\phi_{\bm{k},\mu\ell}(\bm{r}) = \frac{1}{\sqrt{N}}\sum_{\bm{R}_i} e^{i\bm{k}\cdot\bm{R}_i}\phi_{\ell}(\bm{r}-\bm{R}_{i\mu}-\delta\bm{R}_{i\mu}). 
\label{app_ep_eq5}
\end{align}
In this case, the transfer integral is not diagonal with respect to $\bm{k}$ and is given by
\begin{equation}
T_{\mu\ell,\nu m} (\bm{k},\bm{k}') 
= \frac{1}{N}\sum_{\bm{R}_i,\bm{R}_j} e^{-i\bm{k}\cdot\bm{R}_i} e^{i\bm{k}'\cdot\bm{R}_{j}} T_{i\mu\ell,j\nu m} ,
\label{app_ep_eq6} 
\end{equation}
where 
\begin{align}
&T_{i\mu\ell,j\nu m} 
\equiv  \int d\bm{r} \phi^{*}_{\ell}(\bm{r}-\bm{R}_{i\mu}-\delta\bm{R}_{i\mu}) H_e 
\notag \\
&\qquad \qquad \qquad \qquad
\times \phi_{m}(\bm{r}-\bm{R}_{j\nu}-\delta\bm{R}_{j\nu}). 
\label{app_ep_eq7}
\end{align}
Assuming the lattice displacements $\delta \bm{R}_{i\mu}$ are small, we expand the transfer integral 
to the first-order of $\delta \bm{R}_{i\mu}$ as 
\begin{align}
T_{i\mu\ell,j\nu m} 
= T^{(0)}_{i\mu\ell,j\nu m} + \left[ \bm{\nabla} T_{i\mu\ell,j\nu m}\right]\cdot \left[ \delta \bm{R}_{i\mu}-\delta \bm{R}_{j\nu} \right],   
\label{app_ep_eq8}
\end{align}
where the $\gamma$ $(=x,y,z)$ component of $\bm{\nabla} T_{i\mu\ell,j\nu m}$ is given by~\cite{Mo86} 
\begin{align}
\nabla_{\gamma} T_{i\mu\ell,j\nu m} 
= \left. \left( \frac{\partial}{\partial R^{\gamma}}  T_{i\mu\ell,j\nu m} \right) \right|_{\bm{R} =\bm{R}_{i\mu} - \bm{R}_{j\nu}}. 
\label{app_ep_eq9}
\end{align}
Defining the Fourier transformation of $\delta \bm{R}_{i\mu}$ as 
\begin{align}
\delta \bm{R}_{i\mu} = \frac{1}{\sqrt{N}} \sum_{\bm{q}}e^{i\bm{q}\cdot\bm{R}_i}  \bm{u}_{\bm{q},\mu},
\label{app_ep_eq10}
\end{align}
we obtain the transfer integral $T_{\mu\ell,\nu m} (\bm{k},\bm{k}')$ in Eq.~(\ref{app_ep_eq6}) as 
\begin{align}
&T_{\mu\ell,\nu m} (\bm{k},\bm{k}') 
= T^{(0)}_{\mu\ell,\nu m} (\bm{k}) \delta_{\bm{k}',\bm{k}} 
\notag \\
&+\! \frac{1}{\sqrt{N}} \sum_{\bm{q}} \! \left[ \dot{\bm{T}}_{\mu\ell,\nu m}(\bm{k}-\bm{q}) \! \cdot \! \bm{u}_{\bm{q},\mu} \! -\! \dot{\bm{T}}_{\mu\ell,\nu m}(\bm{k}) \! \cdot \! \bm{u}_{\bm{q},\nu} \right] \! \delta_{\bm{k}',\bm{k}-\bm{q}},  
\label{app_ep_eq11}
\end{align}
where we define 
\begin{align}
\dot{\bm{T}}_{\mu\ell,\nu m}(\bm{k})  \equiv  \sum_{\bm{R}_i-\bm{R}_j} e^{-i\bm{k}\cdot(\bm{R}_i-\bm{R}_{j})}  \left[ \bm{\nabla} T_{i\mu\ell,j\nu m} \right]. 
\label{app_ep_eq12}
\end{align}
The first term of $T_{\mu\ell,\nu m} (\bm{k},\bm{k}')$ in Eq.~(\ref{app_ep_eq11}) is given 
by the transfer integral in the undistorted system and the second term corresponds to 
the electron-phonon coupling.  

The displacement $\bm{u}_{\bm{q},\mu}$ is in general characterized by the phonon normal coordinates $Q_{\bm{q}\lambda}$ as 
\begin{align}
\bm{u}_{\bm{q},\mu} = \sum_{\lambda} \frac{\bm{\varepsilon}(\bm{q}\lambda,\mu)}{\sqrt{M_{\mu}}} Q_{\bm{q}\lambda},
\label{app_ep_eq13}
\end{align}
where $M_{\mu}$ is the mass of atom $\mu$ and $\bm{\varepsilon}(\bm{q}\lambda,\mu)$ is 
the polarization vector of the phonon of mode $\lambda$ with the phonon frequency 
$\omega_0(\bm{q}\lambda)$.  
Using the normal coordinates $Q_{\bm{q}\lambda}$, $T_{\mu\ell,\nu m} (\bm{k},\bm{k}')$ becomes
\begin{align}
T_{\mu\ell,\nu m} (\bm{k},\bm{k}') 
= &T^{(0)}_{\mu\ell,\nu m} (\bm{k})  \delta_{\bm{k}',\bm{k}} 
\notag \\
+\frac{1}{\sqrt{N}} \sum_{\bm{q},\lambda} \Bigl[ &\frac{\bm{\varepsilon}(\bm{q}\lambda,\mu)}{\sqrt{M_{\mu}}} \cdot \dot{\bm{T}}_{\mu\ell,\nu m}(\bm{k}-\bm{q})  
\notag \\
-  &\frac{\bm{\varepsilon}(\bm{q}\lambda,\nu)}{\sqrt{M_{\nu}}} \cdot \dot{\bm{T}}_{\mu\ell,\nu m}(\bm{k}) \Bigr] Q_{\bm{q}\lambda} \delta_{\bm{k}',\bm{k}-\bm{q}}. 
\label{app_ep_eq14}
\end{align}
Here, defining the coefficient of $Q_{\bm{q}\lambda}$ in the second term of $T_{\mu\ell,\nu m} (\bm{k},\bm{k}')$ in Eq.~(\ref{app_ep_eq14}) as 
\begin{align}
 g^{\lambda}_{\mu\ell,\nu m}(\bm{k},\bm{q}) 
&\equiv
 \frac{\bm{\varepsilon}(\bm{q}\lambda,\mu)}{\sqrt{M_{\mu}}} \cdot \dot{\bm{T}}_{\mu\ell,\nu m}(\bm{k}-\bm{q})  
 \notag \\
 &\qquad -  \frac{\bm{\varepsilon}(\bm{q}\lambda,\nu)}{\sqrt{M_{\nu}}} \cdot \dot{\bm{T}}_{\mu\ell,\nu m}(\bm{k}),
\label{app_ep_eq15}
\end{align}
we finally write the transfer integral with the small lattice displacement as 
\begin{align}
T_{\mu\ell,\nu m} (\bm{k},\bm{k}') 
&= t_{\mu\ell,\nu m} (\bm{k}) \delta_{\bm{k}',\bm{k}} 
\notag \\
&+\frac{1}{\sqrt{N}} \sum_{\bm{q},\lambda} g^{\lambda}_{\mu\ell,\nu m}(\bm{k},\bm{q}) Q_{\bm{q}\lambda} \delta_{\bm{k}',\bm{k}-\bm{q}} ,
\label{app_ep_eq16}
\end{align}
where we also use $t_{\mu\ell,\nu m} (\bm{k})=T^{(0)}_{\mu\ell,\nu m} (\bm{k})$.  
The second term in Eq.~(\ref{app_ep_eq16}) is derived by the lattice distortion and 
$g^{\lambda}_{\mu\ell,\nu m}(\bm{k},\bm{q})$ corresponds to the electron-phonon coupling 
for the phonon mode $\lambda$.  
When we write $ \bm{\nabla} T_{i\mu\ell,j\nu m} = \bm{\nabla} t_{\mu\ell,\nu m}(\bm{R}_n)$ with $\bm{R}_n = \bm{R}_i-\bm{R}_j$, 
$g^{\lambda}_{\mu\ell,\nu m}(\bm{k},\bm{q})$ becomes 
\begin{align}
g^{\lambda}_{\mu\ell,\nu m}(\bm{k},\bm{q}) 
&=\sum_{\bm{R}_n}  \Bigl[ \bm{\nabla} t_{\mu\ell,\nu m}(\bm{R}_n) \Bigr] 
\label{app_ep_eq17}  \\
&\cdot 
\left[ \frac{\bm{\varepsilon}(\bm{q}\lambda,\mu)}{\sqrt{M_{\mu}}}e^{-i(\bm{k}-\bm{q})\cdot \bm{R}_n}
 - \frac{\bm{\varepsilon}(\bm{q}\lambda,\nu)}{\sqrt{M_{\nu}}} e^{-i\bm{k}\cdot \bm{R}_n} \right].
\notag
\end{align} 

If we assume that the displacement $\bm{u}_{\bm{q},\mu}$ is characterized by a particular normal coordinate $Q_{\bm{q}}$ with   
\begin{align}
\bm{u}_{\bm{q},\mu} =  \frac{\bm{\varepsilon}(\bm{q},\mu)}{\sqrt{M_{\mu}}} Q_{\bm{q}}, 
\label{app_ep_eq18}
\end{align}
the transfer integral $T_{\mu\ell,\nu m} (\bm{k},\bm{k}')$ in Eq.~(\ref{app_ep_eq16}) becomes
 \begin{align}
T_{\mu\ell,\nu m} (\bm{k},\bm{k}') 
&= t_{\mu\ell,\nu m} (\bm{k}) \delta_{\bm{k}',\bm{k}} 
\notag \\
&+\frac{1}{\sqrt{N}} \sum_{\bm{q}} g_{\mu\ell,\nu m}(\bm{k},\bm{q}) Q_{\bm{q}} \delta_{\bm{k}',\bm{k}-\bm{q}} . 
\label{app_ep_eq19}
\end{align}
In the main text, we assume that the displacement $\bm{u}_{\bm{q},\mu}$ is characterized 
only by the normal coordinate of the transverse phonon mode shown in 
Figs.~\ref{TiSe2_cs_fig3}(a)--\ref{TiSe2_cs_fig3}(c).

%%%%%%%%%%%%%%%%%%%%
\subsection{Susceptibility and phonon softening}\label{app_epsus}
Here, we derive the susceptibility $\chi(\bm{q})$ by the second-order perturbation theory 
with respect to $Q_{\bm{q}}$ following Motizuki {\it et al.}~\cite{Mo86}. 
The susceptibility $\chi(\bm{q})$ is used in Sec.~\ref{TiSe2_epsus} to discuss 
the phonon softening.  

We first transform the transfer integral of Eq.~(\ref{app_ep_eq19}) 
from the atomic orbital $\mu\ell$ representation to the band index $a$ representation, i.e., 
$\hat{T}'= \hat{U}^{(0)\dag}\hat{T}\hat{U}^{(0)}$, where the transformation matrix $\hat{U}^{(0)}$ 
is given by the eigenvectors of the undistorted energy bands $\varepsilon^{(0)}_{\bm{k},a}$.  
Using the matrix elements $u^{(0)}_{\mu \ell,a}(\bm{k})$ in $\hat{U}^{(0)}$, the transfer integral 
in the band-index representation is given by 
\begin{align}
&T'_{a,b} (\bm{k},\bm{k}') 
=\sum_{\mu\ell,\nu m} u^{(0)\,*}_{\mu\ell,a}(\bm{k})T_{\mu\ell,\nu m} (\bm{k},\bm{k}') u^{(0)}_{\nu m,b}(\bm{k}')
\notag \\
&=\varepsilon^{(0)}_{\bm{k},a}  \delta_{\bm{k}',\bm{k}}\delta_{a,b}
+\frac{1}{\sqrt{N}} \sum_{\bm{q}}  V_{ep}(a\bm{k},b\bm{k}-\bm{q}) Q_{\bm{q}} \delta_{\bm{k}',\bm{k} - \bm{q}},  
\label{app_sus_eq1}
\end{align}
where $\varepsilon^{(0)}_{\bm{k},a} \delta_{a,b}= \sum_{\mu\ell,\nu m}u^{(0)\,*}_{\mu\ell,a}(\bm{k})t_{\mu\ell,\nu m} (\bm{k}) u^{(0)}_{\nu m,b}(\bm{k})$ and 
\begin{align}
V_{ep}(a\bm{k},b\bm{k} \! - \! \bm{q})  \equiv \! \sum_{\mu\ell,\nu m} u^{(0)\, *}_{\mu\ell,a}(\bm{k}) g_{\mu\ell,\nu m}(\bm{k},\bm{q}) u^{(0)}_{\nu m,b}(\bm{k} \! - \! \bm{q}). 
\label{app_sus_eq2}
\end{align}

Treating the second term of Eq.~(\ref{app_sus_eq1}) as perturbation~\cite{Mo86}, 
we may write the energy in the second-order perturbation theory as 
\begin{align}
\varepsilon^{(2)}_{\bm{k},a}
= \frac{1}{N}\sum_{\bm{q}} \sum_{b}  
\frac{\left| V_{ep}(a \bm{k}, b \bm{k}-\bm{q}) Q_{\bm{q}} \right|^2}{\varepsilon^{(0)}_{\bm{k},a}-\varepsilon^{(0)}_{\bm{k}-\bm{q},b}},
\label{app_sus_eq3}
\end{align}
and the change in the free energy as $\Delta F = 2 \sum_{\bm{k},a} \varepsilon^{(2)}_{\bm{k},a} f(\varepsilon^{(0)}_{\bm{k},a})$. 
Using the relation $V^{*}_{ep}(a\bm{k},b\bm{k}-\bm{q})=V_{ep}(b\bm{k}-\bm{q},a\bm{k})$, we find $\Delta F = \sum_{\bm{q}} \Delta F_{\bm{q}} $ with  
\begin{align}
\Delta F_{\bm{q}} 
= \frac{1}{N}\sum_{\bm{k}} \sum_{a,b}  
\left| V_{ep}(a \bm{k}, b \bm{k}-\bm{q}) Q_{\bm{q}} \right|^2
\notag \\
\times \frac{f(\varepsilon^{(0)}_{\bm{k},a}) -f(\varepsilon^{(0)}_{\bm{k}-\bm{q},b}) }{\varepsilon^{(0)}_{\bm{k},a}-\varepsilon^{(0)}_{\bm{k}-\bm{q},b}}. 
\label{app_sus_eq4}
\end{align}
Defining the susceptibility as 
\begin{align}
\chi(\bm{q})  
\equiv \! -  \frac{2}{N} \sum_{\bm{k}} \sum_{a,b}
\left| V_{ep}(a \bm{k}, b \bm{k}\!-\!\bm{q}) \right|^2 \!
\frac{f(\varepsilon^{(0)}_{\bm{k},a}) \! - \! f(\varepsilon^{(0)}_{\bm{k}-\bm{q},b}) }{\varepsilon^{(0)}_{\bm{k},a} \! - \! \varepsilon^{(0)}_{\bm{k}-\bm{q},b}} , 
\label{app_sus_eq5}
\end{align}
we obtain the $\bm{q}$ component of the change in the free energy as 
\begin{align}
\Delta F_{\bm{q}} = - \frac{1}{2} \chi(\bm{q})  \left| Q_{\bm{q}} \right|^2 .
\label{app_sus_eq6}
\end{align}

The change in the free energy is not only from the electronic energy $\Delta F_{\bm{q}}=\Delta F^{\rm elec}_{\bm{q}}$ 
but also from the elastic energy $\Delta F^{\rm elas}_{\bm{q}}$. 
The change in the elastic energy may be written as 
\begin{align}
\Delta F^{\rm elas}_{\bm{q}} = \frac{1}{2} \omega^2_0(\bm{q})  \left| Q_{\bm{q}} \right|^2 
\label{app_sus_eq7}
\end{align}
with the bare phonon frequency $\omega_0(\bm{q})$.  
The change in the total free energy 
$\Delta F^{\rm tot}_{\bm{q}} = \Delta F^{\rm elas}_{\bm{q}} +\Delta F^{\rm elec}_{\bm{q}}$ may thus be given by 
\begin{align}
\Delta F^{\rm tot}_{\bm{q}} 
&= \frac{1}{2} \omega^2_0(\bm{q})  \left| Q_{\bm{q}} \right|^2 - \frac{1}{2} \chi(\bm{q})  \left| Q_{\bm{q}} \right|^2
\notag \\
&= \frac{1}{2} \omega^2(\bm{q})  \left| Q_{\bm{q}} \right|^2, 
\label{app_sus_eq8}
\end{align}
where we define the effective phonon frequency $\omega(\bm{q})$ as 
\begin{align}
\omega^2(\bm{q})  =  \omega^2_0(\bm{q}) - \chi(\bm{q}). 
\label{app_sus_eq9}
\end{align}
Therefore, the structural instability of the system may be discussed in terms of this phonon 
frequency~\cite{Mo86}, which includes the influence of the electronic system via $\chi(\bm{q})$.  
We discuss the phonon softening using Eq.~(\ref{app_sus_eq9}) in Sec.~\ref{TiSe2_epsus}.

%%%%%%%%%%%%%%%%%%%%
%%%%%%%%%%%%%%%%%%%%
\section{Estimation of the Electron-Phonon Coupling Constant} \label{app_eelc}

%%%%%%%%%%%%%%%%%%%%
\subsection{Derivatives of the transfer integrals}

To estimate the electron-phonon couplings $g^{\lambda}_{\mu\ell,\nu m}(\bm{k},\bm{q})$ defined 
in Eq.~(\ref{app_ep_eq17}), the first derivatives of the transfer integrals 
$\nabla_{\gamma} t_{\alpha,\beta}(\bm{R})$ are required.  
Following Motizuki {\it et al.}~\cite{Mo86}, we use the derivatives of the transfer integrals expressed in terms of the Slater-Koster integrals.  
Here, we write the transfer integral between the $\alpha$ and $\beta$ orbitals located at a distance 
$\bm{R}$ [$=R\times(l,m,n)$] as $t_{\alpha,\beta}(\bm{R})$ and its derivative in the $\gamma$ (=$x$ or $y$ or $z$) direction as 
\begin{align}
\nabla_{\gamma} t_{\alpha,\beta}(\bm{R}) 
= \lim_{\delta \rightarrow 0} \frac{t_{\alpha,\beta}(\bm{R} + \delta \bm{e}_{\gamma})-t_{\alpha,\beta}(\bm{R})}{\delta},
\label{app_dti_eq1}
\end{align}
where $\bm{e}_{\gamma}$ is the unit vector pointing to the $\gamma$ direction. 
For example, the first derivative in the $\gamma=x$ direction of the transfer integral 
\begin{align}
t_{x,yz}(\bm{R}) = lmn\left[ \sqrt{3} t(pd\sigma)-2 t(pd\pi)\right] 
\label{app_dti_eq2}
\end{align}
is given by 
\begin{align}
\nabla_{x} t_{x,yz}(\bm{R}) &=   mn(1-3l^2) \frac{1}{R} \left[ \sqrt{3} t(pd\sigma) -2 t(pd\pi) \right]  
\notag \\
&+ l^2mn \left[ \sqrt{3} t'(pd\sigma) -2 t'(pd\pi) \right], 
\label{app_dti_eq3}
\end{align}
where $t'(pd\sigma) \! = \! [d t(pd\sigma) / dR]$, $t'(pd\pi) \! = \! [d t(pd\pi) / d R]$, 
and $(l,m,n)$ are the direction cosines. 
All the first derivatives $\nabla_{\gamma} t_{\alpha,\beta}(\bm{R})$ expressed 
in terms of the Slater-Koster integrals are tabulated in Ref.~\cite{Mo86}.

%%%%%%%%%%%%%%%%%%%%
\subsection{Overlap integrals and their derivatives estimated by Slater-type orbitals}

To estimate the first derivatives of the transfer integrals $\nabla_{\gamma} t_{\alpha,\beta}(\bm{R})$, 
we need the first derivatives of the Slater-Koster parameters, e.g., $t'(pd\sigma)$ and $t'(pd\pi)$.  
Motizuki and co-workers estimated $t'(pd\sigma)$, etc.,~using the following relation~\cite{YM80,Mo86}: 
\begin{align}
\frac{t'(pd\sigma)}{t(pd\sigma)} = \alpha_{\rm c} \frac{s'(pd\sigma)}{s(pd\sigma)}, \;\; {\rm etc.}, 
\label{app_sto_eq1}
\end{align}
where $s(pd\sigma)$ and $s'(pd\sigma)$ are the Slater-Koster parameter for the overlap integral and its first derivative, respectively. 
Following Motizuki {\it et al.}~\cite{YM80,Mo86}, 
we apply the Slater-type orbitals (STOs)~\cite{Sl30} to estimate the ratio $s'(pd\sigma)/s(pd\sigma)$ and $s'(pd\pi)/s(pd\pi)$ analytically.  

In general, the overlap integral between the $\alpha$ and $\beta$ orbitals located at a distance $\bm{R}$ is given by  
\begin{align}
S(\alpha,\beta) = \int d \bm{r} \phi_{\alpha}(\bm{r})\phi_{\beta}(\bm{r}-\bm{R}), 
\label{app_sto_eq2}
\end{align}
where $\phi_{\alpha}(\bm{r})$ is the atomic wave function of orbital $\alpha$.  
In the STO, we assume $R_{n_{\alpha}}(r) = C_{\alpha} r^{n_{\alpha}-1} e^{-\zeta_{\alpha} r}$ as a radial wave function.  
Thus the atomic orbital $\phi_{\alpha}(\bm{r})$ in the STO is given by~\cite{Sl30,CR63,CRR67} 
\begin{align}
\phi_{\alpha}(\bm{r}) = C_{\alpha} r^{n_{\alpha}-1} e^{-\zeta_{\alpha} r}Y_{l_{\alpha} m_{\alpha}}(\theta,\varphi), 
\label{app_sto_eq3}
\end{align}
where $n_{\alpha}$, $l_{\alpha}$, and $m_{\alpha}$ are the principal, azimuthal, and magnetic quantum numbers 
of the $\alpha$ orbital, respectively, and $Y_{l_{\alpha}m_{\alpha}}(\theta,\varphi)$ is the spherical (tesseral) harmonics.  
$\zeta_{\alpha}$ is the orbital exponent of the $\alpha$ orbital and 
$C_{\alpha} = (2\zeta_{\alpha})^{n_{\alpha}+\frac{1}{2}}/\sqrt{(2n_{\alpha})!}$ is a normalization constant.  
The orbital exponents $\zeta_{\alpha}$ are estimated semi-empirically by Slater as the Slater's rules~\cite{Sl30}.  
However, we use values revised by Clementi {\it et al.}~\cite{CR63,CRR67} based on the Hartree-Fock method, 
where the effective principal quantum number $n^{*}_{\alpha}=3.7$ estimated for the 4$p$ ($n_{\alpha}=4$) orbital 
in the Slater's rules~\cite{Sl30} becomes an integer in the Clementi's estimation~\cite{CR63,CRR67}, and hence the 
overlap integrals can be estimated analytically.  Moreover, in transition-metal and chalcogen atoms, the orbital 
exponents in the Clementi's estimations are larger than those in the semi-empirical Slater's rules~\cite{CR63,CRR67}, 
indicating that the more localized atomic orbitals (and thus smaller overlap integrals) are realized when we 
use the orbital exponents estimated by Clementi {\it et al}.  
Note that we do not write the Bohr radius $a_0$ ($\sim 0.529$~\AA) explicitly in this section; 
we rather assume $a_0$ as the unit of length.   

%%  Fig. 19  %%
\begin{figure}[!t]
\begin{center}
\includegraphics[width=0.75\columnwidth]{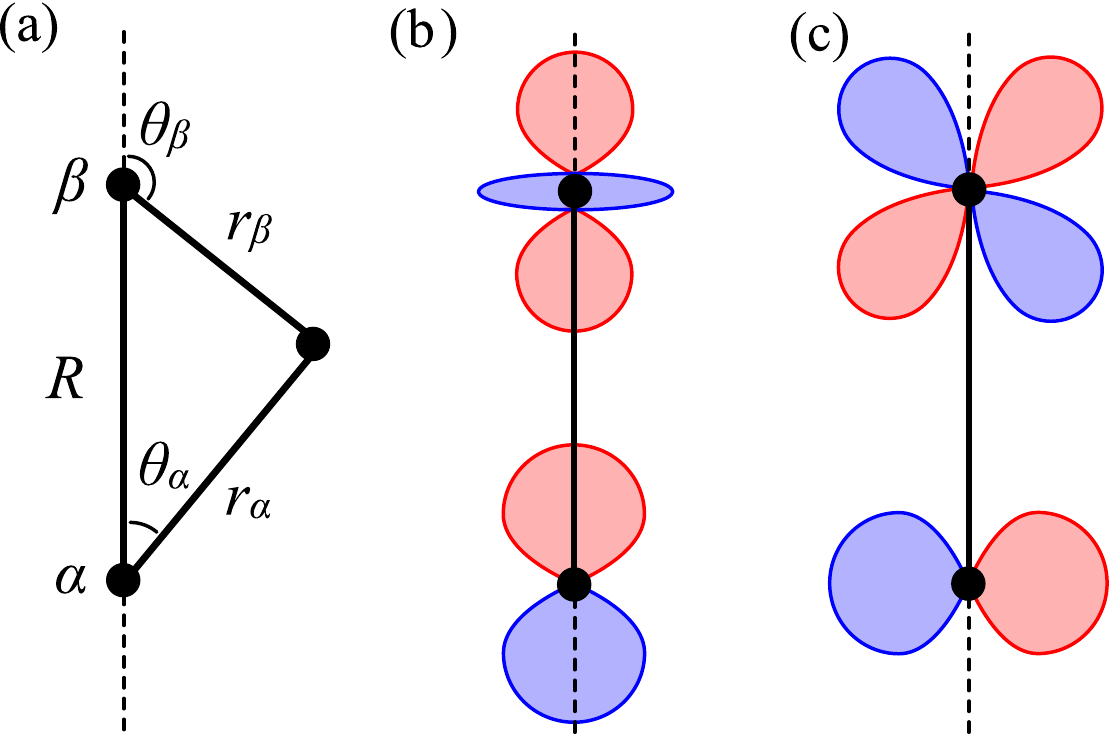}
\caption{(a) Elliptic coordinate system and examples of the orbitals for the overlap integrals (b) $s(pd\sigma)$ and (c) $s(pd\pi)$.}
\label{TiSe2_app_fig1}
\end{center}
\end{figure}
%%%%%%%%

Although there are several ways to estimate the overlap integrals with the STOs, we adopt 
the approach of Mulliken {\it et al.}~\cite{MROetal49}, where the elliptic coordinate system is employed.  
As in Fig.~\ref{TiSe2_app_fig1}(a), we assume $\bm{R}=R\bm{e}_z$ and the elliptic coordinate system defined by 
\begin{align}
\xi &= \frac{r_{\alpha} + r_{\beta}}{R}           \:\:\:\:\:\:   \left( 1\le \xi \le \infty \right), 
\label{app_sto_eq4} \\
\eta &= \frac{r_{\alpha} - r_{\beta}}{R}     \:\:\:\:\:\:   \left( -1 \le \eta \le 1 \right), 
\label{app_sto_eq5} \\
\varphi &= \varphi_{\alpha} = \varphi_{\beta}  \:\:\:\:\:  \left( 0 \le \varphi \le 2\pi \right),
\label{app_sto_eq6}
\end{align}
where $\varphi$ is the azimuthal angle. 
Using the coordinates $\xi$ and $\eta$, the distance $r_{\alpha}$, $r_{\beta}$ and angle $\theta_{\alpha}$, $\theta_{\beta}$ 
in Fig.~\ref{TiSe2_app_fig1}(a) are given by 
\begin{align}
&r_{\alpha} = \frac{R}{2} \left( \xi + \eta \right), \:\:\:\:
\cos \theta_{\alpha} = \frac{\xi \eta+1}{\xi+\eta}, 
\label{app_sto_eq7} \\
&r_{\beta} = \frac{R}{2} \left( \xi - \eta \right), \:\:\:\:\:
\cos \theta_{\beta} = \frac{\xi\eta-1}{\xi-\eta}.
\label{app_sto_eq8}
\end{align}
The volume element in the elliptic coordinate system is given by $d\bm{r} = (R/2)^3 \left( \xi^2 -\eta^2 \right)d\xi d\eta d\varphi$. 
When we estimate the overlap integrals in the elliptic coordinate system, we usually define~\cite{MROetal49} 
\begin{align}
p = \frac{R}{2} \left( \zeta_{\alpha} + \zeta_{\beta} \right) , \;\;\;
t = \frac{\zeta_{\alpha} - \zeta_{\beta} }{\zeta_{\alpha} + \zeta_{\beta} }, 
\label{app_sto_eq9}
\end{align}
for different orbital exponents $\zeta_{\alpha}$ and $\zeta_{\beta}$. 
Then, the overlap integral $S(\alpha,\beta)$ and its derivative $S'(\alpha,\beta)$ [$=dS(\alpha,\beta)/dR$] 
are written in terms of the parameters $p$ and $t$.  

In this paper, we consider the electron-phonon coupling between the nearest-neighbor Ti 3$d$ 
and Se 4$p$ orbitals, and thus we estimate the overlap integrals $s(pd\sigma)$ and $s(pd\pi)$ 
for $n_{\alpha} = 4$ and $n_{\beta} = 3$.  
As shown in Figs.~\ref{TiSe2_app_fig1}(b) and \ref{TiSe2_app_fig1}(c), 
$s(pd\sigma)$ is given by the $\alpha = 4p\sigma$ (4$p_z$) 
and $\beta=3d\sigma$ (3$d_{3z^2-r^2}$) orbitals, and $s(pd\pi)$ is given by the $\alpha = 4p\pi$ (4$p_x$) 
and $\beta=3d\pi$ (3$d_{zx}$) orbitals.  
Using the spherical functions for these orbitals in the elliptic coordinate system, the overlap 
integrals $s(pd\sigma)=S(4 p\sigma, 3 d\sigma)$ and $s(pd\pi)=S(4 p\pi, 3 d\pi)$ are given, respectively, by 
\begin{widetext}
\begin{align}
%s(pd\sigma) %\! = \! 
S(4 p\sigma, 3 d\sigma) %\!
& = C_{pd\sigma} p^8 \int^{\infty}_{1} d\xi \int^{1}_{-1}d\eta 
 (\xi+\eta)^{3}  (\xi-\eta) (\xi \eta+1)\left[ 3 (\xi \eta-1)^2 - (\xi - \eta)^2 \right]e^{-p(\xi+\eta t)} 
 \notag \\
&= C_{pd\sigma} p^8
 [ 
A_7(3B_3-B_1) +A_6(6B_4-3B_2-B_0)-3A_5(B_3+B_1) -3A_4(2B_6+B_2-B_0) \notag \\
&\qquad \quad -3A_3(B_7-B_5-2B_1) +3A_2(B_6+B_4)+A_1(B_7+3B_5-6B_3) + A_0(B_6-3B_4)
 ] 
\label{app_sto_eq10} 
\end{align}
and 
\begin{align}
%s(pd\pi) \! = \! 
S(4 p\pi, 3 d\pi)% \!
 &=  C_{pd\pi} p^8   \int^{\infty}_{1} d\xi \int^{1}_{-1}d\eta 
 (\xi+\eta)^{3}  (\xi-\eta)(\xi^2-1)(1-\eta^2) (\xi \eta-1) e^{-p(\xi+\eta t)} 
 \notag \\
 &=  C_{pd\pi} p^8 
[ 
-A_7(B_3-B_1) -A_6(2B_4-3B_2+B_0)+3A_5(B_3-B_1) +A_4(2B_6-3B_2+B_0) \notag \\
&\qquad \qquad  +A_3(B_7-3B_5+2B_1) -3A_2(B_6-B_4)-A_1(B_7-3B_5+2B_3) + A_0(B_6-B_4)
 ] ,
\label{app_sto_eq11}  
\end{align} 
\end{widetext}
where functions $A_k$ and $B_k$ are defined by 
\begin{align}
A_k(p) &\equiv  \int^{\infty}_{1}\xi^k e^{-p\xi} d\xi = e^{-p}\sum^{k}_{\mu=0}\frac{k!}{(k-\mu)!p^{\mu+1}} 
\label{app_sto_eq12} %\\
\end{align}
and 
\begin{align}
B_k(pt) &\equiv  \int^{1}_{-1}\eta^k e^{-pt\eta} d\eta 
= \sum^{k}_{\mu=0}\frac{k! [ (-1)^{k-\mu}e^{pt}-e^{-pt} ]   }{(k-\mu)!(pt)^{\mu+1}} ,
\label{app_sto_eq13}
\end{align}
respectively, and also $C_{pd\sigma} = (\sqrt{210}/80640) (1+t)^{\frac{9}{2}}(1-t)^{\frac{7}{2}}$ 
and $C_{pd\pi} = (\sqrt{70}/26880) (1+t)^{\frac{9}{2}}(1-t)^{\frac{7}{2}}$. 

Using the relations $R[dA_k(p)/dR]=-pA_{k+1}(p)$ and $R[dB_k(pt))/dR]=- ptB_{k+1}(pt)$, 
we can also estimate the dimensionless derivative parameter 
$R \! \times \! s'(pd\sigma)=RS'(4 p\sigma, 3 d\sigma)$ analytically.  
Therefore we can evaluate the ratio $s'(pd\sigma)/s(pd\sigma)$ 
from the dimensionless parameter 
$R\times[s'(pd\sigma)/s(pd\sigma)]=RS'(4 p\sigma, 3 d\sigma)/S(4 p\sigma, 3 d\sigma)$.

%%%%%%%%%%%%%%%%%%%%
%%%%%%%%%%%%%%%%%%%%
\section{Periodic Lattice Distortion and Hamiltonian of the Triple-$\bm{q}$ Structure} \label{app_tri}

Here, we review the triple-$\bm{q}$ structure in TiSe$_2$, where the transverse phonon modes 
at the three M points are frozen simultaneously.  We also introduce the Hamiltonian in the static 
triple-$\bm{q}$ structure.  

When the transverse phonon modes at the $\bm{q}_{1}$, $\bm{q}_{2}$, and $\bm{q}_{3}$ points are 
frozen simultaneously, the triple-$\bm{q}$ structure is characterized by the static displacement~\cite{Mo86,SYM85} 
\begin{align}
\delta \bm{R}_{i\mu} = \frac{1}{\sqrt{N}} \sum_{\bm{q}_j=\bm{q}_1,\bm{q}_2,\bm{q}_3}e^{i\bm{q}_j\cdot\bm{R}_i}  \langle \bm{u}_{\bm{q}_j,\mu}  \rangle. 
\label{app_tri_eq1}
\end{align} 
Since the transverse phonon modes are softened at the $\bm{q}_{1}$, $\bm{q}_{2}$, and $\bm{q}_{3}$ points, 
the direction of $\langle \bm{u}_{\bm{q}_j,\mu} \rangle$ is perpendicular to its respective wave vector 
$\bm{q}_j$ [see Figs.~\ref{TiSe2_cs_fig3}(a)--\ref{TiSe2_cs_fig3}(c)].  
In practice, $\langle \bm{u}_{\bm{q}_j,\mu}  \rangle$ for Ti atom at $\bm{q}_{1}$, $\bm{q}_{2}$, and $\bm{q}_{3}$ 
are given by
\begin{align}
\frac{\langle \bm{u}_{\bm{q}_1,{\rm Ti}} \rangle}{\sqrt{N}} &=  u\bm{e}_Y,
\label{app_tri_eq2}
\\
\frac{\langle \bm{u}_{\bm{q}_2,{\rm Ti}} \rangle}{\sqrt{N}} &= - \frac{\sqrt{3}}{2}u\bm{e}_X - \frac{1}{2}u\bm{e}_Y,
\label{app_tri_eq3}
%\\
\end{align}
and 
\begin{align}
\frac{\langle \bm{u}_{\bm{q}_3,{\rm Ti}} \rangle}{\sqrt{N}} & =  \frac{\sqrt{3}}{2}u\bm{e}_X - \frac{1}{2}u\bm{e}_Y,
\label{app_tri_eq4}
\end{align}
respectively~\cite{SYM85,Mo86}, where $u$ is the magnitude of the displacement of Ti atoms.  
If we assume the ratio $| \bm{u}_{\bm{q}_j,{\rm Se}}|/| \bm{u}_{\bm{q}_j,{\rm Ti}}| = 1/3$, 
$\langle \bm{u}_{\bm{q}_j,\mu}  \rangle$ for Se atoms at $\bm{q}_{1}$, $\bm{q}_{2}$, and $\bm{q}_{3}$ are given by
\begin{align}
\langle \bm{u}_{\bm{q}_1,{\rm Se1}} \rangle &= \langle \bm{u}_{\bm{q}_1,{\rm Se2}} \rangle =-\langle \bm{u}_{\bm{q}_1,{\rm Ti}} \rangle/3,
\label{app_tri_eq5}
\\
\langle \bm{u}_{\bm{q}_2,{\rm Se1}} \rangle &= \langle \bm{u}_{\bm{q}_2,{\rm Se2}} \rangle = \;\;\; \langle \bm{u}_{\bm{q}_2,{\rm Ti}} \rangle/3,
\label{app_tri_eq6}
%\\
\end{align}
and 
\begin{align}
\langle \bm{u}_{\bm{q}_3,{\rm Se1}} \rangle &= \langle \bm{u}_{\bm{q}_3,{\rm Se2}} \rangle =-\langle \bm{u}_{\bm{q}_3,{\rm Ti}} \rangle/3,
\label{app_tri_eq7}
\end{align}
respectively [see Figs.~\ref{TiSe2_cs_fig3}(a)--\ref{TiSe2_cs_fig3}(c)]. 
Note that the sign of $\langle \bm{u}_{\bm{q}_j,{\rm Se2}} \rangle$ is opposite to the definition 
of Motizuki {\it et al.}~\cite{SYM85,Mo86} since we change the definition of the Se(2) position in the unit cell.  

From Eq.~(\ref{app_ep_eq18}), the lattice displacement $\bm{u}_{\bm{q}_j,\mu}$ is characterized 
by the polarization vector $\bm{\varepsilon}(\bm{q}_j,\mu)$ and normal coordinate $Q_{\bm{q}_j}$ 
of the transverse phonon mode as $\bm{u}_{\bm{q}_j,\mu} = \bm{\varepsilon}(\bm{q}_j,\mu) Q_{\bm{q}_j}/\sqrt{M_{\mu}}$.  
In the static triple-$\bm{q}$ structure, the corresponding expectation value $\langle Q_{\bm{q}_j} \rangle$ 
is given by
\begin{align}
\langle Q_{\bm{q}_j} \rangle = \sqrt{N M^{*}} u \;\;\; (\bm{q}_j = \bm{q}_1, \, \bm{q}_2,\, \bm{q}_3), 
\label{app_tri_eq8}
\end{align}
where $M^{*}$ is the effective mass of the transverse phonon soft mode~\cite{SYM85,Mo86}.  
From the relation between  $\bm{u}_{\bm{q}_j,\mu}$ and $Q_{\bm{q}_j}$, the polarization vector 
$\bm{\varepsilon}(\bm{q}_j,\mu)$ of the corresponding transverse mode is given by 
\begin{align}
\bm{\varepsilon}(\bm{q}_j,\mu) = \sqrt{\frac{M_{\mu}}{M^{*}}} \frac{\braket{\bm{u}_{\bm{q}_j,\mu}}}{\sqrt{N}u}. 
\label{app_tri_eq9}
\end{align}
For example, the polarization vectors of the transverse phonon mode at $\bm{q}_j=\bm{q}_1$ 
are given by $\bm{\varepsilon}(\bm{q}_1,{\rm Ti }) = \sqrt{M_{\rm Ti}/M^{*}}\bm{e}_Y$ and 
$\bm{\varepsilon}(\bm{q}_1,{\rm Se1}) = \bm{\varepsilon}(\bm{q}_1,{\rm Se2}) 
= - (1/3)  \sqrt{M_{\rm Se}/M^{*}}\bm{e}_Y $. 
From the normalization condition $\sum_{\mu}|\bm{\varepsilon}(\bm{q}_j,\mu)|^2=1$, 
the effective mass is given by $M^{*}=M_{\rm Ti}+(2/9)M_{\rm Se}$, where we assume 
$| \bm{u}_{\bm{q}_j,{\rm Se}}|/| \bm{u}_{\bm{q}_j,{\rm Ti}}| = 1/3$.  

When the triple-$\bm{q}$ structure is realized, the band structures are modified through 
the electron-phonon couplings.  
Using Eq.~(\ref{app_tri_eq8}), the Hamiltonian of the electron-phonon coupling in the static 
triple-$\bm{q}$ structure becomes 
\begin{align}
\mathcal{H}_{ep} = \sum_{\bm{k},\bm{q}_j} \sum_{\mu \ell, \nu m}   \bar{g}_{\mu \ell, \nu m} (\bm{k},\bm{q}_j) u c^{\dag}_{\bm{k},\mu \ell } c_{\bm{k}-\bm{q}_j,\nu m },
\label{app_tri_eq10}
\end{align}
where $\bar{g}_{\mu \ell, \nu m} (\bm{k},\bm{q}_j) \equiv \sqrt{M^{*}} g_{\mu \ell, \nu m} (\bm{k},\bm{q}_j)$ and 
is given by  
\begin{align}
\bar{g}_{\mu \ell, \nu m} (\bm{k},\bm{q}_j) 
&= \sum_{\bm{R}_n} \left[ \bm{\nabla} t_{\mu \ell, \nu m} (\bm{R}_n) \right] 
\label{app_tri_eq11} \\
 &\cdot \left[ \bm{n}(\bm{q}_j,\mu)e^{-i(\bm{k}-\bm{q}_j)\cdot \bm{R}_n} - \bm{n}(\bm{q}_j,\nu) e^{-i\bm{k}\cdot \bm{R}_n} \right],
\notag
\end{align}
with
\begin{align}
\bm{n}(\bm{q}_j,\mu) &\equiv \sqrt{M^{*}} \times \frac{\bm{\varepsilon}(\bm{q}_j,\mu)}{\sqrt{M_{\mu}}} = \frac{\braket{\bm{u}_{\bm{q}_j,\mu}}}{\sqrt{N}u}.
\label{app_tri_eq12}
\end{align}
For example, the vectors $\bm{n}(\bm{q}_j,\mu)$ at $\bm{q}_j=\bm{q}_1$ are given by 
$\bm{n}(\bm{q}_1,{\rm Ti }) = \bm{e}_Y$ and 
$\bm{n}(\bm{q}_1,{\rm Se1}) = \bm{n}(\bm{q}_1,{\rm Se2}) = - (1/3) \bm{e}_Y$ 
in the transverse phonon mode [see Figs.~\ref{TiSe2_cs_fig3}(a)--\ref{TiSe2_cs_fig3}(c)].

The Hamiltonian of Eq.~(\ref{app_tri_eq10}) is not diagonal with respect to $\bm{k}$ 
in the original BZ without distortion since the transverse phonon modes at $\bm{q}_{1}$, $\bm{q}_{2}$, 
and $\bm{q}_{3}$ are frozen.  
Thus, to diagonalize the Hamiltonian, we need to introduce the RBZ, which is 1/4 
of the original BZ [see Fig.~\ref{TiSe2_cs_fig2}(b)].  
In order to write the Hamiltonian simply in the RBZ, we introduce the 11$\times$11 
matrices of the transfer integral $[\hat{t} (\bm{k}) ]_{\mu \ell, \nu m} = t_{\mu \ell, \nu m} (\bm{k})$ and 
electron-phonon coupling $[\hat{\bar{g}} (\bm{k},\bm{q}) ]_{\mu \ell, \nu m} = \bar{g}_{\mu \ell, \nu m} (\bm{k},\bm{q})$, 
and an eleven dimensional vector of the annihilation (creation) operator 
$[ \bm{c}^{(\dag)}_{\bm{k}} ]_{\mu \ell} = c^{(\dag)}_{\bm{k},\mu \ell}$.  
Using the matrix and vector formalism, the Hamiltonian of the transfer integral is described as  
\begin{align}
\mathcal{H}_{e} 
= \sum_{\bm{k} \in {\rm RBZ}} \sum^3_{i=0}
    \bm{c}^{\dag}_{\bm{k}-\bm{q}_i} \hat{t}(\bm{k}-\bm{q}_i) \bm{c}_{\bm{k}-\bm{q}_i} 
\label{app_tri_eq13}
\end{align}
within the RBZ, where we define $\bm{q}_0=\bm{0}$. 
Similarly, the Hamiltonian of the electron-phonon coupling in Eq.~(\ref{app_tri_eq10}) is now 
\begin{align}
\mathcal{H}_{ep} 
= \sum_{\bm{k} \in {\rm RBZ}} \sum^{3}_{i=0}\sum^{3}_{j=1}
    \bm{c}^{\dag}_{\bm{k}-\bm{q}_i} [\hat{\bar{g}}(\bm{k}-\bm{q}_i,\bm{q}_j) u ] \bm{c}_{\bm{k}-\bm{q}_i-\bm{q}_j}.
\label{app_tri_eq14}
\end{align}
Notice that due to $\hat{\bar{g}}^{\dag}(\bm{k},\bm{q}_j) 
= \hat{\bar{g}}(\bm{k}-\bm{q}_j,\bm{q}_j) 
=  \hat{\bar{g}}(\bm{k}_j,\bm{q}_j)$, 
$\mathcal{H}_{ep}$ in Eq.~(\ref{app_tri_eq14}) satisfies the Hermitian property.  
When we define a 44 dimensional row vector as 
$\bar{\bm{c}}^{\dag}_{\bm{k}} = (\, \bm{c}^{\dag}_{\bm{k}} \; \bm{c}^{\dag}_{\bm{k}_1} \; \bm{c}^{\dag}_{\bm{k}_2} \; \bm{c}^{\dag}_{\bm{k}_3} \, )$ 
with $\bm{k}_i = \bm{k}-\bm{q}_i$, the Hamiltonians $\mathcal{H}_e$ and $\mathcal{H}_{ep}$ are written as 
\begin{align}
\mathcal{H}^{ep}_{cdw} 
=\mathcal{H}_e + \mathcal{H}_{ep} 
= \sum_{\bm{k} \in {\rm RBZ}} \bar{\bm{c}}^{\dag}_{\bm{k}} \hat{\mathcal{H}}^{ep}_{\bm{k}} \bar{\bm{c}}_{\bm{k}},
\label{TiSe2_tri_eq18}
\end{align}
with the 44$\times$44 matrix 
\begin{align}
\hat{\mathcal{H}}^{ep}_{\bm{k}} = 
\begin{pmatrix}
\hat{t}(\bm{k}) & \hat{\bar{g}}(\bm{k},\bm{q}_1)u & \hat{\bar{g}}(\bm{k},\bm{q}_2)u      &  \hat{\bar{g}}(\bm{k},\bm{q}_3)u \\
                       & \hat{t}(\bm{k}_1)                        & \hat{\bar{g}}(\bm{k}_1,\bm{q}_3)u  & \hat{\bar{g}}(\bm{k}_1,\bm{q}_2)u \\
                       &                                                   & \hat{t}(\bm{k}_2)                             & \hat{\bar{g}}(\bm{k}_2,\bm{q}_1)u \\
                       &          \rm{H.C.}                          &                                                        & \hat{t}(\bm{k}_3) 
\end{pmatrix},
\label{app_tri_eq15}
\end{align}
where we use the relations $\bm{q}_1+\bm{q}_2+\bm{q}_3=\bm{0}$ and 
$\bm{c}^{\dag}_{\bm{k}-2\bm{q}_j}=\bm{c}^{\dag}_{\bm{k}}$.  
Thus we can calculate the energy bands in the presence of the triple-$\bm{q}$ structure by diagonalizing 
the Hamiltonian $\hat{\mathcal{H}}^{ep}_{\bm{k}}$ in the RBZ.

%%%%%%%%%%%%%%%%%%%%
%%%%%%%%%%%%%%%%%%%%
\section{Mean-Field Approximation for the Intersite Coulomb Interaction} \label{app_icmf}

Here, we summarize the details of the mean-field approximation for the intersite Coulomb 
interaction, which leads to the excitonic instability in TiSe$_2$.  
We assume the following intersite Coulomb interaction 
\begin{align}
\mathcal{H}_{ee} 
\! =  \! \frac{1}{N} \! \sum_{\bm{k},\bm{k}',\bm{q}} \sum_{\ell, \nu m} \!
V^{dp}_{\ell, \nu m}(\bm{k}\! -\! \bm{k}')   d^{\dag}_{\bm{k},\ell} d_{\bm{k}',\ell} p^{\dag}_{\bm{k}' \! -  \bm{q},\nu m} p_{\bm{k}\!  - \bm{q},\nu m}. 
\label{app_icmf_eq1} 
\end{align} 
In TiSe$_2$, the top of the valence Se $p$ bands and the bottom of the 
conduction Ti $d$ bands are located in the BZ at the momenta separated by 
$\bm{q}_{j}=\bm{q}_1$, $\bm{q}_2$, and $\bm{q}_3$.  
Thus the order parameter defined by the expectation value 
$\braket{p^{\dag}_{\bm{k} - \bm{q}_j,\nu m} d_{\bm{k},\ell }} \ne 0$ 
is anticipated.  We therefore introduce the following mean-field approximation: 
\begin{align}
\sum_{\bm{q}} & d^{\dag}_{\bm{k},\ell} d_{\bm{k}',\ell} p^{\dag}_{\bm{k}'-\bm{q},\nu m} p_{\bm{k}-\bm{q},\nu m} 
\notag \\
\sim 
&- \sum_{\bm{q}_j} \braket{p^{\dag}_{\bm{k}'-\bm{q}_j,\nu m} d_{\bm{k}',\ell}}d^{\dag}_{\bm{k},\ell} p_{\bm{k}-\bm{q}_j,\nu m}
 \notag \\
& - \sum_{\bm{q}_j} \braket{d^{\dag}_{\bm{k},\ell} p_{\bm{k}-\bm{q}_j,\nu m}} p^{\dag}_{\bm{k}'-\bm{q}_j,\nu m} d_{\bm{k}',\ell}
\notag \\
&+ \sum_{\bm{q}_j} \braket{p^{\dag}_{\bm{k}'-\bm{q}_j,\nu m} d_{\bm{k}',\ell}}\braket{d^{\dag}_{\bm{k},\ell} p_{\bm{k}-\bm{q}_j,\nu m}}, 
\label{app_icmf_eq2}
\end{align}
where $\langle\cdots\rangle$ denotes the grand canonical average at temperature $T$ with respect 
to the mean-field Hamiltonian.
Note that we assume the spin-singlet $d$-$p$ hybridization because we also take into account 
the electron-phonon coupling, which is known to induce the spin-singlet hybridization~\cite{KZFetal15}.  
The spin-triplet hybridization is expected to occur in the presence 
of the Hund's-like exchange interaction~\cite{KO14}.  Here, we introduce the excitonic order parameter 
\begin{align}
\Delta^{dp}_{\ell, \nu m}(\bm{k},\bm{q}_j)
\equiv
 - \frac{1}{N}  \sum_{\bm{k}'}   V^{dp}_{\ell, \nu m}(\bm{k}-\bm{k}') \braket{p^{\dag}_{\bm{k}'-\bm{q}_j,\nu m} d_{\bm{k}',\ell}}, 
\label{app_icmf_eq3} 
\end{align}
and thus the mean-field Hamiltonian $\mathcal{H}^{\rm MF}_{ee}$ is given by 
\begin{align}
\mathcal{H}^{\rm MF}_{ee} &= \mathcal{H}_{cdw} ^{ex} + E^{ex}_{0} , 
\label{app_icmf_eq4}  
\end{align}
where 
\begin{align}
\mathcal{H}_{cdw} ^{ex}
&\equiv \sum_{\bm{k},\bm{q}_j} \sum_{\ell, \nu m}   
 \Delta^{dp}_{\ell, \nu m}(\bm{k},\bm{q}_j) d^{\dag}_{\bm{k},\ell} p_{\bm{k} -\bm{q}_j,\nu m}
+  {\mathrm {H.c.}} 
\label{app_icmf_eq5} 
\end{align}
and 
\begin{align}
E^{ex}_{0} &\equiv - \sum_{\bm{k},\bm{q}_j} \sum_{\ell, \nu m} \Delta^{dp}_{\ell, \nu m}(\bm{k},\bm{q}_j)\braket{d^{\dag}_{\bm{k},\ell} p_{\bm{k} -\bm{q}_j,\nu m}}.
\label{app_icmf_eq6}
\end{align}

Since the mean-field Hamiltonian is not diagonal with respect to $\bm{k}$ in the original BZ, 
we also need to apply the RBZ introduced in the Appendix C.  
We use the 5$\times$6 matrix representation of the order parameter $[\hat{\Delta}(\bm{k},\bm{q}_j) ]_{\ell, \nu m} = \Delta^{dp}_{\ell, \nu m}(\bm{k},\bm{q}_j) $, 
the five dimensional vector representation of the annihilation (creation) operators of the Ti $d$ orbitals $[ \bm{d}^{(\dag)}_{\bm{k}} ]_{\ell} = d^{(\dag)}_{\bm{k},\ell} $, 
and the six dimensional vector representation of the two Se($\nu$) $p$ orbitals $[ \bm{p}^{(\dag)}_{\bm{k}} ]_{\nu m} = p^{(\dag)}_{\bm{k},\nu m}$.  
In this matrix and vector representation, the mean-field Hamiltonian $\mathcal{H}_{cdw} ^{ex}$ in the RBZ 
is written as 
\begin{align}
\mathcal{H}_{cdw} ^{ex}
=\! \sum_{\bm{k} \in {\rm RBZ}} \sum_{i=0}^{3} \sum_{j=1}^{3} 
\bm{d}^{\dag}_{\bm{k}-\bm{q}_i} \hat{\Delta}(\bm{k} \! - \! \bm{q}_i,\bm{q}_j) \bm{p}_{\bm{k}-\bm{q}_i -\bm{q}_j}
\! +  {\mathrm {H.c.}}  
\label{app_icmf_eq7}
\end{align}
When we define the eleven dimensional vector $\bm{c}^{\dag}_{\bm{k}_i} = (\bm{d}^{\dag}_{\bm{k}_i} \: \bm{p}^{\dag}_{\bm{k}_i})$ 
in $\bar{\bm{c}}^{\dag}_{\bm{k}} = (\, \bm{c}^{\dag}_{\bm{k}} \; \bm{c}^{\dag}_{\bm{k}_1} \; \bm{c}^{\dag}_{\bm{k}_2} \; \bm{c}^{\dag}_{\bm{k}_3})$, 
the Hamiltonian of Eq.~(\ref{app_icmf_eq7}) is summarized as
\begin{align}
\mathcal{H}_{cdw} ^{ex}
= \sum_{\bm{k} \in {\rm RBZ}}  \bar{\bm{c}}^{\dag}_{\bm{k}} \hat{\mathcal{H}}^{ex}_{\bm{k}} \bar{\bm{c}}_{\bm{k}}
\label{app_icmf_eq8}
\end{align}
with the 44$\times$44 matrix 
 \begin{widetext}
 \begin{align}
\hat{\mathcal{H}}^{ex}_{\bm{k}} = 
\left(
\begin{array}{c c | c c | c c | c c}
\hat{0}   &   \hat{0}   &   \hat{0}   &   \hat{\Delta}(\bm{k},\bm{q}_1)   &   \hat{0}   &   \hat{\Delta}(\bm{k},\bm{q}_2) & \hat{0}   &   \hat{\Delta}(\bm{k},\bm{q}_3)   \\
\hat{0}   &   \hat{0}   &   \hat{\Delta}^{\dag}(\bm{k}_1,\bm{q}_1)   &   \hat{0}   &   \hat{\Delta}^{\dag}(\bm{k}_2,\bm{q}_2)   &   \hat{0}   &   \hat{\Delta}^{\dag}(\bm{k}_3,\bm{q}_3)   &   \hat{0}   \\
\hline
\hat{0}   &   \hat{\Delta}(\bm{k}_1,\bm{q}_1)   &   \hat{0}   &   \hat{0}   &   \hat{0}   &   \hat{\Delta}(\bm{k}_1,\bm{q}_3)   &   \hat{0}   &   \hat{\Delta}(\bm{k}_1,\bm{q}_2)   \\
\hat{\Delta}^{\dag}(\bm{k},\bm{q}_1)   &   \hat{0}   &   \hat{0}   &   \hat{0}   &   \hat{\Delta}^{\dag}(\bm{k}_2,\bm{q}_3)   &   \hat{0}   &   \hat{\Delta}^{\dag}(\bm{k}_3,\bm{q}_2)   &   \hat{0}   \\ 
\hline           
\hat{0}   &   \hat{\Delta}(\bm{k}_2,\bm{q}_2)   &   \hat{0}   &   \hat{\Delta}(\bm{k}_2,\bm{q}_3)   &   \hat{0}   &   \hat{0}   &   \hat{0}   &   \hat{\Delta}(\bm{k}_2,\bm{q}_1)   \\
\hat{\Delta}^{\dag}(\bm{k},\bm{q}_2)   &   \hat{0}   &   \hat{\Delta}^{\dag}(\bm{k}_1,\bm{q}_3)   &   \hat{0}   &   \hat{0}   &   \hat{0}   &   \hat{\Delta}^{\dag}(\bm{k}_3,\bm{q}_1)   &   \hat{0}   \\          
\hline                     
\hat{0}   &   \hat{\Delta}(\bm{k}_3,\bm{q}_3)   &   \hat{0}   &   \hat{\Delta}(\bm{k}_3,\bm{q}_2)   &   \hat{0}   &   \hat{\Delta}(\bm{k}_3,\bm{q}_1)   &   \hat{0}   &   \hat{0}   \\            
\hat{\Delta}^{\dag}(\bm{k},\bm{q}_3)   &   \hat{0}   &   \hat{\Delta}^{\dag}(\bm{k}_1,\bm{q}_2)   &   \hat{0}   &   \hat{\Delta}^{\dag}(\bm{k}_2,\bm{q}_1)   &   \hat{0}   &   \hat{0}   &   \hat{0}   \\                           
\end{array}
\right). 
\label{app_icmf_eq9}
\end{align}
 \end{widetext}
 
In the same way, we need to introduce the RBZ for the order parameter 
$\Delta^{dp}_{\ell, \nu m}(\bm{k},\bm{q}_j)$ in Eq.~(\ref{app_icmf_eq3}).  
In the RBZ, the order parameter $\Delta^{dp}_{\ell, \nu m}(\bm{k}_i,\bm{q}_j)$ in 
$\hat{\mathcal{H}}^{ex}_{\bm{k}}$ is given by  
\begin{align}
\Delta^{dp}_{\ell, \nu m}(\bm{k}_i,\bm{q}_j)
= - \frac{1}{N}   \sum_{\bm{k}' \in {\rm RBZ}} \sum_{i'=0}^{3}  V^{dp}_{\ell, \nu m}(\bm{k}_i -\bm{k}'_{i'} ) 
\notag \\
\times \braket{p^{\dag}_{\bm{k}'-\bm{q}_{i'}-\bm{q}_j,\nu m} d_{\bm{k}' -\bm{q}_{i'},\ell}}. 
\label{app_icmf_eq10}
\end{align}
Once $\hat{\mathcal{H}}^{ex}_{\bm{k}}$ is diagonalized in the RBZ, the annihilation (creation) 
operator of the $\mu\ell$ atomic orbital is given by the unitary transformation
\begin{align}
c^{(\dag)}_{\bm{k}-\bm{q}_j,\mu\ell} = \sum_{a} u^{(*)}_{\bm{q}_j\mu\ell,a}(\bm{k}) \gamma^{(\dag)}_{\bm{k},a}, 
\label{app_icmf_eq11}
\end{align}
where $\gamma^{(\dag)}_{\bm{k},a}$ is the annihilation (creation) operator of the electron 
in the band $\varepsilon_{\bm{k},a}$, and $u^{(*)}_{\bm{q}_j\mu\ell,a}(\bm{k})$ is the matrix element 
in the transformation matrix $\hat{U}$ ($\hat{U}^{\dag}$) between the atomic orbital $\mu\ell$ 
with $\bm{q}_j$ and band index $a$.  
Using this transformation, the order parameter in Eq.~(\ref{app_icmf_eq10}) becomes 
\begin{align}
\Delta^{dp}_{\ell, \nu m}(\bm{k}_i,\bm{q}_j)
= - \frac{1}{N}   \sum_{\bm{k}' \in {\rm RBZ}} \sum_{i'=0}^{3} \sum_{a}
  V^{dp}_{\ell, \nu m}(\bm{k}_i-\bm{k}'_{i'} )
\notag \\
\times u^{*}_{\bm{q}_{i'}+\bm{q}_j p(\nu) m,a}(\bm{k}') u_{\bm{q}_{i'} d \ell,a}(\bm{k}') f(\varepsilon_{\bm{k}',a}),
\label{app_icmf_eq12} 
\end{align}
where we write Ti atom as $d$ and Se($\nu$) atom as $p(\nu)$ in $u_{\bm{q}_j\mu\ell,a}(\bm{k})$, 
and $f(\varepsilon_{\bm{k},a})=\braket{\gamma^{\dag}_{\bm{k},a} \gamma_{\bm{k},a}}$ is 
the Fermi distribution function. 
Equation~(\ref{app_icmf_eq12}) corresponds to the gap equation of the excitonic order.  The order parameter 
$\hat{\Delta}(\bm{k}_i,\bm{q}_j)$ is optimized self-consistently.  
Finally, the energy term $E^{ex}_{0}$ in the RBZ  is given by
\begin{align}
E^{ex}_{0} 
= - 2 \sum_{\bm{k}\in {\rm RBZ}}\sum_{i=0}^{3} \sum_{j=1}^{3}  \sum_{\ell, \nu m} \sum_a 
\Delta^{dp}_{\ell, \nu m}(\bm{k}_i,\bm{q}_j) 
\notag \\
\times u^{*}_{\bm{q}_i d \ell,a}(\bm{k}) u_{\bm{q}_i+\bm{q}_j p(\nu) m,a}(\bm{k})  f(\varepsilon_{\bm{k},a}). 
\label{app_icmf_eq13} 
\end{align} 
Note that the prefactor 2  in Eq.~(\ref{app_icmf_eq13}) is for the spin degrees of freedom.

%%%%%%%%%%%%%%%%%%%%
%%%%%%%%%%%%%%%%%%%%
\section{Single-Particle Spectrum} \label{app_spec}

Here, we introduce the single-particle excitation spectrum $A(\bm{k},\omega)$ 
in the triple-$\bm{q}$ CDW state.  
The single-particle spectrum $A(\bm{k},\omega)$ is given by the sum of the spectra 
over the atomic orbitals $\mu\ell$ as 
\begin{align}
A(\bm{k},\omega) = - \frac{1}{\pi} \sum_{\mu\ell}  {\rm Im}\: G_{\mu\ell} (\bm{k},\omega) 
\label{app_spec_eq1} 
\end{align}
with the $\mu\ell$ component given as 
\begin{align}
G_{\mu\ell} (\bm{k},\omega) = -i \int^{\infty}_0 dt e^{i(\omega+i\eta)t} \braket{\{ c_{\bm{k},\mu\ell}(t), c^{\dag}_{\bm{k},\mu\ell} \} }, 
\label{app_spec_eq2} 
\end{align}
where 
$c_{\bm{k},\mu\ell}(t) $ is the Heisenberg representation of $c_{\bm{k},\mu\ell}$, $\{A,B\}=AB+BA$, and   
$\eta \rightarrow 0^{+}$.  The finite $\eta$ value corresponds to the broadening factor of the spectrum.  

When the Hamiltonian is diagonalized in the RBZ for the triple-$\bm{q}$ CDW state, 
the annihilation (creation) operators of the component $\mu\ell$ are given by 
$c^{(\dag)}_{\bm{k},\mu\ell} = \sum_{a} u^{(*)}_{\bm{q}_0\mu\ell,a}(\bm{k}) \gamma^{(\dag)}_{\bm{k},a}$. 
Note that the wave-vector $\bm{k}$ in $A(\bm{k},\omega)$ is defined in the unfolded original BZ and hence 
we only consider the $\bm{q}_0$ components in the $u_{\bm{q}_j\mu\ell,a}$.  
By the transformation of the operators, the single-particle Green's function $G_{\mu\ell} (\bm{k},\omega)$ becomes 
\begin{align}
G_{\mu\ell} (\bm{k},\omega) =& -i \sum_{a} |u_{\bm{q}_0\mu\ell,a}(\bm{k})|^{2} \notag \\
&\times \int^{\infty}_0 dt e^{i(\omega+i\eta)t} \braket{\{ \gamma_{\bm{k},a}(t), \gamma^{\dag}_{\bm{k},a} \} }.
\label{app_spec_eq3} 
\end{align}
In the one-body approximation, the integral part in the Green's function of Eq.~(\ref{app_spec_eq3}) is given by 
\begin{align}
-i  \int^{\infty}_0 dt e^{i(\omega+i\eta)t} \braket{\{ \gamma_{\bm{k},a}(t), \gamma^{\dag}_{\bm{k},a} \} }
= \frac{1}{\omega \! - \! \varepsilon_{\bm{k},a} \! + \! i\eta} 
\label{app_spec_eq4} 
\end{align}
and the Green's function $G_{\mu\ell} (\bm{k},\omega)$ is 
\begin{align}
G_{\mu\ell} (\bm{k},\omega) 
= \sum_{a}  \frac{|u_{\bm{q}_0\mu\ell,a}(\bm{k})|^{2}}{\omega-\varepsilon_{\bm{k},a}+i\eta} . 
\label{app_spec_eq5} 
\end{align}
From Eqs.~(\ref{app_spec_eq1}) and (\ref{app_spec_eq5}), the single-particle spectrum $A(\bm{k},\omega)$ is given by 
\begin{align}
A(\bm{k},\omega) &= - \frac{1}{\pi} \sum_{\mu\ell}  \sum_{a} {\rm Im} \left[ \frac{|u_{\bm{q}_0\mu\ell,a}(\bm{k})|^{2}}{\omega-\varepsilon_{\bm{k},a}+i \eta}  \right] 
\notag \\  
&=\sum_{\mu\ell}  \sum_{a} |u_{\bm{q}_0\mu\ell,a}(\bm{k})|^{2} \delta(\omega-\varepsilon_{\bm{k},a} ), 
\label{app_spec_eq6} 
\end{align}
where we use ${\rm Im} \left[ 1/(\omega-\varepsilon_{\bm{k},a}+i\eta)  \right] = -\pi \delta(\omega-\varepsilon_{\bm{k},a})$ 
in the limit of $\eta \rightarrow 0^{+}$ in the second equation. 
In Fig.~\ref{TiSe2_spec_fig1}, we assume a finite broadening parameter $\eta$ in the first equation of Eq.~(\ref{app_spec_eq6}).

\end{appendix}

\bibliography{TiSe2-Paper}

\end{document}